\documentclass[12pt,a4paper]{article} %{article} % Beginn der LaTeX-Datei
\pdfoutput=1
\usepackage{geometry}
\usepackage{amsmath,amssymb}  % erleichtert Mathe 
\usepackage{enumerate}% schicke Nummerierung
\usepackage{amsfonts}   % if you want the fonts
\usepackage{graphicx} % f??r Grafik-Einbindung
\usepackage{soul}

\usepackage[english]{babel}
\usepackage[T1]{fontenc}
\usepackage{lmodern}
\usepackage[utf8]{inputenc}  % f??r Unix-Systeme
\usepackage{bbm}

\usepackage{lscape} % For matrix quer

\usepackage[numbers,sort&compress]{natbib}
\usepackage{slashed}

\usepackage{axodraw4j}

\usepackage{siunitx}

\usepackage{tikz}

\usepackage{float}

\usepackage{url}

\usepackage{hyperref}
\hypersetup{
  colorlinks   = true, %Colours links instead of ugly boxes
  urlcolor     = blue, %Colour for external hyperlinks
  linkcolor    = black, %Colour of internal links
  citecolor   = black %Colour of citations
}

\graphicspath{{plot/}}

\allowdisplaybreaks
\def    \beq            {\begin{equation}}
\def    \eeq            {\end{equation}}
\def    \bea           {\begin{eqnarray}}
\def    \eea           {\end{eqnarray}}

\newcommand{\lsim}
{\;\raisebox{-.3em}{$\stackrel{\displaystyle <}{\sim}$}\;}

\newcommand\al{\alpha}

\newcommand\tb{\tan\beta}

\newcommand\ReDiag{\mathop{%
  \raise .5pt\hbox{[}%
  \widetilde{\mathrm{Re}}%
  \raise .5pt\hbox{]}}}
\newcommand\ReOffDiag{\mathop{%
  \raise .5pt\hbox{$\llbracket$}%
  \widetilde{\mathrm{Re}}%
  \raise .5pt\hbox{$\rrbracket$}}}

\newcommand\DRbar{\ensuremath{\smash{\overline{\mathrm{DR}}}}}

\newcommand\cL{{\cal L}}

\newcommand\SW{s_\mathrm{w}}
\newcommand\CW{c_\mathrm{w}}
\newcommand\MW{M_W}
\newcommand\MZ{M_Z}

\newcommand\MA{M_A}
\newcommand\MHp{M_{H^\pm}}

\newcommand\ino[1]{\tilde\chi_{#1}}

\newcommand\neu[1]{\ino{#1}^0}

\newcommand\refeq[1]{Eq.~(\ref{#1})}
\newcommand\refeqs[1]{Eqs.~(\ref{#1})}
\newcommand\refta[1]{Tab.~\ref{#1}}
\newcommand\refse[1]{Sect.~\ref{#1}}

\newcommand\citere[1]{Ref.~\cite{#1}}
\newcommand\citeres[1]{Refs.~\cite{#1}}

\newcommand\wrt{w.r.t.\ }

\newcommand{\mnSSM}{\ensuremath{\mu\nu\mathrm{SSM}}}

\newcommand{\CP}{{\cal CP}}
\newcommand{\cp}{{\CP}}

\newcommand{\tev}{\,\, \mathrm{TeV}}
\newcommand{\gev}{\,\, \mathrm{GeV}}
\newcommand{\mev}{\,\, \mathrm{MeV}}

\newcommand\FA{\texttt{FeynArts}}
\newcommand\FC{\texttt{FormCalc}}
\newcommand\LT{\texttt{LoopTools}}

\newcommand\fh{\texttt{FeynHiggs}}

\newcommand{\br}{\text{BR}}

\newcommand{\sig}{\sigma}

\def\order#1{\ensuremath{{\cal O}(#1)}}
\def\reffi#1{\mbox{Fig.~\ref{#1}}}

\def\als{\alpha_s}
\def\alt{\alpha_t}
\def\alb{\alpha_b}
\def\Ga{\Gamma}
\def\ga{\gamma}
\def\De{\Delta}
\def\de{\delta}

\definecolor{Orange}{named}{orange}
\definecolor{Purple}{named}{purple}
\definecolor{Lightblue}{cmyk}{0.9,0.1,0.1,0.3}
\definecolor{dgelborange}{cmyk}{0.,0.3,0.5, 0.}
\definecolor{Lila}{rgb}{0.5,0.,1}

\usepackage{color}

\definecolor{white}{rgb}{1,1,1}
\definecolor{darkred}{rgb}{0.3,0,0}
\definecolor{darkgreen}{rgb}{0,0.3,0}
\definecolor{darkblue}{rgb}{0,0,0.3}
\definecolor{pink}{rgb}{0.78,0.09,0.51}
\definecolor{purple}{rgb}{0.28,0.24,0.55}
\definecolor{orange}{rgb}{1,0.6,0.0}
\definecolor{grey}{rgb}{0.4,0.4,0.4}
\definecolor{aquamarine}{rgb}{0.4,0.8,0.65}

\newcommand{\ii}{\text{i}}

\newcommand{\bal}{\begin{align}}
\newcommand{\eal}{\end{align}}

\allowdisplaybreaks
\usepackage{breqn}

\begin{document}
\thispagestyle{empty}

\def\thefootnote{\fnsymbol{footnote}}

\begin{flushright}
\mbox{}
IFT--UAM/CSIC--17-118  \\
arXiv:1712.07475 [hep-ph]
\end{flushright}

\vspace{0.5cm}

\begin{center}

{\large\sc 
{\bf Precise prediction for the Higgs-Boson Masses in the 
\boldmath{$\mu\nu$}SSM}}

\vspace{1cm}

{\sc
T.~Biek\"otter$^{1,2}$%
\footnote{email: thomas.biekotter@csic.es}
, S.~Heinemeyer$^{1,3,4}$%
\footnote{email: Sven.Heinemeyer@cern.ch}%
~and C.~Mu\~noz$^{1,2}$%
\footnote{email: c.munoz@uam.es}%
}

\vspace*{.7cm}

{\sl
$^1$Instituto de F\'isica Te\'orica UAM-CSIC, 
Cantoblanco, 28049, Madrid, Spain

\vspace*{0.1cm}

$^2$Departamento de F{\'i}sica Te{\'o}rica, Universidad Aut{\'o}noma 
de Madrid (UAM), \\
Campus de Cantoblanco, 28049 Madrid, Spain

\vspace*{0.1cm}

$^3$Campus of International Excellence UAM+CSIC, 
Cantoblanco, 28049, Madrid, Spain 

\vspace*{0.1cm}

$^4$Instituto de F\'isica de Cantabria (CSIC-UC), 
39005, Santander, Spain
}

\end{center}

\vspace*{0.1cm}

\begin{abstract}
\noindent
The \mnSSM\ is a simple supersymmetric extension of the Standard Model (SM)
capable of predicting neutrino physics in agreement with experiment. In this
paper we perform the complete one-loop renormalization of the neutral scalar
sector of the \mnSSM\ with one generation of right-handed neutrinos in a
mixed on-shell/\DRbar\ scheme.
The renormalization procedure is discussed in detail, emphasizing
conceptual differences to the minimal (MSSM) and next-to-minimal (NMSSM)
supersymmetric standard model regarding the field renormalization and the
treatment of non-flavor-diagonal soft mass parameters, which have their
origin in the breaking of $R$-parity in the \mnSSM. 
We calculate the full one-loop corrections to the neutral scalar masses
of the \mnSSM. The one-loop contributions are supplemented by available
MSSM higher-order corrections. We obtain numerical results
for a SM-like Higgs boson mass consistent with experimental
bounds. We compare our results to predictions in the NMSSM to obtain a measure
for the significance of genuine \mnSSM-like contributions.
We only find minor corrections due to the smallness of the neutrino Yukawa
couplings, indicating that the
Higgs boson mass calculations in the \mnSSM\ are at the same level of
accuracy as in the NMSSM.
Finally we show that the \mnSSM\ can accomodate a Higgs boson 
that could explain an excess of $\ga\ga$ events at $\sim 96 \gev$ as reported
by CMS, as well as the $2\,\sig$ excess of $b \bar b$
events observed at LEP at a similar mass scale.
\end{abstract}

\def\thefootnote{\arabic{footnote}}
\setcounter{page}{0}
\setcounter{footnote}{0}

\newpage

%%%%%%%%%%%%%%%%%%%%%%%%%%%%%%%%%%%%%%%%%%%%%%%%%%%%%%%%%%%%%%%%%%%%%%%%%%%%%%
%%%%%%%%%%%%%%%%%%%%%%%%%%%%%%%%%%%%%%%%%%%%%%%%%%%%%%%%%%%%%%%%%%%%%%%%%%%%%%

\section{Introduction}
\label{sec:intro}
 
The spectacular  discovery of a boson  with a mass around $\sim 125 \gev$ by
the ATLAS and CMS experiments~\cite{Aad:2012tfa,Chatrchyan:2012xdj} at
CERN constitutes a milestone in the quest for understanding the physics
of electroweak symmetry breaking (EWSB). While within the present experimental
uncertainties the properties of the observed Higgs boson are compatible with
the predictions of the Standard Model (SM)~\cite{Aad:2015zhl},
many other interpretations are possible as well, in particular as a
Higgs boson of an extended Higgs sector. Consequently, any model describing 
electroweak physics needs to provide a state that can be identified
with the observed signal. 

One of the prime candidates for physics beyond the SM is supersymmetry
(SUSY), which doubles the particle degrees of freedom by predicting
two scalar partners for all SM fermions, as well as fermionic partners
to all bosons. The simplest SUSY extension is the Minimal Supersymmetric
Standard Model (MSSM)~\cite{Nilles:1983ge,Haber:1984rc}.
In contrast to the single Higgs doublet of the SM, the Higgs sector of
the MSSM contains two Higgs doublets, which in the $\cp$ conserving
case leads to a physical spectrum consisting of two $\cp$-even, one
$\cp$-odd and two charged Higgs bosons. The light (or the heavy)
$\cp$-even MSSM Higgs boson can be interpreted as the signal discovered
at $\sim 125 \gev$~\cite{Heinemeyer:2011aa}. 
   
Going beyond the MSSM, a well-motivated extension is given by
the Next-to-Minimal Supersymmetric Standard Model (NMSSM), see
e.g.~\cite{Ellwanger:2009dp,Maniatis:2009re} for reviews. In particular
the NMSSM provides a solution for the so-called ``$\mu$ problem'' by
naturally associating an adequate scale to the $\mu$ parameter appearing
in the MSSM superpotential~\cite{Ellis:1988er,Miller:2003ay}.
In the NMSSM a new singlet superfield is introduced, which only couples to the
Higgs- and sfermion-sectors, giving rise to an effective $\mu$-term,
proportional to the vacuum expectation value (vev) of the scalar singlet.
Assuming $\cp$ conservation, as we do throughout the paper, the states in the
NMSSM Higgs sector can be classified as three $\cp$-even Higgs bosons,
$h_i$ ($i = 1,2,3$), two $\cp$-odd Higgs bosons, $a_j$ ($j = 1,2$),
and the charged Higgs boson pair $H^\pm$. In addition, the SUSY
partner of the singlet Higgs (called the singlino) extends the
neutralino sector to a total of five neutralinos. In the NMSSM the
lightest but also the second lightest $\cp$-even neutral Higgs boson
can be interpreted as the signal observed at about $125~\gev$, see,
e.g., \cite{King:2012is,Domingo:2015eea}.

A natural extension of the NMSSM is the \mnSSM, in which the 
singlet superfield is interpreted as a right-handed neutrino 
superfield~\cite{LopezFogliani:2005yw,Escudero:2008jg} 
(see \citeres{Munoz:2009an,Munoz:2016vaa,Ghosh:2017yeh} for reviews). 
The \mnSSM\ is the simplest extension of the MSSM that can provide 
massive neutrinos through a see-saw mechanism at the electroweak scale.
In this paper we will focus on the \mnSSM\ with one family of 
right-handed neutrino superfields, and the case of three families will
be studied in a future publication.%
\footnote{The \mnSSM\ with three families of right-handed neutrinos
extends the $\cp$-even and $\cp$-odd scalar sector and the
neutral fermion sector by two additional
particles each, in particular
allowing a more viable reproduction of neutrino
data~\cite{LopezFogliani:2005yw,Escudero:2008jg,Ghosh:2008yh,Bartl:2009an,Fidalgo:2009dm,Ghosh:2010zi}.}
~The $\mu$ problem is solved analogously to the NMSSM by the coupling of the right-handed neutrino 
superfield to  the Higgs sector, and a trilinear coupling of the 
right-handed neutrino generates an effective Majorana mass at the 
electroweak scale. The unique feature of the \mnSSM\ is the introduction 
of a Yukawa coupling for the right-handed neutrino of the order of the
electron Yukawa coupling that induces the explicit
breaking of $R$-parity. One of the consequences is that there is no lightest stable 
SUSY particle anymore. Nevertheless, the model can still provide a dark matter
candidate with a gravitino that has a life time longer  
than the age of the observable universe~\cite{Choi:2009ng,GomezVargas:2011ph,Albert:2014hwa,Gomez-Vargas:2016ocf}. 
Since the lightest particle beyond the SM is not stable, it can 
carry electrical charge or even 
be coloured.
The explicit violation of lepton number and lepton flavor 
can modify the 
spectrum of the neutral and charged fermions in comparison to the NMSSM.
The three families of charged leptons will mix with the chargino and 
the Higgsino and form five massive charged fermions. However, the mixing will naturally be tiny since the breaking of $R$-parity is governed by the 
small neutrino Yukawa couplings. In the neutral fermion sector the three 
left-handed neutrinos mix with the right-handed neutrino and the four 
MSSM-like neutralinos. When just one family of right-handed neutrino is 
considered (as we do in this paper), the mass matrix of the
neutral fermions is of rank six, so  
just one light neutrino mass is generated at tree-level, while the other 
two light-neutrino masses will be generated by quantum corrections.
For the Higgs sector the breaking of $R$-parity has dramatic 
consequences. The three left-handed and the right-handed sneutrinos 
will mix with the doublet Higgses and form six massive $\cp$-even 
and five massive $\cp$-odd states, assuming that there is no 
$\cp$-violation. Additionally, since the vacuum
of the model is not protected anymore by lepton number, the sneutrinos 
will acquire a vev after spontaneous EWSB.
While the vev of the right-handed sneutrino can easily take values up to 
the TeV-scale, the stability of the vacuum together with the smallness 
of the neutrino Yukawa couplings force the vevs of the left-handed 
sneutrinos to be several orders of magnitude 
smaller~\cite{Escudero:2008jg,LopezFogliani:2005yw}.
As in the NMSSM, the couplings of the 
doublet-like Higgses to the gauge-singlet right-handed sneutrino provide 
additional contributions to the tree-level mass of the SM-like Higgs 
boson, relaxing the prediction of the MSSM, that it is bounded from above by the $Z$ boson mass. Still it was shown in the
NMSSM~\cite{Drechsel:2016ukp}
that a consistent 
treatment of the quantum corrections is necessary for accurate
Higgs mass predictions (see also
\citeres{Goodsell:2014bna,Goodsell:2014pla,Staub:2015aea}).
In this paper we will investigate if this is also
the case in the \mnSSM\ and if its unique couplings generate significant 
corrections to the SM-like Higgs mass, that go beyond the corrections 
arising in the NMSSM.

The experimental accuracy of the measured mass of the observed Higgs
boson has already reached the level of a precision observable, with an
uncertainty of less than $300~\mev$~\cite{Aad:2015zhl}.
In the MSSM the masses of the $\cp$-even Higgs bosons can be
predicted at lowest order in terms of two SUSY parameters
characterising the MSSM Higgs sector, e.g.\ $\tb$, the ratio of
the vevs of the two doublets, and the mass of the
$\cp$-odd Higgs boson, $\MA$, or the charged Higgs boson, $\MHp$. This
results in particular in an upper bound on the mass of the light
$\cp$-even Higgs boson given by the $Z$-boson mass.
However, these relations receive large higher-order corrections. 
Beyond the one-loop level, the dominant two-loop corrections of
$\order{\alt\als}$~\cite{Heinemeyer:1998jw,Heinemeyer:1998kz,Heinemeyer:1998np,Zhang:1998bm,Espinosa:1999zm,Degrassi:2001yf}
and \order{\alt^2}~\cite{Espinosa:2000df,Brignole:2001jy} as well as
the corresponding corrections of 
$\order{\alb\als}$~\cite{Brignole:2002bz,Heinemeyer:2004xw} and
\order{\alt\alb}~\cite{Brignole:2002bz} are known since more than a
decade. (Here we use $\al_f = (Y^f)^2/(4\pi)$, with $Y^f$ denoting the
fermion Yukawa coupling.) These corrections, together with a
resummation of leading and subleading logarithms from the top/scalar
top sector~\cite{Hahn:2013ria} (see also~\cite{Draper:2013oza,Lee:2015uza} 
for more details on this type of approach),
a resummation of leading contributions from the bottom/scalar bottom
sector~\cite{Brignole:2002bz,Heinemeyer:2004xw,Hempfling:1993kv,Hall:1993gn,Carena:1994bv,Carena:1999py}
(see also~\cite{Noth:2008tw, Noth:2010jy}) and momentum-dependent two-loop
contributions~\cite{Borowka:2014wla, Borowka:2015ura} (see
also~\cite{Degrassi:2014pfa}) are included in the public
code~\fh~\cite{Heinemeyer:1998yj,Hahn:2009zz,Heinemeyer:1998np,Degrassi:2002fi,Frank:2006yh,Hahn:2013ria,Bahl:2016brp,Bahl:2017aev,feynhiggs-www}. 
A (nearly) full two-loop EP calculation, including even the leading
three-loop corrections, has also been
published~\cite{Martin:2005eg,Martin:2007pg}, which is, however, not
publicly available as a computer code. Furthermore, another leading
three-loop calculation of \order{\alt\als^2}, depending on the various
SUSY mass hierarchies, has been
performed~\cite{Harlander:2008ju,Kant:2010tf}, resulting in the code 
{\tt H3m} and is now available as a stand-alone
  code~\cite{Harlander:2017kuc}. 
The theoretical uncertainty on the lightest
$\cp$-even Higgs-boson mass within the MSSM from unknown higher-order
contributions is still at the level of about $2-3~\gev$ for scalar top
masses at the TeV-scale, where the actual uncertainty
depends on the considered parameter region~\cite{Degrassi:2002fi,Heinemeyer:2004gx,Hahn:2013ria,Buchmueller:2013psa}.

In the NMSSM the status of the higher-order corrections to the Higgs-boson
masses (and mixings) is the following. 
Full one-loop calculations including the momentum dependence have been
performed in the \DRbar\ renormalization scheme in 
\citere{Degrassi:2009yq,Staub:2010ty}, or in a mixed on-shell (OS)-\DRbar\ 
scheme in \citere{Ender:2011qh,Graf:2012hh,Drechsel:2016jdg}.
Two-loop corrections of \order{\alt\als,\alt^2} have been included in the
NMSSM in the leading logarithmic approximation (LLA) in
\citeres{Yeghian:1999kr,Ellwanger:1999ji}.  
In the EP approach at the two-loop level, the dominant 
\order{\alt\als, \alb\als} in the \DRbar\ scheme became available in 
\citere{Degrassi:2009yq}. The two-loop corrections involving only
superpotential couplings such as Yukawa and singlet interactions were given in
\cite{Goodsell:2014pla}. A two-loop calculation of the \order{\alt\als}
corrections with the top/stop sector renormalized in the OS scheme or in the
\DRbar\ scheme were provided in \citere{Muhlleitner:2014vsa}.
A consistent combination of a full one-loop calculation with all corrections
beyond one-loop in the MSSM approximation was given in
\citere{Drechsel:2016jdg}, which is included in the (private) version of
\fh\ for the NMSSM. 
A detailed comparison of the various higher-order corrections up to the
two-loop level involving a \DRbar\ renormalization was performed in
\citere{Staub:2015aea}, and involving an OS renormalization of the top/stop
sector for the \order{\alt\als} corrections in \citere{Drechsel:2016htw}.
Accordingly, at present the theoretical uncertainties from
unknown higher-order corrections in the NMSSM are expected to be still
larger than for the MSSM.

In this paper we go one step beyond and investigate the scalar sector of
the \mnSSM, containing (mixtures of) Higgs bosons and scalar neutrinos.
As a first step we present the renormalization at the one-loop level of the
neutral scalar sector in detail. Here a crucial point is that the NMSSM
part of the \mnSSM\ is treated exactly in the same way as in
\citere{Drechsel:2016jdg}. Consequently, differences (at the one-loop
level) appearing for, e.g., mass relations or couplings can be directly
attributed to the richer structure of the \mnSSM. As for the NMSSM in
\citere{Drechsel:2016jdg}, the full one-loop calculation is supplemented
with higher-order corrections in the MSSM limit (as provided by \fh~\cite{Heinemeyer:1998yj,Hahn:2009zz,Heinemeyer:1998np,Degrassi:2002fi,Frank:2006yh,Hahn:2013ria,Bahl:2016brp,Bahl:2017aev,feynhiggs-www}).\footnote{A
corresponding calculation using a pure \DRbar\ renormalization could
in principle be performed using \texttt{SARAH} and
\texttt{SPheno}~\cite{Goodsell:2014bna}.}
In our numerical analysis we evaluate several ``representative''
scenarios using the full one-loop results together with the MSSM-type
higher-order contributions. Differences found w.r.t.\ the NMSSM can be
interpreted in a two-fold way. On the one hand, if non-negligible
differences are found, they might serve as a probe to distinguish the
two models experimentally. On the other hand, they indicate the level of
theoretical uncertainties of the Higgs-boson/scalar neutrino mass
calculation in the 
\mnSSM, which should be brought to the same level of accuracy as in the
(N)MSSM. 

The paper is organized as follows. In \refse{sec:mnSSM} we describe
the \mnSSM, including the details for all sectors relevant in this paper.
The full one-loop renormalization of the neutral scalar potential 
is presented in \refse{sec:renopot}. 
We will establish a convenient set of free parameters and fix their 
counterterms in a mixed OS-\DRbar\ scheme. The counterterms are 
calculated and applied in the renormalized $\cp$-even and
$\cp$-odd one-loop scalar
self-energies in \refse{sec:getmasses}. 
In this work we focus on the application to the renormalized
$\cp$-even
self-energies, but the calculation of the renormalized $\cp$-odd
ones constitutes a good additional test for the counterterms. 
We also describe the 
incorporation of higher-order contributions taken over from the MSSM.
Our numerical analysis, including an analysis of differences w.r.t.\ 
NMSSM, is presented in \refse{sec:numanal}. 
We conclude in section \refse{sec:concl}.

%%%%%%%%%%%%%%%%%%%%%%%%%%%%%%%%%%%%%%%%%%%%%%%%%%%%%%%%%%%%%%%%%%%%%%%%%%%%%%%
%%%%%%%%%%%%%%%%%%%%%%%%%%%%%%%%%%%%%%%%%%%%%%%%%%%%%%%%%%%%%%%%%%%%%%%%%%%%%%%

\section{\protect\boldmath The model: \mnSSM\ with one generation of right handed neutrinos.}
\label{sec:mnSSM}

In the three-family notation of the \mnSSM\ with one generation of
right-handed neutrinos the superpotential is written as
\begin{align}\label{superpotential}
W = & \;
\epsilon_{ab} \left(
Y^e_{ij} \, \hat H_d^a\, \hat L^b_i \, \hat e_j^c +
Y^d_{ij} \, \hat H_d^a\, \hat Q^{b}_{i} \, \hat d_{j}^{c} 
+
Y^u_{ij} \, \hat H_u^b\, \hat Q^{a}_{i} \, \hat u_{j}^{c}
\right)
\nonumber\\
&+   
\epsilon_{ab} \left(
Y^{\nu}_{i} \, \hat H_u^b\, \hat L^a_i \, \hat \nu^c 
-
%\epsilon_{ab}
\lambda \, \hat \nu^c\, \hat H_u^b \hat H_d^a
\right)
+
\frac{1}{3}
\kappa 
\hat \nu^c\hat \nu^c\hat \nu^c
\ .
\end{align}
where $\hat H_d^T=(\hat H_d^0, \hat H_d^-)$ and 
$\hat H_u^T=(\hat H_u^+, \hat H_u^0)$ are the MSSM-like doublet Higgs
superfields, $\hat Q_i^T=(\hat u_i, \hat d_i)$ and 
$\hat L_i^T=(\hat \nu_i, \hat e_i)$ are the left-chiral
quark and lepton superfield doublets,
and $\hat u_{j}^{c}$, $\hat d_{j}^{c}$, $\hat e_j^c$ and $\hat{\nu}^c$ are
 the right-chiral quark and lepton superfields.
$i$ and $j$ are family indices running from one to three and 
$a,b=1,2$ are indices of the fundamental representation of SU(2) 
with $\epsilon_{ab}$ the totally antisymmetric tensor and 
$\varepsilon_{12}=1$. The colour indices are undisplayed. 
$Y^u$, $Y^d$ and $Y^e$ are the usual Yukawa couplings
also present in the MSSM. 
The right-handed neutrino is a gauge singlet, which permits us to 
write 
the gauge-invariant  
trilinear self coupling $\kappa$ and the trilinear 
coupling with the Higgs doublets $\lambda$ in the second row, 
which are analogues to the couplings of the singlet in the 
superpotential of the trilinear NMSSM.
The $\mu$-term is generated dynamically after the spontaneous EWSB,
when the right-handed sneutrino obtains a vev.
The $\kappa$-term forbids a global U(1) symmetry and we
avoid the existence of a Goldstone boson in the $\cp$-even sector. 
The remarkable difference to the
NMSSM is the additional Yukawa coupling $Y^\nu_i$, which induces
explicit breaking of $R$-parity through the $\lambda$- and
$\kappa$-term,
and which justifies the interpretation of the singlet
superfield as a right-handed neutrino superfield.
It should be pointed out
that in this case lepton number is not conserved anymore, and also the
flavor symmetry in the leptonic sector is broken. 
A more complete motivation of this superpotential can be found in
\citere{LopezFogliani:2005yw,Escudero:2008jg,Ghosh:2017yeh}. 

Working in the framework of low-energy SUSY the
corresponding soft SUSY-breaking Lagrangian can be written as
\bea
-\mathcal{L}_{\text{soft}}  =&&
\epsilon_{ab} \left(
T^e_{ij} \, H_d^a  \, \widetilde L^b_{iL}  \, \widetilde e_{jR}^* +
T^d_{ij} \, H_d^a\,   \widetilde Q^b_{iL} \, \widetilde d_{jR}^{*} 
+
T^u_{ij} \,  H_u^b \widetilde Q^a_{iL} \widetilde u_{jR}^* 
+ \text{h.c.}
\right)
\nonumber \\
&+&
\epsilon_{ab} \left(
T^{\nu}_{i} \, H_u^b \, \widetilde L^a_{iL} \widetilde \nu_{R}^* 
- 
T^{\lambda} \, \widetilde \nu_{R}^*
\, H_d^a  H_u^b
+ \frac{1}{3} T^{\kappa} \, \widetilde \nu_{R}^*
\widetilde \nu_{R}^*
\widetilde \nu_{R}^*
\
+ \text{h.c.}\right)
\nonumber \\
&+& 
\left(m_{\widetilde{Q}_L}^2\right)_{ij}   
\widetilde{Q}_{iL}^{a*}
\widetilde{Q}^a_{jL}
+\left(m_{\widetilde{u}_R}^{2}\right)_{ij} \widetilde{u}_{iR}^*  
\widetilde u_{jR}
+ \left(m_{\widetilde{d}_R}^2\right)_{ij}  \widetilde{d}_{iR}^*  
\widetilde d_{jR}
+
\left(m_{\widetilde{L}_L}^2\right)_{ij}    
\widetilde{L}_{iL}^{a*}  
\widetilde{L}^a_{jL}
\nonumber\\
&+&
\left( m_{H_d\widetilde{L}_L}^2 \right)_{i}H_d^{a*} \widetilde{L}_{iL}^a
+
m_{\widetilde{\nu}_R}^2 \widetilde{\nu}_{R}^*
\widetilde\nu_{R} 
+
\left(m_{\widetilde{e}_R}^2\right)_{ij}  \widetilde{e}_{iR}^* 
\widetilde e_{jR}
+ 
m_{H_d}^2 {H^a_d}^*
H^a_d + m_{H_u}^2 {H^a_u}^*
H^a_u
\nonumber \\
&+&  \frac{1}{2}\, \left(M_3\, {\widetilde g}\, {\widetilde g}
+
M_2\, {\widetilde{W}}\, {\widetilde{W}}
+M_1\, {\widetilde B}^0 \, {\widetilde B}^0 + \text{h.c.} \right)\ ,
\label{2:Vsoft}
\eea
In the first four lines
the fields denote the scalar component of the
corresponding superfields.
In the last line the fields denote the fermionic
superpartners of the gauge bosons.
The scalar trilinear parameters 
$T^{e,\nu,d,u,\lambda,\kappa}$ correspond to the trilinear couplings 
in the superpotential. The soft mass parameters 
$m_{\widetilde{Q}_L,\widetilde{u}_R,\widetilde{d}_R,
\widetilde{L}_L,\widetilde{e}_R}^2$ are hermitian $3\times 3$ matrices in family space. 
$m_{H_d,H_u,\widetilde{\nu}_R}^2$ are the soft masses
of the doublet Higgs fields and the right-handed sneutrino, and 
$ m_{H_d\widetilde{L}_L}^2$ is a 3-dimensional
vector in family space allowed
by gauge symmetries since the left-handed lepton fields and the down-type 
Higgs field share the same quantum numbers. In the last row 
the parameters $M_{3,2,1}$ define Majorana masses for the gluino, wino 
and bino, where the summation
over the gauge-group indices in the 
adjoint representation
is undisplayed. While all the soft parameters except
$m_{H_d}^2$, $m_{H_u}^2$ and $m_{\widetilde{\nu}_R}^2$ can 
in general be complex,
they are assumed to be real in the following to avoid 
$\cp$-violation.
Additionally, we will neglect flavor mixing at tree-level in the squark
and the quark sector, so the soft masses will be diagonal and we write
$m_{\widetilde{Q}_{iL} }^2$, $m_{\widetilde{u}_{iR}}^2$ and
$m_{\widetilde{d}_{iR}}^2$, as well as for the soft trilinears
$T^u_i=A^u_i
Y^u_i$, $T^d_i=A^d_i Y^d_i$, where the summation convention on repeated
indices is not implied, and the quark Yukawas $Y^u_{ii}=Y^u_i$ and
$Y^d_{ii}=Y^d_i$ are diagonal. For the sleptons we define 
$T^e_{ij}=A^e_{ij}Y^e_{ij}$ and $T^\nu_{i}=A^\nu_i Y^\nu_i$, again without summation over repeated indices.

Some care has to be taken with the parameters
$(m_{\widetilde{L}_L}^2)_{ij}$
contributing to the tree-level neutral scalar potential, because
these parameters cannot be set flavor-diagonal a priori. The reason is that
during the renormalization procedure (see \refse{sec:condis})
the non-diagonal elements receive a
counterterm. Of course, the tree-level value of the non-diagonal
elements can and should be set to zero to avoid too large flavor mixing. This assures that the contributions generated by virtual corrections
will always be small.

Similarly to the off-diagonal elements of the squared
sfermion mass matrices,
the parameters $( m_{H_d\widetilde{L}_L}^2 )_{i}$ are usually
not included in the tree-level Lagrangian of the \mnSSM.
In the latter case because
they contribute to the minimization equations of the left-handed
sneutrinos and spoil the electroweak seesaw mechanism that
generates neutrino masses of the correct order of
magnitude. Theoretically, the absence of these
parameters mixing different fields
at tree level, $( m_{H_d\widetilde{L}_L}^2 )_{i}$,
$(m_{\widetilde{L}_L}^2)_{ij}$,
$(m_{\widetilde{Q}_L}^2)_{ij}$, etc., 
can be justified by the diagonal structure of the K\"ahler metric in
certain supergravity models, or when the dilaton field
is the source of SUSY breaking in string
constructions~\cite{Ghosh:2017yeh}.
Notice also that when the
down-type Higgs doublet superfield is interpreted as a fourth family
of leptons, the parameters
$m_{H_d\widetilde{L}_L}^2$ can be seen as
non-diagonal elements
of $m_{\widetilde{L}_L}^2$~\cite{Lopez-Fogliani:2017qzj}.
Nevertheless, we include them in the soft
SUSY-breaking Lagrangian in this paper,
because these terms are generated at (one-)loop level, and
in our renormalization approach we need the functional dependence
of the scalar potential on $m_{H_d\widetilde{L}_L}^2$.

After the electroweak symmetry breaking the neutral scalar fields will acquire
a vev. This includes the left- and right-handed
sneutrinos, because they are not protected by lepton number conservation as in
the MSSM and the NMSSM. We define the decomposition 
\begin{align}
\label{eq:vevdecompH}
H_d^0 =& \frac{1}{\sqrt 2} \left(H_{d}^{\mathcal{R}} + v_d + \ii\ H_{d}^{\mathcal{I}}\right)\ , \\
H^0_u =& \frac{1}{\sqrt 2} \left(H_{u}^{\mathcal{R}} + v_u +\ii\ H_{u}^{\mathcal{I}}\right)\ , \\
\widetilde{\nu}_{R} =&
      \frac{1}{\sqrt 2} \left(\widetilde{\nu}^{\mathcal{R}}_{R}+ v_{R} + \ii\ \widetilde{\nu}^{\mathcal{I}}_{R}\right), \\
\label{eq:vevdecompL}
  \widetilde{\nu}_{iL} =& \frac{1}{\sqrt 2} \left(\widetilde{\nu}_{iL}^{\mathcal{R}} 
  + v_{iL} +\ii\ \widetilde{\nu}_{iL}^{\mathcal{I}}\right)\ ,
\end{align}
which is valid assuming $\cp$-conservation, as we will do
throughout this paper.

%%%%%%%%%%%%%%%%%%%%%%%%%%%%%%%%%%%%%%%%%%%%%%%%%%%%%%%%%%%%%%%%%%%%%%%%%%%%%%%

\subsection{The \boldmath{\mnSSM} Higgs potential}

The neutral scalar potential $V_H$ of the \mnSSM\ with one generation of right-handed neutrinos is given at tree-level with all parameters chosen to be real by the soft terms and the $F$- and $D$-term contributions of the superpotential. We find
\begin{equation}
V^{(0)} = V_{\text{soft}} + V_F  +  V_D\ , 
\label{finalpotential}
\end{equation}
with
\bea
V_{\text{soft}}  =&&
%&=& 
\left(
T^{\nu}_{i} \, H_u^0\,  \widetilde \nu_{iL} \, \widetilde \nu_{R}^* 
- T^{\lambda} \, \widetilde \nu_{R}^*\, H_d^0  H_u^0
+ \frac{1}{3} T^{\kappa} \, \widetilde \nu_{R}^* \widetilde \nu_{R}^* 
\widetilde \nu_{R}^*\
+
\text{h.c.} \right)
\\
&+&
%&&+
\left(m_{\widetilde{L}_L}^2\right)_{ij} \widetilde{\nu}_{iL}^* \widetilde\nu_{jL}
% \widetilde{L}_{Li}^{a^{^*}}  \widetilde{L}^a_{Lj} 
+
\left( m_{H_d\widetilde{L}_L}^2 \right)_{i}H_d^{0*}\widetilde{\nu}_{iL}
+
m_{\widetilde{\nu}_R}^2 \widetilde{\nu}_{R}^* \widetilde\nu_{R} +
m_{H_d}^2 {H^0_d}^* H^0_d + m_{H_u}^2 {H^0_u}^* H^0_u \nonumber
\ ,
\label{akappa}
\\
\nonumber
\\
V_{F}  =&&
 \lambda^2 H^0_{d}H_d^0{^{^*}}H^0_{u}H_u^0{^{^*}}
 +
\lambda^2\tilde{\nu}^{*}_{R}\tilde{\nu}_{R}H^0_{d}H_d^0{^*}
 +
\lambda^2
\tilde{\nu}^{*}_{R}\tilde{\nu}_{R}  H^0_{u}H_u^0{^*}   
\nonumber\\                                              
&+&
\kappa^2\left(\tilde{\nu}^*_{R}\right)^2
	\left(\tilde{\nu}_{R}\right)^2
%  \nonumber\\
%   &-& 
- \left(\kappa\lambda\left(\tilde{\nu}^{*}_{R}\right)^2 H_d^{0*}H_u^{0*}                                      
 -Y^{\nu}_{i}\kappa\tilde{\nu}_{iL}\left(\tilde{\nu}_{R}\right)^2H^0_{u}
 \right.
 \nonumber\\
 &+&
 \left.
 Y^{\nu}_{i}\lambda\tilde{\nu}_{iL} H_d^{0*}H_{u}^{0*}H^0_{u}
% \nonumber \\
% &+&
+{Y^{\nu}_{i}}\lambda \tilde{\nu}_{iL}^{*}\tilde{\nu}_{R}\tilde{\nu}_{R}^* H^0_{d}
 + \text{h.c.}\right) 
\nonumber \\
 &+& 
Y^{\nu}_{i}{Y^{\nu}_{i}} \tilde{\nu}^{*}_{R}
\tilde{\nu}_{R}H^0_{u}H_u^0{^*}                                                
 +
Y^{\nu}_{i}{Y^{\nu}_{j}}\tilde{\nu}_{iL}\tilde{\nu}_{jL}^{*}\tilde{\nu}_{R}^{*}
                                  \tilde{\nu}_{R}  
 +
Y^{\nu}_{i}{Y^{\nu}_{j}}\tilde{\nu}_{i}\tilde{\nu}_{j}^* H^0_{u}H_u^{0*}\, ,
\\
\nonumber
\\
V_D  =&&
\frac{1}{8}\left(g_1^{2}+g_2^{2}\right)\left(\widetilde\nu_{iL}\widetilde{\nu}_{iL}^* 
+H^0_d {H^0_d}^* - H^0_u {H^0_u}^* \right)^{2}\, .
\label{dterms}
\eea
Using the decomposition from
\refeqs{eq:vevdecompH} - (\ref{eq:vevdecompL}) 
the linear and bilinear terms
in the fields define the tadpoles $T_{\varphi}$ and the scalar $\cp$-even
and $\cp$-odd neutral mass matrices $m_{\varphi}^2$ and
$m_{\sigma}^2$ after electroweak symmetry breaking,
\begin{align}
V_H=\cdots - T_{\varphi_i}\varphi_i + \frac{1}{2} \varphi^T m_{\varphi}^2
  \varphi + \frac{1}{2} \sigma^T m_{\sigma}^2 \sigma + \cdots \; .
\end{align}
where we collectively denote with 
$\varphi^T=(H_d^{\mathcal{R}},H_u^{\mathcal{R}},
\widetilde{\nu}_{R}^{\mathcal{R}},\widetilde{\nu}_{iL}^{\mathcal{R}})$ 
and 
$\sigma^T=(H_d^{\mathcal{I}},H_u^{\mathcal{I}},
\widetilde{\nu}_{R}^{\mathcal{I}},\widetilde{\nu}_{iL}^{\mathcal{I}})$ 
the $\cp$-even and $\cp$-odd scalar fields.
The linear terms are only allowed for $\cp$-even fields and given by:
\begin{align}
T_{H_d^{\mathcal{R}}}=&-m_{H_d}^2v_d -
\left( m_{H_d\widetilde{L}_L}^2 \right)_{i} v_{iL} -
\frac{1}{8}\left(g_1^2+g_2^2\right)v_d \left( v_d^2 + v_{iL}v_{iL} - v_u^2\right)
\notag \\
&-\frac{1}{2}\lambda\left(v_R^2+v_u^2\right)\left(\lambda v_d-v_{iL}Y^\nu_i\right)+
\frac{1}{\sqrt{2}}T^\lambda v_R v_u+
\frac{1}{2}\kappa\lambda v_R^2 v_u \; , \label{eq:tp1}\\[4pt]
T_{H_u^{\mathcal{R}}}=&-m_{H_u}^2v_u+
\frac{1}{8}\left( g_1^2+g_2^2\right)v_u\left( v_d^2+v_{iL}v_{iL}-v_u^2\right)
\notag \\
&-\frac{1}{2}\lambda^2\left( v_d^2+v_R^2\right)+
\frac{1}{\sqrt{2}}T^\lambda v_d v_R+
\lambda v_d v_u v_{iL}Y^\nu_i+
\frac{1}{2}\kappa\lambda v_d v_R^2-
\frac{1}{2}\kappa v_R^2 v_{iL}Y^\nu_i
\notag \\
&-\frac{1}{2}v_u\left( v_{iL}Y^\nu_i\right)^2-
\frac{1}{\sqrt{2}}v_R v_{iL}T^\nu_i-
\frac{1}{2}v_R^2v_u Y^\nu_i Y^\nu_i \; , \label{eq:tp2}\\[4pt]
T_{\widetilde{\nu}_R^{\mathcal{R}}}=&-m_{\widetilde{\nu}_R}^2 v_R-
\frac{1}{\sqrt{2}}T^\kappa v_R^2-\kappa^2 v_R^3+
\frac{1}{\sqrt{2}}T^\lambda v_d v_u-
\frac{1}{2}\lambda^2 v_R\left( v_d^2+v_u^2\right)
\notag \\
&+\lambda v_d v_R v_{iL}Y^\nu_i +
\kappa \lambda v_d v_R v_u-
\kappa v_R v_u v_{iL} Y^\nu_i-
\frac{1}{2}v_R\left( v_{iL}Y^\nu_i\right)^2 
\notag \\
&-\frac{1}{\sqrt{2}}v_u v_{iL}T^\nu_i-
\frac{1}{2}v_Rv_u^2 Y^\nu_i Y^\nu_i \; , \label{eq:tp3}\\[4pt]
T_{\widetilde{\nu}_{iL}^{\mathcal{R}}}=&-
\left(m_{\widetilde{L}_L}^2\right)_{ij}v_{jL}-
\left( m_{H_d\widetilde{L}_L}^2 \right)_{i}v_d-
\frac{1}{8}\left( g_1^2+g_2^2\right)v_{iL}\left( v_d^2+v_{jL}v_{jL}-v_u^2\right)
\notag \\
&+\frac{1}{2}\lambda v_d v_R^2 Y^\nu_i-
\frac{1}{\sqrt{2}}v_R v_u T^\nu_i-
\frac{1}{2}\kappa v_R^2v_uY^\nu_i+
\frac{1}{2}\lambda v_d v_u^2 Y^\nu_i
\notag \\
&-\frac{1}{2}v_R^2Y^\nu_i v_{jL}Y^\nu_j-
\frac{1}{2}v_u^2Y^\nu_i v_{jL}Y^\nu_j \label{eq:tp4} \; .
\end{align}
The tadpoles vanish in the true vacuum of the model. During the
renormalization procedure they will be treated as OS parameters,
i.e., finite corrections will be canceled by their corresponding
counterterms. This guarantees that the vacuum is
stable w.r.t.\ quantum corrections.

The bilinear terms 
\begin{align}
m_{\varphi}^2= 
  \left( \begin{array}{cccc} 
  m_{H_{d}^{\mathcal{R}}H_{d}^{\mathcal{R}}}^{2}& 
m_{H_{d}^{\mathcal{R}}H_{u}^{\mathcal{R}}}^{2} & 
m_{H_{d}^{\mathcal{R}}\widetilde{\nu}_{R}^{\mathcal{R}}}^{2} & 
m_{H_{d}^{\mathcal{R}}\widetilde{\nu}_{jL}^{\mathcal{R}}}^{2}
\\
m_{H_{u}^{\mathcal{R}}H_{d}^{\mathcal{R}}}^{2} & 
m_{H_{u}^{\mathcal{R}}H_{u}^{\mathcal{R}}}^{2}& 
m_{H_{u}^{\mathcal{R}}\widetilde{\nu}_{R}^{\mathcal{R}}}^{2}  & 
m_{H_{u}^{\mathcal{R}}\widetilde{\nu}_{jL}^{\mathcal{R}}}^{2}
\\
 m_{\widetilde{{\nu}}^{\mathcal{R}}_{R} H_{d}^{\mathcal{R}}}& 
m_{\widetilde{{\nu}}^{\mathcal{R}}_{R} H_{u}^{\mathcal{R}}}&   
m_{\widetilde{{\nu}}^{\mathcal{R}}_{R} \widetilde{{\nu}}^{\mathcal{R}}_{R}}^{2} &   
m_{\widetilde{{\nu}}^{\mathcal{R}}_{R} \widetilde{{\nu}}^{\mathcal{R}}_{jL}}^{2} 
\\
m_{\widetilde{\nu}_{iL}^{\mathcal{R}} H_{d}^{\mathcal{R}}}^{2}  & 
 m_{\widetilde{\nu}_{iL}^{\mathcal{R}} H_{u}^{\mathcal{R}}}^{2} & 
m_{\widetilde{\nu}_{iL}^{\mathcal{R}} \widetilde{\nu}^{\mathcal{R}}_{R}}^{2} &   
m_{\widetilde{\nu}_{iL}^{\mathcal{R}} \widetilde{\nu}_{jL}^{\mathcal{R}}}^{2} 
         \end{array} \right)\ ,  
\label{matrixscalar1}
\end{align}
and
\begin{align}
m^2_{\sigma}= 
  \left( \begin{array}{cccc} 
  m_{H_{d}^{\mathcal{I}}H_{d}^{\mathcal{I}}}^{2} & 
m_{H_{d}^{\mathcal{I}}H_{u}^{\mathcal{I}}}^{2} & 
m_{H_{d}^{\mathcal{I}}\widetilde{\nu}_{R}^{\mathcal{I}}}^{2} & 
m_{H_{d}^{\mathcal{I}}\widetilde{\nu}_{jL}^{\mathcal{I}}}^{2}
\\
m_{H_{u}^{\mathcal{I}}H_{d}^{\mathcal{I}}}^{2} & 
m_{H_{u}^{\mathcal{I}}H_{u}^{\mathcal{I}}}^{2}& 
m_{H_{u}^{\mathcal{I}}\widetilde{\nu}_{R}^{\mathcal{I}}}^{2}  & 
m_{H_{u}^{\mathcal{I}}\widetilde{\nu}_{jL}^{\mathcal{I}}}^{2}
\\
 m_{\widetilde{{\nu}}^{\mathcal{I}}_{R} H_{d}^{\mathcal{I}}}^2& 
m_{\widetilde{{\nu}}^{\mathcal{I}}_{R} H_{u}^{\mathcal{I}}}^2&   
m_{\widetilde{{\nu}}^{\mathcal{I}}_{R} \widetilde{{\nu}}^{\mathcal{I}}_{R}}^{2} &   
m_{\widetilde{{\nu}}^{\mathcal{I}}_{R} \widetilde{{\nu}}^{\mathcal{I}}_{jL}}^{2} 
\\
m_{\widetilde{\nu}_{iL}^{\mathcal{I}} H_{d}^{\mathcal{I}}}^{2}  & 
 m_{\widetilde{\nu}_{iL}^{\mathcal{I}} H_{u}^{\mathcal{I}}}^{2} & 
m_{\widetilde{\nu}_{iL}^{\mathcal{I}} \widetilde{\nu}^{\mathcal{I}}_{R}}^{2} &   
m_{\widetilde{\nu}_{iL}^{\mathcal{I}} \widetilde{\nu}_{jL}^{\mathcal{I}}}^{2}
         \end{array} \right)\ , 
\label{matrixscalar2} 
\end{align}
are $6\times 6$ matrices in family space whose rather lengthy entries
are given in the appendix in \refse{app:cpeven} and \refse{app:cpodd}.
We transform to the mass eigenstate basis
of the $\cp$-even scalars through a
unitary transformation defined by the matrix $U^H$,
that diagonalizes the mass matrix $m_{\varphi}^2$,
\bea
U^H
m^2_{\varphi}\ {U^H}^{^T}=
m^2_{h}
\ ,
\label{eq:scalarhiggs}
\eea
with 
\begin{equation}
\varphi = {U^H}^{^T} h\, ,
\label{physscalarhiggses}
\end{equation}
where the $h_i$ are the $\cp$-even scalar fields in the mass eigenstate basis. Without $\cp$-violation in the scalar sector the matrix $U^H$ is real. Similarly, for the $\cp$-odd scalar we define the rotation matrix $U^A$, that diagonalizes the mass matrix $m_{\sigma}^2$,
\bea
U^A
m^2_{\sigma}\ {U^A}^{^T}=
m^2_{A}
\ , \qquad \text{with } \sigma={U^A}^T A \; .
\label{eq:scalaroddhiggs}
\eea
Because of the smallness of the neutrino Yukawa couplings $Y^\nu_i$, which also implies that the left-handed sneutrino vevs $v_{iL}$ have to be small, so that the tadpole coefficients vanish at tree-level \cite{Escudero:2008jg}, the mixing of the left-handed sneutrinos with the doublet fields and the singlet will be small. 

It is a well known fact that the quantum corrections to the Higgs
potential are highly significant in supersymmetric models, see
e.g.\ \citeres{Heinemeyer:2004ms,Heinemeyer:2004gx,Djouadi:2005gj} 
for reviews. As in the NMSSM~\cite{Ellwanger:2009dp}, the upper bound on 
the lowest Higgs mass squared at tree-level is relaxed through additional
contributions from the 
singlet~\cite{Escudero:2008jg};
\begin{equation}
M_Z^2 \left( \cos^2 2\beta +
\frac{2\lambda^2}{g_1^2+g_2^2}\sin^2 2\beta \right) \; .
\end{equation}
Nevertheless, quantum corrections were 
still shown to contribute significantly especially in the prediction of 
the SM-like Higgs boson
mass~\cite{Draper:2016pys,Ellwanger:1993hn,Elliott:1993uc,Elliott:1993bs,Ender:2011qh,Drechsel:2016htw,Drechsel:2016ukp,Drechsel:2016jdg}.
In this paper we will investigate how
important the unique loop corrections of the \mnSSM\ beyond the NMSSM 
are in realistic scenarios. Before that we briefly describe the other
relevant sectors of the \mnSSM.

%%%%%%%%%%%%%%%%%%%%%%%%%%%%%%%%%%%%%%%%%%%%%%%%%%%%%%%%%%%%%%%%%%%%%%%%%%%%%%%
%%%%%%%%%%%%%%%%%%%%%%%%%%%%%%%%%%%%%%%%%%%%%%%%%%%%%%%%%%%%%%%%%%%%%%%%%%%%%%%

\subsection{Squark sector}

The numerically most important one-loop corrections to the scalar
potential are expected from the stop/top-sector, analogous to the
(N)MSSM~\cite{ELLIS199183,Ellwanger:1993hn,Elliott:1993ex,Elliott:1993uc,Elliott:1993bs,Pandita:1993tg} due to the huge Yukawa
coupling of the (scalar) top. The tree-level mass matrices of the squarks differ
slightly from the ones in the MSSM. Neglecting flavor mixing in the
squark sector, one finds for the up-type squark mass matrix
$M^{\widetilde{u}_i}$ of flavor $i$, 
\begin{align}
    M^{\widetilde{u}_i}_{11}&=m_{\widetilde{Q}_{iL}}^2 + \frac{1}{24}(3g_2^{2}-g_1^2)(v_d^2+v_{jL}v_{jL}-v_u^2)+\frac{1}{2}v_u^2{Y^{u}_i}^2
    \label{eq:usquarks11} \\
    M^{\widetilde{u}_i}_{12}&=\frac{1}{2}(\sqrt{2}A^u_i v_u + v_RY^u v_{jL}Y^\nu_j -\lambda v_d v_R) \label{eq:usquarks12} \\
    M^{\widetilde{u}_i}_{22}&=m_{\widetilde{u}_{iR}}^2+\frac{1}{6}g_1^2(v_d^2+v_{jL}v_{jL}-v_u^2)+\frac{1}{2}v_u^2{Y^u_i}^2 \; .
    \label{eq:usquarks22}
\end{align}
It should be noted that in the non-diagonal element explicitly
appear the neutrino 
Yukawa couplings. This term arises in the F-term contributions of the
squark potential through the quartic coupling of up-type quarks and one 
left-handed and the right-handed sneutrino after EWSB.
The mass eigenstates $\widetilde{u}_{i1}$ and $\widetilde{u}_{i2}$ are
obtained by the unitary transformation
\begin{equation}
\begin{pmatrix}
 \widetilde{u}_{i1} \\
 \widetilde{u}_{i2}
\end{pmatrix}
=U^{\widetilde{u}}_i
\begin{pmatrix}
 \widetilde{u}_{iL} \\
 \widetilde{u}_{iR}
\end{pmatrix}\; ,
\qquad U^{\widetilde{u}}_i{U^{\widetilde{u}}_i}^\dagger=\mathbbm{1} \; .
\end{equation}
Similarly, for the down-type squarks it is
\begin{align}
    M^{\widetilde{d}_i}_{11}&=m_{\widetilde{Q}_{iL}}^2 - \frac{1}{24}(3g_2^{2}+g_1^2)(v_d^2+v_{jL}v_{jL}-v_u^2)-\frac{1}{2}v_d^2{Y^{d}_i}^2 
    \label{eq:dsquarks11} \\
    M^{\widetilde{d}_i}_{12}&=\frac{1}{2}(\sqrt{2}A^d_i v_d -\lambda v_d v_R) \\
    M^{\widetilde{d}_i}_{22}&=m_{\widetilde{d}_{iR}}^2-\frac{1}{12}g_1^2(v_d^2+v_{jL}v_{jL}-v_u^2)+\frac{1}{2}v_d^2{Y^d_i}^2 \; .
    \label{eq:dsquarks22}
\end{align}
The mass eigenstates $\widetilde{d}_{i1}$ and $\widetilde{d}_{i2}$ are obtained by the unitary transformation
\begin{equation}
\begin{pmatrix}
 \widetilde{d}_{i1} \\
 \widetilde{d}_{i2}
\end{pmatrix}
=U^{\widetilde{d}}_i
\begin{pmatrix}
 \widetilde{d}_{iL} \\
 \widetilde{d}_{iR}
\end{pmatrix}\; ,
\qquad U^{\widetilde{d}}_i{U^{\widetilde{d}}_i}^\dagger=\mathbbm{1} \; .
\end{equation}

%%%%%%%%%%%%%%%%%%%%%%%%%%%%%%%%%%%%%%%%%%%%%%%%%%%%%%%%%%%%%%%%%%%%%%%%%%%%%%%

\subsection{Charged scalar sector}

Since $R$-parity, lepton number and lepton-flavor are broken, the six charged left- and right-handed sleptons mix with each other and with the two charged scalars from the Higgs doublets. In the basis $C^T=
({H^-_d}^*,{H^+_u},\widetilde{e}_{iL}^*,\widetilde{e}_{jR}^*)$ we find
the following mass terms in the Lagrangian:
\begin{equation}
\cL_{C} \; = \; 
-{C^*}^T {m}^2_{H^+} C\, ,
\label{matrix122}
\end{equation}
where ${m}^2_{H^+}$ assuming $\cp$ conservation is a symmetric
matrix of dimension 8,
\begin{align}
m_{H^+}^2= 
  \left( \begin{array}{cccc} 
  m_{H_{d}^-{H^{-}_d}^{*}}^{2} & m_{H_{d}^- H_{u}^+}^{2} & 
m_{{H_d^-} \widetilde{e}^*_{jL}}^2 & m_{{H_d^-} \widetilde{e}^*_{jR}}^2  \\
 m_{{H_{u}^+}^* {H_d^-}^*}^{2}  & m_{{H^{+}_u}^{*} H_{u}^{+}}^{2} & 
m_{{H_u^+}^*\widetilde{e}^*_{jL}}^{2} & m_{{H_u^+}^*\widetilde{e}^*_{jR}}^{2} \\
 m_{\widetilde{e}_{iL} {{H^{-}_d}}^{*}}^2 & m_{\widetilde{e}_{iL} H_u^+}^2 &   
m_{\widetilde{e}_{iL} \widetilde{e}_{jL}^{*}}^{2} &   
m_{\widetilde{e}_{iL} \widetilde{e}_{jR}^{*}}^{2}\\
 m_{\widetilde{e}_{iR} {{H^{-}_d}}^{*}}^2 & m_{\widetilde{e}_{iR} H_u^+}^2 & 
  m_{\widetilde{e}_{iR} \widetilde{e}_{jL}^{*}}^{2} &   
m_{\widetilde{e}_{iR} \widetilde{e}_{jR}^{*}}^{2}
         \end{array} \right)\ .  
         \label{matrixcharged2}
\end{align}
The entries are given
in appendix \ref{app:charged}.
The mass matrix is diagonalized by an orthogonal matrix $U^+$:
\bea
%\left(m^2_{S_{\alpha} S_{\beta}}\right)^{\text{diag}}
U^+
m^2_{H^+}\ {U^+}^{^T}
=
\left(m^2_{H^+}\right)^{\text{diag}}
\ ,
\label{scalarhiggs22}
\eea
where the diagonal elements of $\left(m^2_{H^+}\right)^{\text{diag}}$ are the squared masses of the mass eigenstates
\begin{equation}
H^+=U^+\; C \; ,
\end{equation}
which include the charged Goldstone boson $H^+_1=G^\pm_0$.

%%%%%%%%%%%%%%%%%%%%%%%%%%%%%%%%%%%%%%%%%%%%%%%%%%%%%%%%%%%%%%%%%%%%%%%%%%%%%%%

\subsection{Charged fermion sector}

The charged leptons mix with the charged gauginos and the charged higgsinos. Following the notation of \citere{Ghosh:2017yeh} we write the relevant part of the Lagrangian in terms of two-component spinors 
$(\chi^-)^T = \left({({e_{iL})}^{c}}^{^*}, \widetilde{W}^-, \widetilde{H}^-_d\right)$
and 
$(\chi^+)^T = \left(({e_{jR}})^c, \widetilde{W}^+, \widetilde{H}^+_u\right)$:
\begin{equation}
\cL_{\chi^\pm} \; = \; -({\chi^{-}})^T 
%\mathcal{M}_{\mathrm{n}}
%\mathcal{M}_{\chi^0} 
%\mathcal{M}_{\nu} 
{m}_{e}
\chi^+ + \mathrm{h.c.}\, .
\label{matrixcharginos0}
\end{equation}
The $5\times5$ mixing matrix $m_e$ is defined by
\begin{equation}
m_e=
\begin{pmatrix}
    \frac{v_d Y^e_{11}}{\sqrt{2}} & \frac{v_d Y^e_{12}}{\sqrt{2}} & \frac{v_d Y^e_{13}}{\sqrt{2}} & \frac{g_2 v_{1L}}{\sqrt{2}} &
    	-\frac{v_R Y^\nu_1}{\sqrt{2}} \\
    \frac{v_d Y^e_{21}}{\sqrt{2}} & \frac{v_d Y^e_{22}}{\sqrt{2}} & \frac{v_d Y^e_{23}}{\sqrt{2}} & \frac{g_2 v_{2L}}{\sqrt{2}} & 
    	-\frac{v_R Y^\nu_2}{\sqrt{2}} \\
    \frac{v_d Y^e_{31}}{\sqrt{2}} & \frac{v_d Y^e_{32}}{\sqrt{2}} & \frac{v_d Y^e_{33}}{\sqrt{2}} & \frac{g_2 v_{3L}}{\sqrt{2}} &
    	-\frac{v_R Y^\nu_3}{\sqrt{2}} \\
    0 & 0 & 0 & M_2 & \frac{g_2 v_{u}}{\sqrt{2}} \\
    -\frac{v_{iL}Y^e_{1i}}{\sqrt{2}} & -\frac{v_{iL}Y^e_{2i}}{\sqrt{2}} & -\frac{v_{iL}Y^e_{3i}}{\sqrt{2}} & \frac{g_2 v_{d}}{\sqrt{2}} & \frac{\lambda v_R}{\sqrt{2}}
\end{pmatrix} \; .
\end{equation}
It is diagonalized by two unitary matrices $U^e_L$ and $U^e_R$:
\begin{equation}
{U_R^e}^{^*} m_{e} {U_L^e}^{^\dagger} = m_{e}^{\text{diag}}
\, ,
\label{diagmatrixneutralinosn}
\end{equation}
where $m_{e}^{\text{diag}}$ contains the masses of the charged fermions in the mass eigenstate base
\begin{eqnarray}
\chi^{+} = {U_L^e}^{^\dagger} \lambda^+\, ,
\\
\chi^{-} = {U_R^e}^{^\dagger} \lambda^-\, .
\label{physcharginos}
\end{eqnarray}
The smallness of the left-handed sneutrino vevs in comparison to the doublet ones assures the decoupling of the three leptons from the Higgsino and the wino.

%%%%%%%%%%%%%%%%%%%%%%%%%%%%%%%%%%%%%%%%%%%%%%%%%%%%%%%%%%%%%%%%%%%%%%%%%%%%%%%

\subsection{Neutral fermion sector}

The three left-handed neutrinos and the right-handed neutrino mix with
the neutral Higgsinos and gauginos. Again, following
\citere{Ghosh:2017yeh} we write the relevant part of the Lagrangian in
terms of two-component spinors
$({\chi^{0}})^T=\left({(\nu_{iL})^{c}}^{^*},\widetilde B^0, 
\widetilde W^{0},\widetilde H_{d}^0,\widetilde H_{u}^0,\nu_{R}^*\right)$
as
\begin{equation}
\cL_{\chi^0} \; = \;-\frac{1}{2} ({\chi^{0}})^T 
%\mathcal{M}_{\mathrm{n}}
%\mathcal{M}_{\chi^0} 
%\mathcal{M}_{\nu} 
{m}_{\nu} 
\chi^0 + \mathrm{h.c.}\, ,
\label{matrixneutralinos}
\end{equation}
where ${m}_{\nu}$ is the $8\times 8$ symmetric mass matrix. The neutral fermion mass matrix is determined by
\begin{equation}
{m}_{\nu}=
\begin{pmatrix}
 0 & 0 & 0 & -\frac{g_1 v_{1L}}{2} & \frac{g_2 v_{1L}}{2} &
 0 & \frac{v_{R} Y^\nu_1}{\sqrt{2}} & \frac{v_{u}Y^\nu_1}{\sqrt{2}} \\
 0 & 0 & 0 & -\frac{g_1 v_{2L}}{2} & \frac{g_2 v_{2L}}{2} &
 0 & \frac{v_{R} Y^\nu_2}{\sqrt{2}} & \frac{v_{u}Y^\nu_2}{\sqrt{2}} \\
 0 & 0 & 0 & -\frac{g_1 v_{3L}}{2} & \frac{g_2 v_{3L}}{2} &
 0 & \frac{v_{R} Y^\nu_3}{\sqrt{2}} & \frac{v_{u}Y^\nu_3}{\sqrt{2}} \\
 -\frac{g_1 v_{1L}}{2} & -\frac{g_1 v_{2L}}{2} & -\frac{g_1 v_{3L}}{2} &
 M_1 & 0 & -\frac{g_1 v_{d}}{2} & \frac{g_1 v_{u}}{2} & 0 \\
  \frac{g_2 v_{1L}}{2} & \frac{g_2 v_{2L}}{2} & \frac{g_2 v_{3L}}{2} &
 0 & M_2 & \frac{g_2 v_{d}}{2} & -\frac{g_2 v_{u}}{2} & 0 \\
 0 & 0 & 0 & -\frac{g_1 v_{d}}{2} & \frac{g_2 v_{d}}{2} & 0 &
 -\frac{\lambda v_{R}}{\sqrt{2}} & -\frac{\lambda v_{u}}{\sqrt{2}} \\
 \frac{v_R Y^\nu_1}{\sqrt{2}} & \frac{v_R Y^\nu_2}{\sqrt{2}} &
 \frac{v_R Y^\nu_3}{\sqrt{2}} & \frac{g_1 v_u}{2} & -\frac{g_2 v_u}{2} &
 -\frac{\lambda v_R}{\sqrt{2}} & 0 &
 \frac{-\lambda v_d + v_{kL}Y^\nu_k}{\sqrt{2}} \\
 \frac{v_u Y^\nu_1}{\sqrt{2}} & \frac{v_u Y^\nu_2}{\sqrt{2}} &
 \frac{v_u Y^\nu_3}{\sqrt{2}} & 0 & 0 & -\frac{\lambda v_u}{\sqrt{2}} &
 \frac{-\lambda v_d +v_{iL}Y^\nu_i}{\sqrt{2}} &
 \sqrt{2}\kappa v_R
\end{pmatrix}~.
\end{equation}
Because of the Majorana nature of the neutral fermions we can 
diagonalize ${m}_{\nu}$ with the help of just a single - but complex - 
unitary matrix $U^V$,
\begin{equation}
{U^V}^{^*} m_{\nu}\ {U^V}^{^\dagger} = m_{\nu}^{\text{diag}}\, ,
\label{diagmatrixneutralinos}
\end{equation}
with
\begin{equation}
\chi^{0} = {U^V}^{^\dagger} \lambda^0\, ,
\label{physneutralinos}
\end{equation}
where $\lambda^0$ are the two-component spinors in the mass basis. The eigenvalues of the diagonalized mass matrix $m_{\nu}^{\text{diag}}$ are the masses of the neutral fermions in the mass eigenstate basis. It turns out that the matrix $m_\nu$ is of rank six, so it can only generate a single neutrino mass at tree-level.\footnote{Including three generations of right-handed neutrinos, three light tree-level neutrino masses are generated.} The remaining two light neutrino masses can be generated by loop-effects.

%%%%%%%%%%%%%%%%%%%%%%%%%%%%%%%%%%%%%%%%%%%%%%%%%%%%%%%%%%%%%%%%%%%%%%%%%%%%%%%

\section{Renormalization of the Higgs potential at One-Loop}
\label{sec:renopot}

The first step in renormalizing the neutral scalar potential is to choose
the set of free parameters. These free parameters will receive a counter
term fixed by consistent renormalization conditions to cancel all
ultraviolet divergences that are produced by higher-order corrections. 

At tree-level the relevant part of the Higgs potential $V_H$ is given by
the tadpole coefficients \refeqs{eq:tp1}-(\ref{eq:tp4}) and the $\cp$-even and
$\cp$-odd mass matrix elements in \refeqs{matrixscalar1}
and (\ref{matrixscalar2}). The following
parameters appear in the Higgs potential:
\begin{itemize}
	\setlength\itemsep{0.25em}
	\renewcommand\labelitemi{--}
	\item Scalar soft masses: $m_{H_d}^2$, $m_{H_u}^2$, $m_{\widetilde{\nu}_R}^2$,
			$\left( m_{\widetilde{L}_L}^2 \right)_{ij}$,
			$\left( m_{H_d\widetilde{L}_L}^2 \right)_{i}$
			$\quad$ (12 parameters)
	\item Vacuum expectation values: $v_d$, $v_u$, $v_R$, $v_{iL}$
			$\quad$ (6 parameters)
	\item Gauge couplings: $g_1$, $g_2$ $\quad$ (2 parameters)
	\item Superpotential parameters: $\lambda$, $\kappa$, $Y^\nu_i$
			$\quad$ (5 parameters)
	\item Soft trilinear couplings: $T^\lambda$, $T^\kappa$, $T^\nu_i$
			$\quad$ (5 parameters)
\end{itemize}
The complexity of the \mnSSM\ Higgs scalar
sector becomes evident when we
compare the numbers of free parameters (30) with the one in the real
MSSM (7) \cite{Frank:2006yh} and the NMSSM (12)
\cite{Drechsel:2016ukp}. While the number of free parameters is fixed,
we are free to replace some of the parameters by physical parameters. We
chose to make the following replacements: 

The soft masses $m_{H_d}^2$, $m_{H_u}^2$, $m_{\widetilde{\nu}_R}^2$, and
the diagonal elements of the matrix $m_{\widetilde{L}_L}^2$ will
be replaced  
by the tadpole coefficients. The substitution is defined by
the tadpole~\refeqs{eq:tp1}-(\ref{eq:tp4})
solved for the soft mass parameters just
mentioned. This will give us the possiblity
to define the
renormalization scheme in a way that the true vacuum is not spoiled by
the higher-order corrections. The Higgs doublet vevs $v_d$ and $v_u$ will be replaced by the MSSM-like parameters
$\tan\beta$ and $v$ according to 
\begin{equation}\label{eq:tanbetadef}
\tan\beta=\frac{v_u}{v_d} \qquad \text{and} \qquad
v^2=v_d^2+v_u^2+v_{iL}v_{iL} \; .
\end{equation}
Note that the definition of $v^2$ differs from the one in the MSSM by
the term $v_{iL}v_{iL}$. This allows to maintain the relations
between $v^2$ and the gauge boson masses as they are in the
MSSM. Numerically, the difference in the definition of $v^2$ is
negligible, since the $v_{iL}$ are of the order of $10^{-4}\gev$ in
realistic scenarios. Analytically, however, maintaining the
functional form of $\tan\beta$ as it is in the (N)MSSM is convenient
to facilitate the comparison of the quantum corrections in the
\mnSSM\ and the NMSSM. In particular, we can still express the one-loop
counterterm of $\tan\beta$ without having to include the counterterms
for the left-handed sneutrino vevs.
For the vev of the
right-handed sneutrino we chose to make the same substitution as was
done in previous calculations in the NMSSM \cite{Drechsel:2016ukp} 
\begin{equation}\label{eq:mudef}
\mu=\frac{v_R \lambda}{\sqrt{2}} \; ,
\end{equation}
where we make use of the fact that when the sneutrino obtains the vev,
the $\mu$-term of the MSSM is dynamically generated.
The gauge couplings $g_1$ and $g_2$ will be replaced by the
gauge boson masses $\MW$ and $\MZ$ via the definitions
\begin{equation}\label{eq:gaguebosonmasses}
\MW^2=\frac{1}{4}g_2^2v^2 \qquad \text{and} \qquad
\MZ^2=\frac{1}{4}\left( g_1^2+g_2^2\right) v^2 \; .
\end{equation}
This is reasonable because the gauge boson masses are well measured
physical observables, so we can define them as OS parameters.
Interestingly, the mass counterterm for $\MW^2$ drops out at one-loop,
but it will contribute in the definition of the counterterm for $v^2$,
so it is not a redundant parameter.
For the soft trilinear couplings we chose to adopt the redefinitions
\begin{equation}
T^\lambda = A^\lambda \lambda \; ,  \qquad T^\kappa = A^\kappa \kappa
\; , \qquad T^\nu_i=A^\nu_i Y^\nu_i \; .
\end{equation} 
The reparametrization from the initial to the physical set of independent parameters is summarized in \refta{tab:setpara}.
\begin{table}
\centering
\begin{tabular}{c c c c c}
Soft masses & VEVs & Gauge cpl. & Superpot. & Soft trilinears \\
\hline
$m_{H_d}^2$, $m_{H_u}^2$, $m_{\widetilde{\nu}_R}^2$,
${m_{\widetilde{L}_L}^2}_{ij}$,
			${m_{H_d\widetilde{L}_L}^2}_i$ 
 & $v_d$, $v_u$, $v_R$, $v_{iL}$ &
$g_1$, $g_2$ &
$\lambda$, $\kappa$, $Y^\nu_i$ &
$T^\lambda$, $T^\kappa$, $T^\nu_i$ \\
% & & & & \\
$\downarrow$ & $\downarrow$ & $\downarrow$ & $\downarrow$ & $\downarrow$ \\
 $T_{H_d^{\mathcal{R}}}$, $T_{H_u^{\mathcal{R}}}$,
 $T_{\widetilde{\nu}_{R}^{\mathcal{R}}}$,
 $T_{\widetilde{\nu}_{iL}^{\mathcal{R}}}$,&
 $\tan\beta$, $v$, $\mu$, $v_{iL}$ &
 $\MW$, $\MZ$ &
 $\lambda$, $\kappa$, $Y^\nu_i$ &
 $A^\lambda$, $A^\kappa$, $A^\nu_i$ \\
 ${m_{\widetilde{L}_L}^2}_{i\neq j}$,
 ${m_{H_d\widetilde{L}_L}^2}_i$ & & & & 
\end{tabular}
\caption{Set of independent parameters initially entering the tree-level Higgs potential of the \mnSSM\ in the first row, and final choice of free parameters after the substitutions mentioned in the text.}
\label{tab:setpara}
\end{table}

In the following we will regard the entries of the neutral scalar mass matrix as functions of the final set of parameters,
\begin{align}
m_{\varphi}^2&=m_{\varphi}^2\left( \MZ^2, v^2, \tan\beta, \lambda, \dots \right) \; , \\
m_{\sigma}^2&=m_{\sigma}^2\left( \MZ^2, v^2, \tan\beta, \lambda, \dots \right) \; ,
\end{align}
and we define their renormalization as
\begin{align}
m_{\varphi}^2&\to m_{\varphi}^2+\delta m_{\varphi}^2 \; , \\
m_{\sigma}^2&\to m_{\sigma}^2+\delta m_{\sigma}^2 \; .
\end{align}
The mass counterterms $\delta m_{\varphi}^2$ and $\delta m_{\sigma}^2$
enter the renormalized one-loop scalar self-energies. They have to be
expressed as a linear combination of the counterterms of the independent
parameters. We define their one-loop renormalization as
\begin{equation}
  \begin{split}
T_{H_d^{\mathcal{R}}} &\to
T_{H_d^{\mathcal{R}}}+\delta T_{H_d^{\mathcal{R}}} \; , \\
T_{H_u^{\mathcal{R}}} &\to
T_{H_u^{\mathcal{R}}}+\delta T_{H_u^{\mathcal{R}}} \; , \\
T_{\widetilde{\nu}_{R}^{\mathcal{R}}} &\to
T_{\widetilde{\nu}_{R}^{\mathcal{R}}}+\delta T_{\widetilde{\nu}_{R}^{\mathcal{R}}}
\; , \\
T_{\widetilde{\nu}_{iL}^{\mathcal{R}}} &\to
T_{\widetilde{\nu}_{iL}^{\mathcal{R}}}+ \delta T_{\widetilde{\nu}_{iL}^{\mathcal{R}}}
\; , \\
{m_{\widetilde{L}_L}^2}_{i\neq j} &\to
{m_{\widetilde{L}_L}^2}_{i\neq j}+ \delta {m_{\widetilde{L}_L}^2}_{i\neq j}
\; , \\
{m_{H_d\widetilde{L}_L}^2}_i &\to
{m_{H_d\widetilde{L}_L}^2}_i+\delta {m_{H_d\widetilde{L}_L}^2}_i \; ,
  \end{split}
\qquad
  \begin{split}
\tan\beta &\to \tan\beta+\delta \tan\beta \; , \\
v^2 &\to v^2+\delta v^2 \; , \\
\mu &\to \mu +\delta \mu \; , \\
v_{iL}^2 &\to v_{iL}^2+\delta v_{iL}^2 \; , \\
\MW^2 &\to \MW^2 +\delta \MW^2 \; , \\
\MZ^2 &\to \MZ^2 +\delta \MZ^2 \; , \\
  \end{split}
 \qquad
  \begin{split}
\lambda &\to \lambda + \delta \lambda \; , \\
\kappa &\to \kappa + \delta \kappa \; , \\
Y^\nu_i &\to Y^\nu_i + \delta Y^\nu_i \; , \\
A^\lambda &\to A^\lambda+\delta A^\lambda \; , \\
A^\kappa &\to A^\kappa + \delta A^\kappa \; , \\
A^\nu_i &\to A^\nu_i + \delta A^\nu_i \; .
  \end{split}
\end{equation}
Since the \mnSSM\ is a renormalizable theory, the divergent parts of the
counterterms are fixed to cancel the UV divergences.
The finite pieces, and thus the meaning of the parameters have to be
fixed by renormalization conditions. 
We will adopt a mixed renormalization scheme, where tadpoles
and gauge boson masses are fixed OS, and the other parameters are
fixed in the $\overline{\text{DR}}$ scheme. The exact renormalization
conditions will be given in \refse{sec:condis}. The dependence of the
mass counterterms $\delta m_{\varphi}^2$ and $\delta
m_{\sigma}^2$ on the counterterms of the free parameters is given
at one-loop by 
\begin{equation}
\delta m_{\varphi}^2=\sum_{X\in\text{Free param.}} \left(
\frac{\partial}{\partial X}  m_{\varphi}^2 \right)\delta X \; , 
\qquad
\delta m_{\sigma}^2=\sum_{X\in\text{Free param.}} \left(
\frac{\partial}{\partial X}  m_{\sigma}^2 \right)\delta X \; . 
\label{eq:masscounterderive}
\end{equation}
In our calculation the mixing matrices are defined in a way to diagonalize the renormalized mass matrices, so they do not have to be renormalized, because they are defined exclusively by renormalized quantities. The expressions for the counterterms of the scalar mass matrices in the mass eigenstate basis are then simply
\begin{equation}\label{eq:masscounterrot}
\delta m_{h}^2=U^H \delta m_{\varphi}^2 {U^H}^T \; , 
\qquad
\delta m_{A}^2=U^A \delta m_{\sigma}^2{U^A}^T \; .
\end{equation}
It should be noted at this point that the counterterm matrices in the mass eigenstate basis $\delta m_{h}^2$ and $\delta m_{A}^2$ are not diagonal, as they would be in a purely OS renormalization procedure, which is often used in theories with flavor mixing \cite{Grimus:2016hmw}.

In the following chapter we will discuss the field renormalization, which is necessary to obtain finite scalar self-energies at arbitrary momentum.

%%%%%%%%%%%%%%%%%%%%%%%%%%%%%%%%%%%%%%%%%%%%%%%%%%%%%%%%%%%%%%%%%%%%%%%%%%%%%%%

\subsection{Field renormalization}

We write the renormalization of the neutral scalar-component fields as
\begin{equation}\label{eq:fieldrenodef}
\begin{pmatrix}
H_d \\ H_u \\ \widetilde{\nu}_R \\ \widetilde{\nu}_{iL}
\end{pmatrix}
\to \sqrt{Z} 
\begin{pmatrix}
H_d \\ H_u \\ \widetilde{\nu}_R \\ \widetilde{\nu}_{iL}
\end{pmatrix}=
\left( \mathbbm{1} + \frac{1}{2} \delta Z \right)
\begin{pmatrix}
H_d \\ H_u \\ \widetilde{\nu}_R \\ \widetilde{\nu}_{iL}
\end{pmatrix} \; ,
\end{equation}
where $\sqrt{Z}$ and $\delta Z$ are $6\times6$ dimensional matrices and the equal sign is valid at one-loop. It should be emphasized that in contrast to the MSSM and the NMSSM these matrices cannot be made diagonal even in the interaction basis. The reason is that the \mnSSM\ explicitly breaks lepton number and lepton flavor, so the fields $H_d$ and $\widetilde{\nu}_{iL}$ share exactly the same quantum numbers and kinetic mixing terms are already generated at one-loop order.

For the $\cp$-even and $\cp$-odd neutral scalar fields the 
definition in \refeq{eq:fieldrenodef} implies the following field
renormalization in the mass eigenstate basis: 
\begin{equation}
h \to \left( \mathbbm{1} + \frac{1}{2} \delta Z^H \right) h \; , 
\qquad
A \to \left( \mathbbm{1} + \frac{1}{2} \delta Z^A \right) A \; ,
\end{equation}
with
\begin{equation}\label{eq:fieldrenorotate}
\delta Z^H = U^H \left(\delta Z \right) {U^H}^T \; \qquad \text{and} \qquad
\delta Z^A = U^A \left(\delta Z \right) {U^A}^T \; .
\end{equation}
As renormalization conditions for the field renormalization counterterms we chose to adopt the $\overline{\text{DR}}$ scheme. We calculate the UV-divergent part of the derivative of the scalar $\cp$-even self-energies in the interaction basis and define
\begin{equation}\label{eq:drfieldcounterdef}
\delta Z_{ij}=-\left.\frac{d}{dp^2}\Sigma_{\varphi_i \varphi_j} \right|^{\rm div} \; .
\end{equation}
Here $\mbox{}^{\rm div}$ denotes taking the divergent part only,
proporional to $\De$, 
\begin{equation}
\Delta = \frac{1}{\varepsilon}-\gamma_E+\ln{4\pi} \; ,
\end{equation}
where loop integral are solved in $4-2\varepsilon$ dimensions and
$\gamma_E=0.5772\dots$ is the Euler-Mascharoni constant.
Since the field renormalization constants contribute only via divergent
parts, they do not contribute to the finite result after canceling 
divergences in the self-energies.
As regularization
scheme we chose dimensional reduction~\cite{SIEGEL1979193,CAPPER1980479}, 
which was shown to be SUSY
conserving at one-loop~\cite{Stockinger:2005gx}. In contrast to the
OS renormalization scheme our field renormalization matrices are
hermitian.
%, since we do not account for $\cp$-violation.
This holds also
true for the field renormalization in the mass eigenstate basis, because
as already mentioned the rotations in \refeq{eq:scalarhiggs} and
\refeq{eq:scalaroddhiggs} diagonalize the renormalized tree-level scalar
mass matrices, so \refeqs{eq:fieldrenorotate} do not introduce
non-hermitian parts into the field renormalization, that would have to
be canceled by a renormalization of the mixing matrices $U^H$ and $U^A$
themselves. 

In appendix \ref{app:fieldcounters} we list our field renormalization
counterterms $\delta Z_{ij}$ in terms of the divergent quantity $\De$.
Note that the field counterterms mixing the down-type Higgs and the left-handed sleptons are proportional to the neutrino Yukawa couplings $Y^\nu_i$, while the counterterms mixing different flavors of left-handed sneutrinos contain terms proportional to non-diagonal lepton Yukawa couplings $Y^e$ and  terms proportional to $Y^\nu_i Y^\nu_j$. This is why their numerical impact is negligible, but they are needed for a consistent renormalization of the scalar self-energies.

%%%%%%%%%%%%%%%%%%%%%%%%%%%%%%%%%%%%%%%%%%%%%%%%%%%%%%%%%%%%%%%%%%%%%%%%%%%%%%%

\subsection{Renormalization conditions for free parameters}
\label{sec:condis}

In this section we describe our choice for the renormalization
conditions, where we stick to the one-loop level everywhere. 
We start
with the OS conditions for the gauge boson mass parameters and the
tadpole coefficients followed by our definitions for the
\DRbar\ renormalized parameters.

The SM gauge boson masses are renormalized OS requiring
\begin{equation}
\text{Re}\left[{\hat{\Sigma}_{ZZ}}^T\left( \MZ^2 \right) \right]=0 \qquad
\text{and} \qquad
\text{Re}\left[{\hat{\Sigma}_{WW}}^T\left( \MW^2 \right) \right]=0 \; ,
\end{equation}
where $\hat{\Sigma}^T$ stands for the transverse part of the renormalized gauge boson self-energy. For their mass counterterms these conditions yield
\begin{equation}
\delta \MZ^2 = \text{Re}\left[ {\Sigma}_{ZZ}^T\left( \MZ^2 \right) \right] \qquad
\text{and} \qquad
\delta \MW^2 = \text{Re}\left[ {\Sigma}_{WW}^T\left( \MW^2 \right) \right] \; .
\end{equation}
Here the ${\Sigma}^T$ (without the hat) denote the transverse part of the unrenormalized gauge boson self-energies.

\begin{figure}[t]
  \centering
  \includegraphics[width=0.6\textwidth]{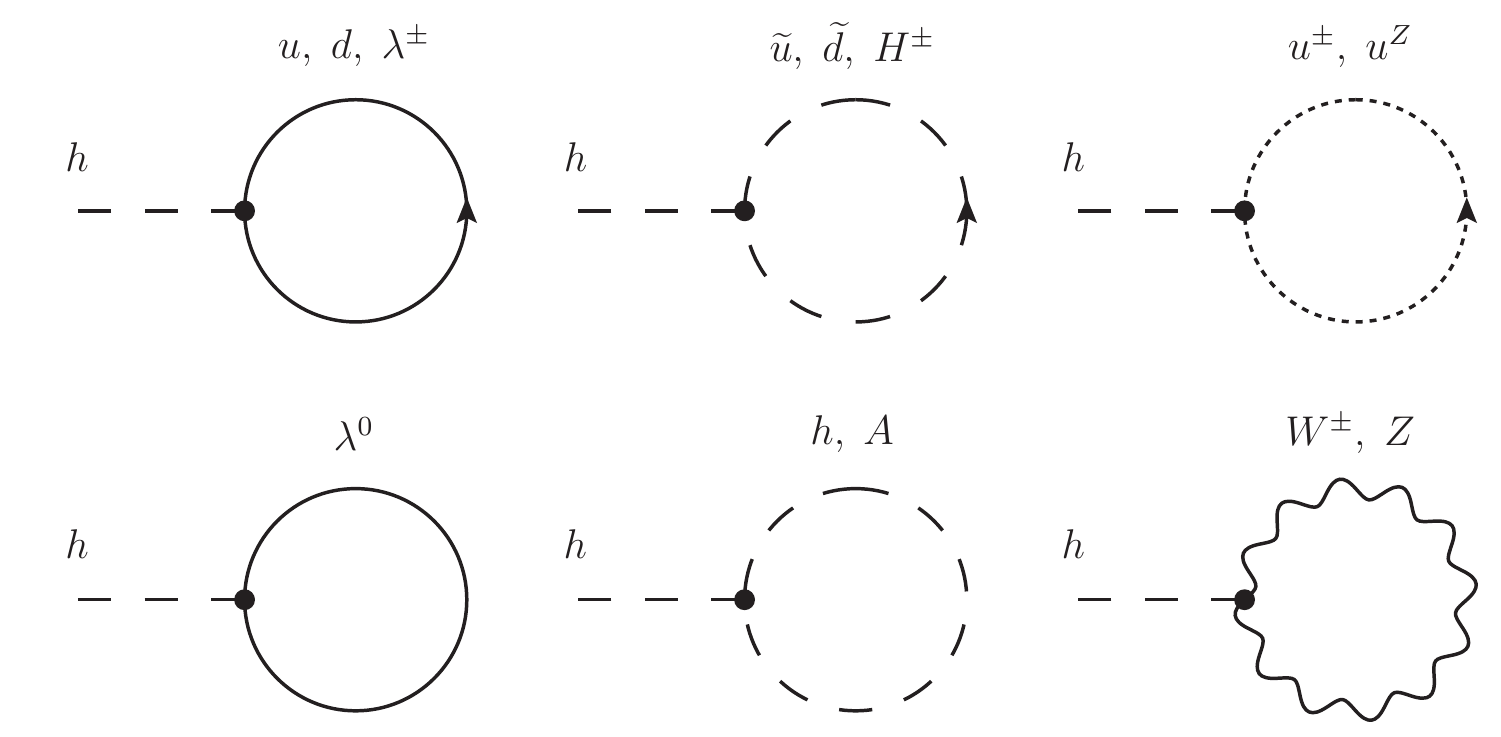}
  \caption{Generic Feynman diagrams for the tadpoles $T_{h_i}$.}
  \label{fig:tadpoles}
\end{figure}

For the tadpole coefficients $T_{\varphi_i}$ the OS conditions read
\begin{equation}
%T_{\varphi_i}+T_{\varphi_i}^{(1)}+\delta T_{\varphi_i}=0 \; ,
T_{\varphi_i}^{(1)}+\delta T_{\varphi_i}=0 \; ,
\end{equation}
where $T_{\varphi_i}^{(1)}$ are the one-loop contributions to the linear terms of the scalar potential, stemming from tadpole diagrams shown in \reffi{fig:tadpoles}. The tadpole diagrams are calculated in the mass eigenstate basis $h$. The one-loop tadpole contributions in the interaction basis $\varphi$ are then obtained by the rotation
\begin{equation}
T_\varphi^{(1)} = {U^H}^T T_h^{(1)} \; .
\end{equation}
Accordingly we find for the one-loop tadpole counterterms
\begin{equation}
\delta T_{\varphi_i}=-T_{\varphi_i}^{(1)} \; .
\end{equation}
%%%%%%%%%%%%%%%%%%%%%%%%%%%% F I G U R E %%%%%%%%%%%%%%%%%%%%%%%%%%%%%%%%%%%%%%
%%%%%%%%%%%%%%%%%%%%%%%%%%%% F I G U R E %%%%%%%%%%%%%%%%%%%%%%%%%%%%%%%%%%%%%%

For practical purposes we
decided to renormalize all remaining parameters in the $\DRbar$
scheme (reflecting the fact that there are no physical observables that
could be directly related to them). 
The counterterms of each parameter were obtained by
calculating the divergent parts of one-loop corrections to different
scalar and fermionic two- and three-point functions. 
We state the determination of the counterterms in the (possible) order 
in which they can be successively derived.
We start with the counterterms that
were obtained by renormalizing certain neutral fermion self-energies. 

\paragraph{\protect\boldmath Renormalization of $\mu$:}
%%%%%%%%%%%%%%%%%%%%%%%%%%%% F I G U R E %%%%%%%%%%%%%%%%%%%%%%%%%%%%%%%%%%%%%%
\begin{figure}[t]
  \centering
  \includegraphics[width=0.8\textwidth]{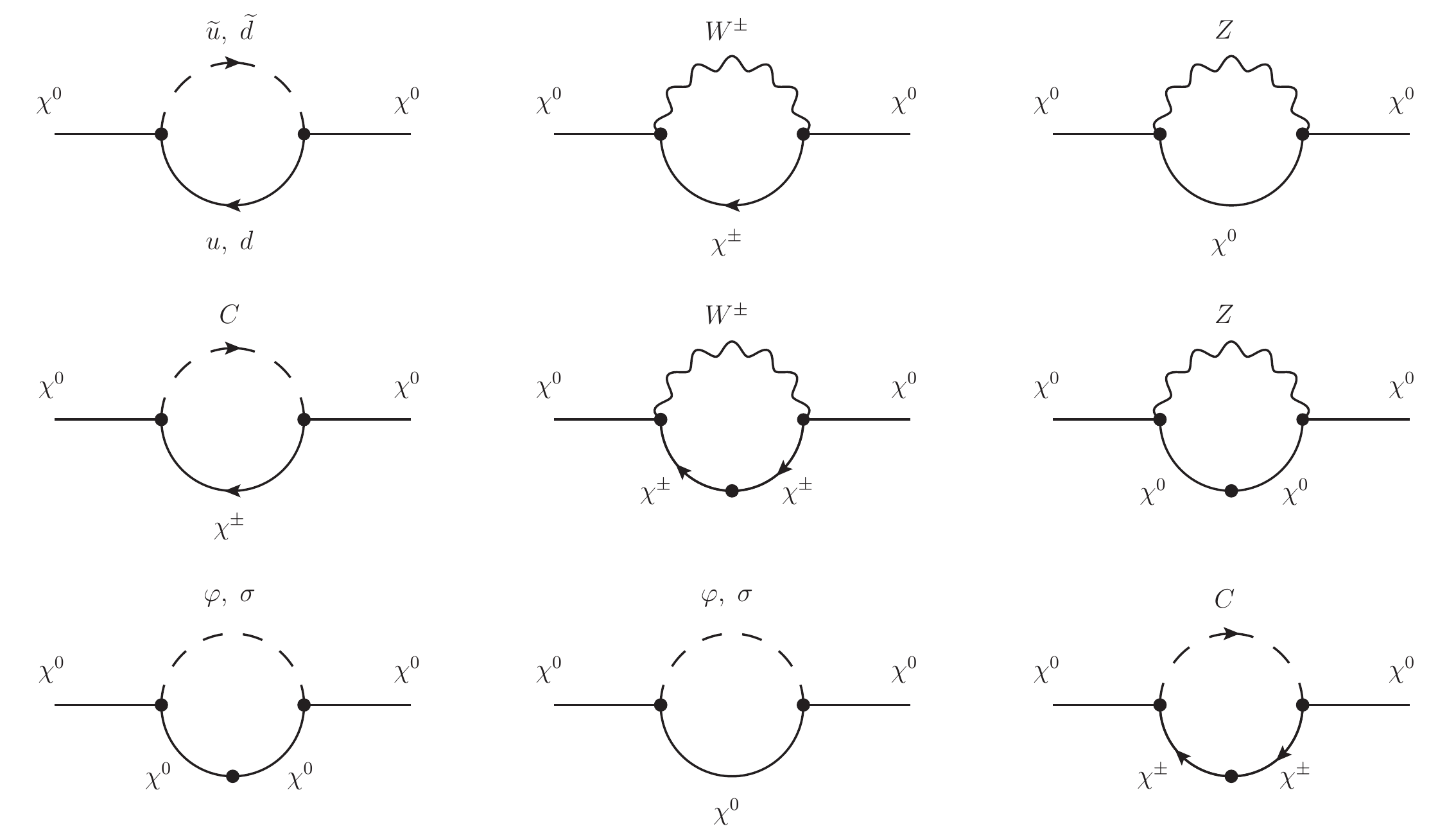}
  \caption{Diagrams contributing to the neutral
    fermion self-energies in the interaction basis.}
\label{fig:neufermdiags}
\end{figure}
%%%%%%%%%%%%%%%%%%%%%%%%%%%% F I G U R E %%%%%%%%%%%%%%%%%%%%%%%%%%%%%%%%%%%%%%
The $\mu$ parameter appears isolated in the Majorana-type mass matrix of
the neutral fermions 
\begin{equation}
\left(m_\nu\right)_{67}=-\frac{\lambda v_R}{\sqrt{2}}=-\mu \; ,
\end{equation}
which is the element mixing the down-type and the up-type Higgsinos
$\widetilde{H}_d$ and $\widetilde{H}_u$.
The entries
$\left(m_\nu\right)_{ij}$ get one-loop corrections via the neutral
fermion self-energies ${\sum}_{\neu{i} \neu{j}}$, that for Majorana
fermions can be decomposed as\footnote{Left-handed components and
  right-handed components are the same for Majorana fields.} 
\begin{equation}
\Sigma_{\neu{i} \neu{j}}\left( p^2 \right) = \slashed{p}\Sigma_{\neu{i} \neu{j}}^F\left( p^2 \right)+\Sigma_{\neu{i} \neu{j}}^S\left( p^2 \right) \; .
\end{equation}
The part $\Sigma_{\neu{i} \neu{j}}^F$ is renormalized through field
renormalization and the part $\Sigma_{\neu{i} \neu{j}}^S$ is
renormalized by both the field renormalization and a mass counter
term. Since we are interested in the mass renormalization we focus on
$\Sigma_{\neu{i} \neu{j}}^S$ and write for the renormalized self-energy
at zero momentum 
\begin{equation}\label{eq:neufermreno}
\hat{\Sigma}_{\neu{i} \neu{j}}^S\left(0\right)=\Sigma_{\neu{i} \neu{j}}^S\left(0\right)-\frac{1}{2}
	\left( \delta Z_{ki}^\chi \left( m_\nu \right)_{kj} +
			\left( m_\nu \right)_{ik} \delta Z_{kj}^\chi \right)
		-\delta\left(m_\nu\right)_{ij} \; .
\end{equation}
The field renormalization constants can be obtained by calculating the divergent part of $\Sigma_{\neu{i} \neu{j}}^F$:
\begin{equation}
\delta Z_{ij}^\chi=-\left.\Sigma_{\neu{i}\neu{j}}^F\right|^{\rm div} \; ,
\end{equation}
where we make use of the fact that there are no divergences proportional
to $p^2$ in our case. The divergent parts of the self-energies of the
neutral fermions are calculated diagrammatically in the interaction
basis, where diagrams with mass insertions have to be included. In
\reffi{fig:neufermdiags} we show the generic diagrams potentially
contributing to the divergent part of the self-energies.
Diagrams with a scalar mass insertion or more than
one fermionic mass insertion are power-counting finite, so we do
not depict them.
The diagram shown in \reffi{fig:neufermdiags} with a mass
insertion on the chargino propagator
can be divergent depending on the expressions for the couplings of the
charginos.

We checked that our results for the field renormalization counterterms
for the neutral fermions are consistent with the one-loop anomalous
dimensions $\gamma_{ij}^{(1)}$ of the corresponding superfields, i.e., 
\begin{equation}
\delta Z_{ij}^\chi=\frac{\gamma_{ij}^{(1)}\Delta}{16 \pi^2} \; .
\end{equation}
To extract $\delta \mu$ we now just have to identify 
\begin{equation}
\delta\left(m_\nu\right)_{67}=-\delta\mu \; ,
\end{equation}
and calculate the divergent part of $\Sigma_{\widetilde{H}_d \widetilde{H}_u}^S$, which again is not momentum dependent. $\delta\mu$ is then given by
\begin{equation}
\delta\mu=\frac{1}{2}\mu\left(
	-\left(
		\delta Z_{66}^\chi+
		\delta Z_{77}^\chi
	\right)+
	\frac{1}{\lambda}\left(
	\delta Z_{16}^\chi Y^\nu_1+
	\delta Z_{26}^\chi Y^\nu_2+
	\delta Z_{36}^\chi Y^\nu_3 \right)\right) -
	\left.\Sigma_{\widetilde{H}_d \widetilde{H}_u}^S\right|^{\rm div} \; ,
\end{equation}
where we made us of the fact that the matrix $\delta Z_{ij}^\chi$ is real and symmetric and that components mixing left-handed neutrinos and the down-type Higgsino are the only non-diagonal elements contributing here.

Explicit formulas for the counterterms of the parameters renormalized in the $\DRbar$ scheme are listed in the appendix \ref{app:paracounters}. 
For the $\DRbar$ counterterms we checked that in the limit $Y^\nu_i\rightarrow 0$ our results coincide with the one in the 
NMSSM~\cite{Ellwanger:2009dp}.

\begin{figure}[tb]
  \centering
  \includegraphics[width=0.6\textwidth]{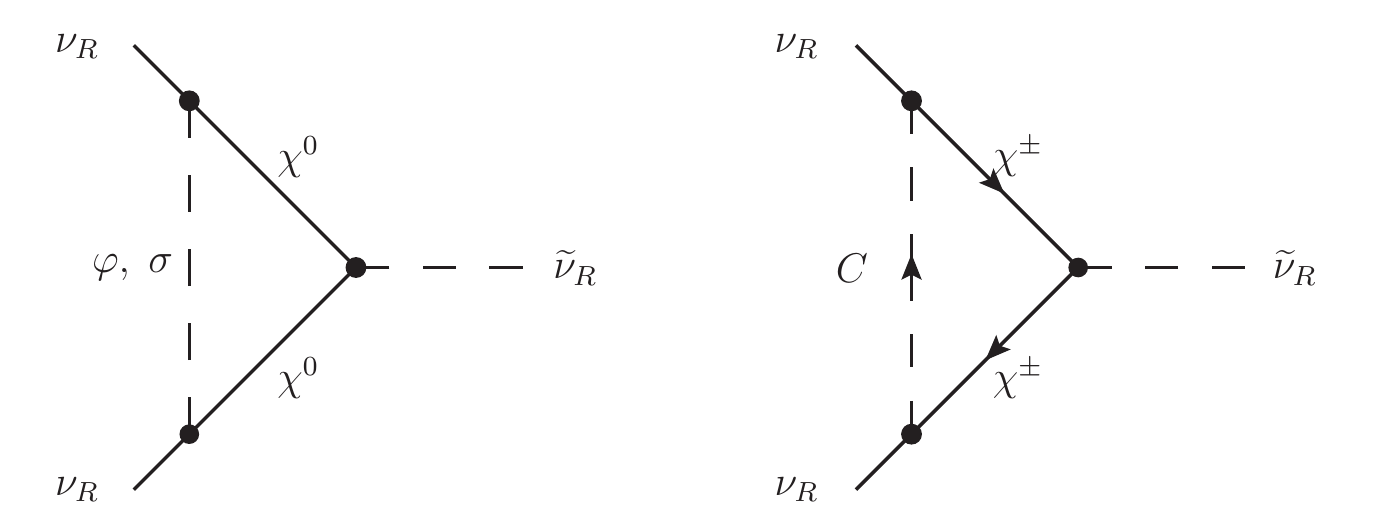}
  \caption{Potentially divergent one-particle irreducible diagrams
    contributing to the three-point vertex between two right-handed
    neutrinos and one right-handed sneutrino.}
  \label{fig:kappadiags}
\end{figure}

\paragraph{\protect\boldmath Renormalization of $\kappa$:}
The parameters $\kappa$ appears isolated at tree-level in the three-point vertex that couples the right-handed neutrino to the right-handed sneutrino,
\begin{equation}
\Gamma_{\nu_R \nu_R \widetilde{\nu}_R}^{(0)}=-\sqrt{2}\kappa \; .
\end{equation}
The divergences induced to this coupling at one-loop have to be absorbed by the field renormalization of the right-handed neutrino and sneutrino and the counterterm for $\kappa$, which is the only parameter in the tree-level expression. We find
\begin{equation}
\delta \kappa = 
\frac{1}{\sqrt{2}}
\left.\Gamma{\nu_R \nu_R \widetilde{\nu}_R}^{(1)}\right|^{\rm div} -
\frac{1}{2}\kappa\left( \delta Z_{33}+2 \delta Z^\chi_{88}\right) \; ,
\end{equation}
where $\Gamma{\nu_R \nu_R \widetilde{\nu}_R}^{(1)}\rvert^{\rm div}$ is the divergent part of the corresponding one-loop three-point function, and the terms containing the field renormalization is trivial, because there is only one singlet-like superfield so that no non-diagonal field renormalization constants appear. The divergent one-loop contributions to the vertex are calculated diagrammatically in the interaction basis. The only contributing generic diagrams are shown in \reffi{fig:kappadiags}.

All other topologies, including diagrams with one or more
mass insertion, are finite, and there are no diagrams with gauge bosons
instead of scalars  in the loop, because there are three gauge-singlet
fields on the outer legs. It turns out that the sum over the diagrams
shown in \reffi{fig:kappadiags} is also finite, so that 
$\Gamma_{\nu_R \nu_R\widetilde{\nu}_R}^{(1)}\rvert^{\rm div}$ vanishes. 
%%%%%%%%%%%%%%%%%%%%%%%%%%%% F I G U R E %%%%%%%%%%%%%%%%%%%%%%%%%%%%%%%%%%%%%%
%%%%%%%%%%%%%%%%%%%%%%%%%%%% F I G U R E %%%%%%%%%%%%%%%%%%%%%%%%%%%%%%%%%%%%%%

\paragraph{\protect\boldmath Renormalization of $\lambda$:}
Having calculated $\delta \mu$ and $\delta \kappa$ we can extract the counterterm for $\lambda$ in the neutral fermion sector. $\lambda$ appears in the mass matrix element
\begin{equation}
\left( m_\nu \right)_{88}=\frac{2\kappa\mu}{\lambda} \; .
\end{equation}
Making use of \refeq{eq:neufermreno} we find
\begin{equation}
\delta\lambda=
\lambda\left( \delta Z^\chi_{88}+\frac{\delta\kappa}{\kappa}
+\frac{\delta\mu}{\mu} \right)-\frac{\lambda^2}{2\mu\kappa}
\left.\Sigma_{\nu_R \nu_R}^S\right|^{\rm div} \; ,
\end{equation}
where we calculated the divergent part of the right-handed neutrino self-energie $\Sigma_{\nu_R \nu_R}^S\rvert^{\rm div}$ diagrammatically in the interaction basis using the diagrams already shown in \reffi{fig:neufermdiags}.

\begin{figure}[t]
  \centering
  \includegraphics[width=0.7\textwidth]{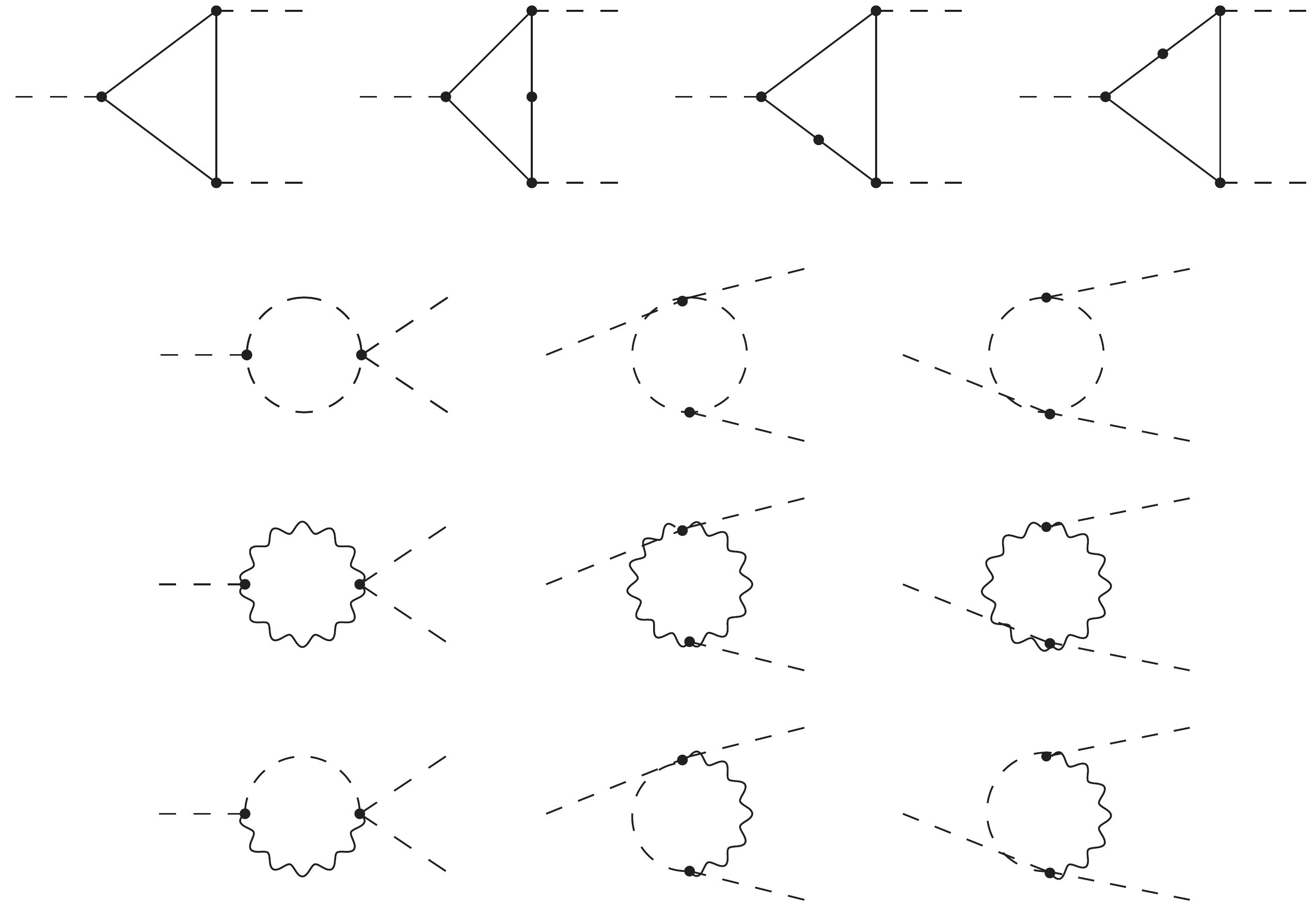}
\caption{Potentially divergent one-particle irreducible topologies
  contributing to a scalar three-point vertex at one-loop in the
  interaction basis. Diagrams with scalar mass insertions or more than
  one fermionic mass insertions are finite.} 
\label{fig:triplescalardiags}
\end{figure}

\paragraph{\protect\boldmath Renormalization of $A_\kappa$:}
The counterterm for the parameter $A_\kappa$ can be extracted from the one-loop corrections to the scalar three-point vertex of right-handed sneutrinos when $\delta\kappa$ is known and using the one-loop relation
\begin{equation}
\left[ \frac{\delta\mu}{\mu} - \frac{\delta\lambda}{\lambda} \right]^{\rm div}
=\left.\frac{1}{2}\delta Z_{33}\right|^{\rm div} \; ,
\end{equation}
which was found in the NMSSM~\cite{Sperling:2013eva} and confirmed
for this work also in the \mnSSM.
For the trilinear singlet vertex we have at tree-level 
\begin{equation}
\Gamma_{\widetilde{\nu}_R\widetilde{\nu}_R\widetilde{\nu}_R}^{(0)}=
-\sqrt{2}\kappa\left(A_\kappa+\frac{6\kappa\mu}{\lambda}\right) \; .
\end{equation}
The tree-level vertex does not depend on the momentum, so the one-loop counterterm for $A_\kappa$ can be calculated through
\begin{equation}
\delta A_\kappa = \frac{1}{\sqrt{2}\kappa}\left( 
\left.\Gamma_{\widetilde{\nu}_R\widetilde{\nu}_R\widetilde{\nu}_R}^{(1)}\right|^{\rm div} +
\frac{3}{2} \delta Z_{33}
\Gamma_{\widetilde{\nu}_R\widetilde{\nu}_R\widetilde{\nu}_R}^{(0)} \right)
-A_\kappa \frac{\delta \kappa}{\kappa}-
\frac{6\kappa\mu}{\lambda} \left( 2\frac{\delta\kappa}{\kappa}+
\frac{1}{2}\delta Z_{33} \right) \; .
\end{equation}
Here $\Gamma_{\widetilde{\nu}_R\widetilde{\nu}_R\widetilde{\nu}_R}^{(1)}\rvert^{\rm div}$ is the divergent part of the one-loop corrections to the three-point vertex, which was calculated diagrammatically in the interaction basis. The number of contributing diagrams is rather high, so for simplicity we just show the topologies of the diagrams contributing, that potentially lead to divergences, in \reffi{fig:triplescalardiags}.
%%%%%%%%%%%%%%%%%%%%%%%%%%%% F I G U R E %%%%%%%%%%%%%%%%%%%%%%%%%%%%%%%%%%%%%%
%%%%%%%%%%%%%%%%%%%%%%%%%%%% F I G U R E %%%%%%%%%%%%%%%%%%%%%%%%%%%%%%%%%%%%%%
In the case of the vertex
$\Gamma_{\widetilde{\nu}_R\widetilde{\nu}_R\widetilde{\nu}_R}$ we can
neglect the diagrams with gauge bosons, because the right-handed
sneutrinos are gauge singlets.  

\paragraph{\protect\boldmath Renormalization of $A_\lambda$:}
The counterterm for the parameter $A_\lambda$ is like in the previous case extracted from the one-loop corrections to a scalar three-point function. Here we consider $\Gamma_{H_d H_u \widetilde{\nu}_R}$, the coupling between the two doublet-type Higgses and the right-handed sneutrino. At tree-level it is
\begin{equation}
\Gamma_{H_d H_u \widetilde{\nu}_R}^{(0)}=
\frac{A_\lambda \lambda}{\sqrt{2}}+
\sqrt{2}\kappa\mu \; ,
\end{equation}
so we will make use of the fact that we already know the counterterms for $\lambda$, $\kappa$ and $\mu$.

The final expression defining $\delta A_\lambda$ will also contain the
tree-level expressions for the couplings where the down-type Higgs is
replaced by one of the left-handed sneutrinos. They are induced by the
non-diagonal field renormalization of $H_d$ and
$\widetilde{\nu}_{iL}$ and enter the
renormalization of $\Gamma_{H_d H_u \widetilde{\nu}_R}$ at one-loop. We
find
\begin{align}
\delta A_\lambda&=
-\frac{\sqrt{2}}{\lambda}
\left.\Gamma_{H_d H_u \widetilde{\nu}_R}^{(1)}\right|^{\rm div}-
\frac{1}{\sqrt{2}\lambda}\left(
\delta Z_{11}\Gamma_{H_d H_u \widetilde{\nu}_R}^{(0)}+
\delta Z_{14}\Gamma_{\widetilde{\nu}_{1L}H_u\widetilde{\nu}_R}^{(0)}+
\delta Z_{15}\Gamma_{\widetilde{\nu}_{2L}H_u\widetilde{\nu}_R}^{(0)}
    \right. \notag \\  +& \left.
\delta Z_{16}\Gamma_{\widetilde{\nu}_{3L}H_u\widetilde{\nu}_R}^{(0)}+
\delta Z_{22}\Gamma_{H_d H_u \widetilde{\nu}_R}^{(0)}+
\delta Z_{33}\Gamma_{H_d H_u \widetilde{\nu}_R}^{(0)} \right)-
\frac{A_\lambda}{\lambda}\delta\lambda-
\frac{2\kappa}{\lambda}\delta\mu-
\frac{2\mu}{\lambda}\delta\kappa \; ,
\end{align}
with
\begin{equation}
\Gamma_{\widetilde{\nu}_{iL}H_u\widetilde{\nu}_R}^{(0)}=
\frac{-Y^\nu_i\left( A^\nu_i+\frac{2\kappa\mu}{\lambda} \right)}{\sqrt{2}}
\; .
\end{equation}

\paragraph{\protect\boldmath Renormalization of $v^2$:}
The SM-like vev is renormalized via the
renormalization of the electromagnetic coupling in the Thompson limit,
which can be done when the counterterms for the gauge boson masses are
fixed. We follow here the approach of \citere{Drechsel:2016ukp} used in
the NMSSM to be able to compare the results in both models as best as
possible.

The renormalization of the electromagnetic coupling is defined by
\begin{equation}
e \to e \left( 1+ \delta Z_e \right) \; ,
\end{equation}
and the counterterm $\delta Z_e$ can be calculated via
\begin{equation}
\left.\delta Z_e\right|^{\rm div}=
\left[
\frac{1}{2} \left( \frac{\partial \Sigma_{\gamma\gamma}^T}{\partial p^2} \left( 0 \right) \right) +
\frac{\SW}{\CW \MZ^2} \Sigma_{\gamma Z}^T \left( 0 \right)
\right]^{\rm div} \; ,
\end{equation}
where $\Sigma_{\gamma\gamma}^T(0)$ is the transverse part of the photon
self-energy and $\Sigma_{\gamma Z}^T$ is the transverse part of the
mixed photon-Z boson self-energy. $\SW$ and $\CW$ are defined as 
$\SW = \sqrt{1 - \CW^2}$ with $\CW = \MW/\MZ$. $v^2$ and $e$ are related by
\begin{equation}
v^2=\frac{2\SW^2 \MW^2}{e^2} \; ,
\end{equation}
so the counterterm $\delta v^2$ can be obtained through
\begin{equation}\label{eq:dv2def}
\delta v^2 = \frac{4 \SW^2 \MW^2}{e^2}\left.\left(
\frac{\delta \SW^2}{\SW^2}+\frac{\delta \MW^2}{\MW^2}-
2\delta Z_e \right)\right|^{\rm div} \; ,
\end{equation}
where
\begin{equation}\label{eq:dsw2def}
\SW^2  \rightarrow \SW^2 + \delta \SW^2 \; , \quad
\text{with }  \quad
\delta \SW^2=-\CW^2 \left( \frac{\delta \MW^2}{\MW^2}
-\frac{\delta \MZ^2}{\MZ^2} \right) .
\end{equation}
Here we take only the divergent parts of the counterterms 
$\delta \MZ^2$,
$\delta \MW^2$ and $\delta Z_e$, so that $\delta v^2$ is
renormalized in the $\DRbar$ scheme. 
This implies that the counterterm $\delta Z_e$ is not a
free parameter, even if we calculated it as if it
would be to determine $\delta v^2$. Instead $\delta Z_e$ is a
dependent parameter defined by $\delta v^2$ in the $\DRbar$ scheme
and $\delta M_Z^2$ and $\delta M_W^2$ in the OS scheme through
\refeq{eq:dv2def} and \refeq{eq:dsw2def},
\begin{equation}
\delta Z_e = \frac{1}{2 \SW^2}
\left( \CW^2 \frac{\delta \MZ^2}{\MZ^2}
        +\left( \SW^2-\CW^2 \right)\frac{\delta \MW^2}{\MW^2}
        -\frac{e^2}{4\MW^2}\delta v^2 \right) \; .
\end{equation}

\paragraph{\protect\boldmath Renormalization of $v_{iL}^2$:}
The counterterms for the three vevs of the left-handed sneutrinos
$v_{iL}$ can be extracted from the divergent part of the one-loop
self-energies $\Sigma_{\widetilde{B}\nu_{iL}}$ between the bino and the
corresponding left-handed neutrino. The tree-level mass matrix entries
we renormalize are defined by
\begin{equation}
\left( m_\nu \right)_{4i}=-\frac{g_1 v_{iL}}{2} \; ,
\end{equation}
so it is necessary to have the counterterm of the gauge 
coupling $g_1$, whose renormalization we define as 
$g_1\rightarrow g_1+\delta g_1$. We then can obtaine 
$\delta g_1$ from $\delta \MW^2$, $\delta \MZ^2$
and $\delta v^2$ through the definitions of the gauge boson masses in
\refeq{eq:gaguebosonmasses},
\begin{equation}
\delta g_1 = \frac{2}{g_1 v^2}\left( \delta M_Z^2-\delta M_W^2 \right)
	-\frac{g_1}{2}\frac{\delta v^2}{v^2} \; .
\end{equation}
Renormalizing the self-energies
$\Sigma_{\widetilde{B}\nu_{iL}}$ using \refeq{eq:neufermreno} we
find the following expression for the $\delta v_{iL}^2$:
\begin{equation}
\delta v_{iL}^2=\frac{4 v_{iL}}{g_1}
\left.\Sigma_{\widetilde{B}\nu_{iL}}^S\right|^{\rm div}-
v_{iL}\left( \delta Z_{44}^\chi v_{iL} +
\delta Z_{ij}^\chi v_{jL} 
+ \delta Z_{i6}^\chi v_d \right) -
2v_{iL}^2\left.\frac{\delta g_1}{g_1}\right|^{\rm div} \; ,
\end{equation}
where again the divergent contributions of $\Sigma_{\widetilde{B}\nu_{iL}}^S$ are calculated diagrammatically in the interaction basis.

\paragraph{\protect\boldmath Renormalization of $Y^\nu_i$:}
The counterterm for the neutrino Yukawas $Y^\nu_i$ can be extracted in the neutral fermion sector as well. We decide to use the renormalization of the tree-level masses
\begin{equation}
\left( m_\nu \right)_{i7}=\frac{\mu Y^\nu_i}{\lambda} \; ,
\end{equation}
that mix the left-handed neutrinos and the up-type Higgsino. Since we already found $\delta\lambda$ and $\delta\mu$  we can get $\delta Y^\nu_i$ from the divergent part of the one-loop self-energies $\Sigma_{\nu_{iL}\widetilde{H}_u}^S$,
\begin{equation}
\delta Y^\nu_i = \frac{1}{2}\left( \delta Z_{16}^\chi \lambda -
\delta Z_{77}^\chi Y^\nu_i - \delta Z_{ij}^\chi Y^\nu_j \right)
- \left( \frac{\delta\mu}{\mu}-\frac{\delta\lambda}{\lambda} \right)
+\frac{\lambda}{\mu} \left.\Sigma_{\nu_{iL}\widetilde{H}_u}^S\right|^{\rm div} \; .
\end{equation}

\paragraph{\protect\boldmath Renormalization of $\tb$:}
We adopted the usual definition for $\tb$ as in the MSSM (see
\refeq{eq:tanbetadef}). 
If we define the renormalization for the vevs of the doublet fields as
\begin{equation}\label{eq:vudreno}
 v_d^2 \rightarrow v_d^2 + \delta v_d^2 \; , \quad
 v_u^2 \rightarrow v_u^2 + \delta v_u^2 \; ,
\end{equation}
the counterterm for $\tb$ can be written at one-loop as a linear
combination of the counterterms for the vevs of
the doublet Higgses, 
\begin{equation}\label{eq:dtanbetadef}
\delta \tb = \frac{1}{2}\tb\left( \frac{\delta v_u^2}{v_u^2} -
\frac{\delta v_d^2}{v_d^2} \right) \; .
\end{equation}
Note that our renormalization of $v_u^2$ and $v_d^2$ in \refeq{eq:vudreno} 
includes the contributions from the field renormalization constants inside 
the counterterms $\delta v_u^2$ and $\delta v_d^2$. This approach is 
equivalent as defining 
\begin{equation}
 v_d \rightarrow \sqrt{Z_{11}}
 	\left(v_d + \delta \hat{v}_d\right) \; , \quad
 v_u \rightarrow \sqrt{Z_{22}}
 	\left(v_u + \delta \hat{v}_u\right) \; ,
\end{equation}
and writing the counterterm of $\tan\beta$ as
\begin{equation}\label{eq:tanfieldreno}
 \delta \tan\beta = \frac{1}{2}\tan\beta
 	\left( \delta Z_{22} - \delta Z_{11} \right) +
 	\tan\beta \left( \frac{\delta\hat{v}_u}{v_u} -
 		\frac{\delta\hat{v}_d}{v_d} \right) \; .
\end{equation}
This notation was convenient in the MSSM and the NMSSM, because the 
second bracket in \refeq{eq:tanfieldreno} is finite at 
one-loop~\cite{Chankowski:1992er,Dabelstein:1994hb,Ender:2011qh,Drechsel:2016ukp}
and can be set to zero in the $\DRbar$ scheme,   
so that $\delta \tan\beta$ can be expressed exclusively by the
field renormalization constants. In contrast, 
in the \mnSSM\ we find
\begin{equation}
\left.\left( \frac{\delta\hat{v}_u}{v_u} -
 		\frac{\delta\hat{v}_d}{v_d} \right)\right|^{\rm div}
=-\frac{\Delta \lambda v_{iL} Y^\nu_i }{32 \pi^2 v_d} \; .
\end{equation}
There are several possibilities to extract the counterterms $\delta
v_d^2$ and $\delta v_u^2$. 
A convenient choice is to extract $\delta v_d^2$
from the renormalization of the entry of the neutral fermion
mass matrix
mixing the up-type Higgsino and
the right-handed neutrino, 
\begin{equation}
\left( m_\nu \right)_{78}=\frac{-\lambda v_d + v_{iL}Y^\nu_i}{\sqrt{2}} \; ,
\end{equation}
because in this case no non-diagonal field renormalization counterterms are needed. Calculating the divergent part of $\Sigma_{\widetilde{H}_u v_R}^S$ and using the counterterms previously calculated we can extract $\delta v_d^2$ via the expression
\begin{align}
\delta v_d^2 =& -\frac{2\sqrt{2}v_d}{\lambda}
\left.\Sigma_{\widetilde{H}_u v_R}^S\right|^{\rm div}+
\frac{v_d}{\lambda}\left( \delta Z_{77}^\chi+ \delta Z_{88}^\chi \right)
\left( -v_d\lambda + v_{iL}Y^\nu_i \right) -
2v_d^2\frac{\delta\lambda}{\lambda} \notag \\
&+\frac{v_d}{\lambda} Y^\nu_i \left( \frac{\delta v_L^2}{v_L} \right)_i
+\frac{2v_d}{\lambda}v_{iL}\delta Y^\nu_i \; . \label{eq:dvd2def}
\end{align}
Since all counterterms appearing in \refeq{eq:dvd2def}
are renormalized in the $\DRbar$ scheme also
$\delta v_d^2$ has no finite part.
There are now two ways to determine $\delta v_u^2$. Firstly, we could similarly to $\delta v_d^2$ extract the counterterm $\delta v_u^2$ by renormalizing the up-type Higgsino self-energy $\Sigma_{\widetilde{H}_u \widetilde{H}_u}^S$. Alternatively, we can deduce $\delta v_u^2$ from the definition of $v^2$ in \refeq{eq:tanbetadef} and simply write
\begin{equation}\label{eq:dvu2def}
\delta v_u^2 = \delta v^2 -\delta v_d^2 - \delta v_{1L}^2 - \delta v_{2L}^2 - \delta v_{3L}^2 \; .
\end{equation}
We verified that both options yield the same result, which constitutes a
consistency test for the counterterms $\delta v_{iL}^2$, which are
unique for the \mnSSM. Inserting $\delta v_d^2$ from \refeq{eq:dvd2def}
and $\delta v_u^2$ from \refeq{eq:dvu2def} into \refeq{eq:dtanbetadef}
finally gives the counterterm for $\tb$.
We checked that the final expression for $\tb$ in \refeq{eq:dtbdrbar}
agrees with the NMSSM result in the limit
$Y^\nu_i\rightarrow 0$.

The renormalization of $\tb$ in the $\DRbar$ scheme is manifestly process-independent and has shown to give stable numerical results in the MSSM \cite{Frank:2002qf,Freitas:2002um} and the NMSSM \cite{Drechsel:2016ukp,Ender:2011qh}.

\paragraph{\protect\boldmath Renormalization of $A^\nu_i$:}
The soft trilinears $A^\nu_i$ can be renormalized through the calculation of the radiative corrections to the corresponding scalar vertex in the interaction basis. The tree-level expression for the interaction between the up-type Higgs, one left-handed sneutrinos and the right-handed sneutrino is given by
\begin{equation}
\Gamma_{H_u \widetilde{\nu}_R \widetilde{\nu}_{iL}}^{(0)}=
-\left( \frac{A^\nu_i}{\sqrt{2}}+\frac{\sqrt{2}\kappa
\mu}{\lambda} \right) Y^\nu_i \; .
\end{equation}
The renormalized one-loop corrected vertex will define the counterterm for $A^\nu_i$ since the counterterms for $\kappa$, $\mu$ and $\lambda$ were already determined. We showed in \reffi{fig:triplescalardiags} the topologies of the diagrams that have to be calculated in the interaction basis to get the divergent part of one-loop corrections
$\Gamma_{H_u \widetilde{\nu}_R \widetilde{\nu}_{iL}}^{(1)}$.
As in the case of the renormalization of $A^\lambda$ the renormalization of the scalar vertex will contain the tree-level expressions of all the vertices with the same quantum numbers of the external fields, because of the non-diagonal field renormalization. Solved for $\delta A^\nu_i$ the renormalization of the vertex leads to
\begin{align}
\delta A^\nu_i=&
\frac{\sqrt{2}}{Y^\nu_i}\left.\Gamma_{H_u \widetilde{\nu}_R
\widetilde{\nu}_{iL}}^{(1)}\right|^{\rm div}+
\frac{1}{\sqrt{2}Y^\nu_i}\left( \left( \delta Z_{22}+ \delta Z_{33}
\right) \Gamma_{H_u \widetilde{\nu}_R
\widetilde{\nu}_{iL}}^{(0)}+
\delta Z_{1,3+i} \Gamma_{H_u \widetilde{\nu}_R H_d}^{(0)} \right. \notag \\ 
&\left.+\delta Z_{3+j,3+i}\Gamma_{H_u \widetilde{\nu}_R
\widetilde{\nu}_{jL}}^{(0)}\right)-
\frac{A^\nu_i}{Y^\nu_i}\delta Y^\nu_i -
\frac{2\mu}{\lambda}\delta \kappa -
\frac{2\kappa}{\lambda}\delta \mu -
\frac{2\kappa\mu}{\lambda Y^\nu_i}\delta Y^\nu_i +
\frac{2\kappa\mu}{\lambda^2}\delta\lambda \; ,
\end{align}
with
\begin{equation}
\Gamma_{H_u \widetilde{\nu}_R H_d}^{(0)} =
\frac{\lambda A^\lambda}{\sqrt{2}}+\sqrt{2}\kappa\mu \; .
\end{equation}

\paragraph{\protect\boldmath Renormalization of ${m_{H_d\widetilde{L}_L}^2}_i$:}
The soft scalar masses appear in the bilinear terms of the Higgs
potential. They can be renormalized by calculating radiative corrections
to scalar self-energies. It proved to be convenient 
to calculate the $\cp$-odd scalar self-energies in the mass basis,
and then to rotate the self-energies back to the interaction basis. 

\begin{figure}
  \centering
  \includegraphics[width=0.76\textwidth]{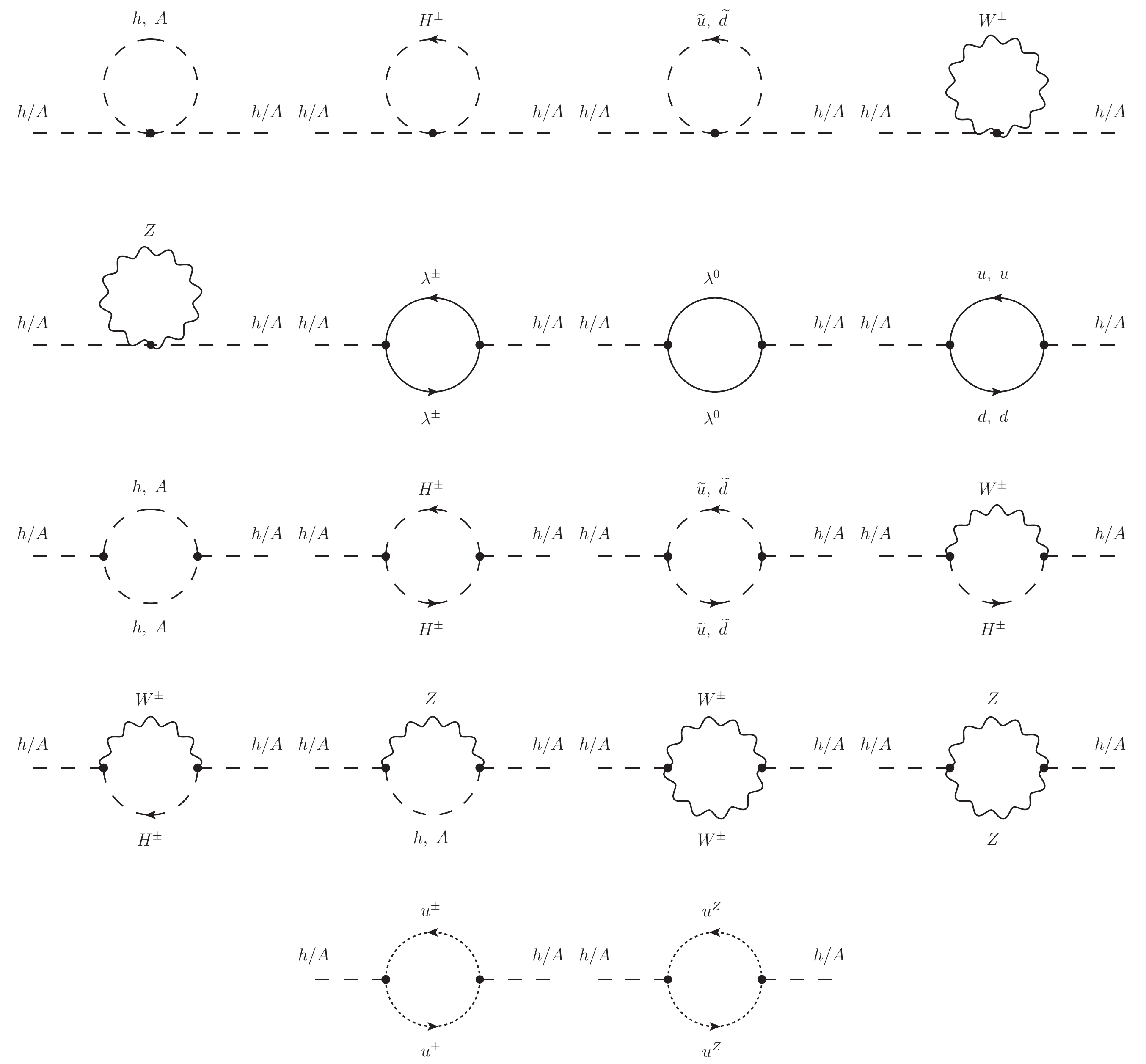}
  \caption{Generic diagrams for the $\cp$-even (h) and 
  $\cp$-odd (A) scalar self-energies in the mass
    eigenstate basis.}
\label{fig:scalarselfs}
\end{figure}

We find ${m_{H_d\widetilde{L}_L}^2}_i$ at tree-level in
\begin{equation}
m_{\widetilde{\nu}_{iL}^{\mathcal{I}} H_{d}^{\mathcal{I}}}^{2}=
\left(m_{H_d\widetilde{L}_L}^2\right)_i-
\frac{1}{2}v_R^2\lambda Y^\nu_i-
\frac{1}{2}v_u^2\lambda Y^\nu_i \; .
\end{equation}
The general form of the renormalized scalar self-energies at one-loop is
\begin{align}
\hat{\Sigma}_{X_i X_j}\left( p^2 \right) =&
\Sigma_{X_i X_j}\left( p^2 \right)+
\frac{1}{2} p^2 \left( \delta Z_{ji} + \delta Z_{ij} \right) \notag \\
-&\frac{1}{2} \left( \delta Z_{ki} \left( m_{X}^2 \right)_{kj} +
\left( m_{X}^2 \right)_{ik} \delta Z_{kj} \right) -
\delta \left( m_{X}^2 \right)_{ij} \; ,
\end{align}
where $X=(\varphi,\sigma)$ represents either the $\cp$-even or the $\cp$-odd scalar fields and we made use of the fact that the field renormalization constants $\delta Z$ and the mass matrix $m_{X}^2$
are real.
%neglecting $\cp$-violation.
Demanding that the renormalized self-energies $\hat{\Sigma}_{A_i A_j}$ are finite in the mass eigenstate basis we can define the divergent parts of the mass counterterms via
\begin{equation}
\left.\delta \left( m_{A}^2 \right)_{ij}\right|^{\rm div} =
\left.\Sigma_{A_i A_j}\left( 0 \right) \right|^{\rm div} -
\frac{1}{2}\left( \left(\delta Z^A\right)_{ji} m_{A_j}^2 +
m_{A_i}^2 \left(\delta Z^A\right)_{ij} \right)\; 
\label{eq:oddcounters} ,
\end{equation}
where the field counterterms in the mass eigenstate basis were defined in \refeq{eq:fieldrenorotate} and the masses $m_{A_i}^2$ are the eigenvalues of the diagonal $\cp$-odd scalar mass matrix $m_{A}^2$. In \reffi{fig:scalarselfs} we show the diagrams that have to be calculated to get the quantum corrections to scalar self-energies at one-loop in the mass eigenstate basis.

%%%%%%%%%%%%%%%%%%%%%%%%%%%% F I G U R E %%%%%%%%%%%%%%%%%%%%%%%%%%%%%%%%%%%%%%
%%%%%%%%%%%%%%%%%%%%%%%%%%%% F I G U R E %%%%%%%%%%%%%%%%%%%%%%%%%%%%%%%%%%%%%%

We calculated all diagrams in the 't Hooft-Feynman gauge, in which the Goldstone bosons $A_1$ and $H^\pm_1$ and the ghost fields $u^\pm$ and $u^Z$ have the same masses as the corresponding gauge bosons.
Calculating the $\cp$-odd self-energies $\Sigma_{A_i A_i}$ diagrammatically, we get the mass counterterms in mass eigenstate basis through the \refeq{eq:oddcounters}. Now inverting the rotation in 
\refeq{eq:masscounterrot} we can get the mass counterterms for the $\cp$-odd self-energies in the interaction basis via 
\begin{equation}\label{eq:oddcountersrot}
\left.\delta m_{\sigma}^2\right|^{\rm div}=
{U^A}^T \left.\delta m_{A}^2\right|^{\rm div} U^A \; ,
\end{equation}
Recognizing that
\begin{equation}
\left(\delta m_{\sigma}^2\right)_{3+i,1}=
\delta m_{\widetilde{\nu}_{iL}^{\mathcal{I}} H_{d}^{\mathcal{I}}}^{2} \; ,
\end{equation}
and that $m_{\widetilde{\nu}_{iL}^{\mathcal{I}} H_{d}^{\mathcal{I}}}^{2}$
 depends on $(m_{H_d\widetilde{L}_L}^2)_i$, we can extract
$\delta (m_{H_d\widetilde{L}_L}^2)_i$ through
\begin{align}
\delta \left(m_{H_d\widetilde{L}_L}^2\right)_i=&
\left.\left(\delta m_{\sigma}^2\right)_{3+i,1}\right|^{\rm div}+
\frac{2\mu Y^\nu_i}{\lambda}\delta\mu +
\lambda \left( v_d^2+v_u^2\right) Y^\nu_i \cos^3\beta\sin\beta
\; \delta\tb \notag \\
&+\frac{1}{2}\lambda Y^\nu_i \sin^2\beta \; \delta v^2 -
\frac{1}{2}\lambda\sin^2\beta Y^\nu_i\left( \delta v_{1L}^2 +
\delta v_{2L}^2 + \delta v_{3L}^2 \right) \notag \\
+& \left( \frac{\mu^2}{\lambda}+\frac{1}{2}\lambda\left(
v_d^2+v_u^2 \right) \sin^2\beta \right) \delta Y^\nu_i \notag \\
-&\left( \frac{\mu^2 Y^\nu_i}{\lambda^2}-\frac{1}{2}\left( v_d^2+v_u^2 \right)Y^\nu_i\sin^2\beta\right)\delta\lambda \; .
\end{align}

\paragraph{\protect\boldmath Renormalization of ${m_{\widetilde{L}_L}^2}_{ij}$:}
Since we neglect $\cp$-violation the counterterms for the 
non-diagonal elements of the hermitian matrix
${m_{\widetilde{L}_L}^2}_{ij}$ 
are symmetric under the exchange of the indices $i$ and $j$.
Then we can extract the counterterms for the non-diagonal elements in the same way as the ones for ${m_{H_d\widetilde{L}_L}^2}_i$ in the 
$\cp$-odd scalar sector. They appear in the tree-level mass matrix in 
\begin{equation}
m_{\widetilde{\nu}_{iL}^{\mathcal{I}} \widetilde{\nu}_{jL}^{\mathcal{I}}}^{2}=
\left(m_{\widetilde{L}}^2\right)_{ij}+
\frac{1}{2}\left(v_R^2+v_u^2\right)Y^\nu_iY^\nu_j \quad \text{for }
i\neq j \; .
\end{equation}
Hence, the counterterms
$\delta\left({m_{\widetilde{L}_L}^2}\right)_{ij}$ for $i\neq j$ 
are given by
\begin{align}
\delta\left({m_{\widetilde{L}_L}^2}\right)_{ij}=&
\left.\left(\delta m_{\sigma}^2\right)_{3+i,3+j}\right|^{\rm div} -
\frac{1}{2}\left( v_R^2+v_u^2 \right)\left(Y^\nu_i\delta Y^\nu_j -
Y^\nu_j\delta Y^\nu_i \right) \\
&-\frac{2\mu Y^\nu_i Y^\nu_j}{\lambda^2}\left( \frac{\delta\mu}{\mu}-
\frac{\delta\lambda}{\lambda} \right)-
\frac{1}{2}Y^\nu_i Y^\nu_j  \sin^2\beta\; \delta v^2 \notag \\
&-\left( v_d^2+v_u^2 \right)Y^\nu_i Y^\nu_j \cos^3\beta\sin\beta
\;\delta\tb+
\frac{1}{2}Y^\nu_i Y^\nu_j\sin^2\beta
\left( \delta v_{1L}^2 + \delta v_{3L}^2 + \delta v_{3L}^2 \right)
\; . \notag
\end{align} 
\paragraph{\protect\boldmath \FA\ modelfile:}
The diagrams and their amplitudes
that had to be calculated to obtain the counterterms, as described
in this section,
were generated using the Mathematica package
\FA~\cite{Hahn:2000kx} and further evaluated with the package
\FC~\cite{Hahn:1998yk}. The \FA\ model file for the
\mnSSM\ was created with the Mathematica program \texttt{SARAH}
\cite{Staub:2009bi}. We modified the model file to neglect $\cp$-violation 
by choosing all relevant parameters to be real. We also neglected 
flavor-mixing in the squark- and the quark-sector 
in this work. The \FA\ model file can be provided by the authors upon
request.
The calculation of renormalized two- and three-point
functions of the neutral scalars of the \mnSSM\ at one-loop accuracy is
thereby fully automated. 
(as it is in the MSSM~\cite{Fritzsche:2013fta}).

In \refse{sec:numanal} we will present our predictions for the Higgs
masses in the \mnSSM\ compared to the ones of the NMSSM.
To be able to make this comparison, we had to calculate the
NMSSM-predictions in the same renormalization scheme and using
the same conventions as were used in the \mnSSM.
This is why we calculated the one-loop self-energies in the NMSSM 
with our own NMSSM-modelfile for \FA/\FC\ created with
\texttt{SARAH} using the same procedure as for the \mnSSM.
We verified that the results
calculated in the NMSSM with our modelfile are equal to the results
calculated with the modelfile presented in \citere{Paehr:276451}, which
was a good check that the generation of the modelfiles for the NMSSM
and the \mnSSM\ was correct.

%%%%%%%%%%%%%%%%%%%%%%%%%%%%%%%%%%%%%%%%%%%%%%%%%%%%%%%%%%%%%%%%%%%%%%%%%%%%%%%
%%%%%%%%%%%%%%%%%%%%%%%%%%%%%%%%%%%%%%%%%%%%%%%%%%%%%%%%%%%%%%%%%%%%%%%%%%%%%%%

\section{Loop corrected Higgs boson masses}
\label{sec:getmasses}

In the previous section we have derived an
OS/\DRbar\ renormalization scheme for the \mnSSM\ Higgs sector. This
can be applied (via the future \FA\ model file, once the
counterterms are implemented) to any higher-order
correction in the \mnSSM. As a first application, we evaluate the full
one-loop corrections to the $\cp$-even scalar sector in the \mnSSM.
Due to the still missing implementation of counterterms
in the \FA\ model file,
the calculation of the renormalized scalar self-energies is done in
two steps. Firstly, the unrenormalized self-energies are calculated
using \FA\ and \FC, and subsequently the self-energies are renormalized
subtracting (by hand) the field renormalization and
mass counterterms, as will
be described in the next section.

\subsection{Evaluation at one-loop} 

Here we describe the final form of the renormalized $\cp$-even scalar
self-energies 
$\hat{\Sigma}_{hh}$ and how the loop corrected physical masses of the
Higgs boson masses are evaluated.

The one-loop renormalized self-energies in the mass eigenstate basis are
given by
\begin{equation}\label{eq:renomself}
\hat{\Sigma}_{h_i h_j}^{(1)}\left( p^2 \right) = \Sigma_{h_i h_j}^{(1)}\left( p^2 \right) + \delta Z^H_{ij}
\left( p^2 - \frac{1}{2}\left( m_{h_i}^2 + m_{h_j}^2 \right) \right)
- \left(\delta m_{h}^2\right)_{ij} \; ,
\end{equation}
with the field renormalization constants $\de Z^H$ 
and the mass counter terms $\delta m_{h}^2$
in the mass eigenstate basis
defined by the rotations in \refeq{eq:fieldrenorotate}
and \refeq{eq:masscounterrot}.
$\Sigma_{h_i h_j}$ is the unrenormalized self-energy obtained by
calculating the diagrams shown in \reffi{fig:scalarselfs} 
with the $\cp$-even states $h$ on the external legs.
The self-energies were calculated in the Feynman
gauge, so that gauge-fixing terms do not yield counterterm contributions
in the Higgs sector at one-loop. The loop integrals were regularized
using dimensional reduction \cite{SIEGEL1979193,CAPPER1980479} and
numerically evaluated for arbitrary real momentum using
\LT~\cite{Hahn:1998yk}. The contributions from complex values of $p^2$
were approximated using a Taylor expansion with respect to the imaginary part
of $p^2$ up to first order. 

In \refeq{eq:renomself} we already made use of the fact that $\delta
Z^H$ is real and symmetric in our renormalization scheme. The mass
counterterms are defined as functions of the counterterms of the free
parameters following \refeq{eq:masscounterderive} and
\refeq{eq:masscounterrot}. They contain finite contributions from the
tadpole counterterms and from the counterterm for the gauge boson mass
$\MZ^2$. The matrix $\delta m_{h}^2$ is real and symmetric.

The renormalized self-energies enter the inverse propagator matrix
\begin{equation}\label{eq:renoselfdef}
\hat{\Gamma}_{h}=\ii \left[ p^2\; \mathbbm{1} -
\left( m_{h}^2 - \hat{\Sigma}_{h}\left(p^2\right) \right)
\right] \; , \qquad  \text{with }
\left(\hat{\Sigma}_{h}\right)_{ij} = 
\hat{\Sigma}_{h_i h_j} \; .
\end{equation}
The loop-corrected scalar masses squared are the zeroes of the
determinant of the inverse propagator matrix. The determination of
corrected masses has to be done numerically when we want to account for
the momentum-dependence of the renormalized self-energies. This is done
by an iterative method that has to be carried out for each of the six
squared loop-corrected masses \cite{Fuchs:2015jwa}.

%%%%%%%%%%%%%%%%%%%%%%%%%%%%%%%%%%%%%%%%%%%%%%%%%%%%%%%%%%%%%%%%%%%%%%%%%%%%%%%

\subsection{Inclusion of higher orders}\label{sec:higherorders}

In \refeq{eq:renoselfdef} we did not include the superscript $^{(1)}$ in
the self-energies. Restricting the numerical evaluation to a pure
one-loop calculation would lead to very large theoretical uncertainties.
These can be avoided by the inclusion of corrections beyond the one-loop
level. Here we follow the approach of \citere{Drechsel:2016jdg} and
supplement the \mnSSM\ one-loop results by higher-order corrections in
the MSSM limit as provided by \fh\ (version 2.13.0)~\cite{Heinemeyer:1998yj,Hahn:2009zz,Heinemeyer:1998np,Degrassi:2002fi,Frank:2006yh,Hahn:2013ria,Bahl:2016brp,feynhiggs-www}.
In this way the leading and subleading two-loop corrections are
included, as well as a resummation of large logarithmic terms, see the
discussion in \refse{sec:intro}, 
\begin{equation}
\hat{\Sigma}_{h}\left( p^2 \right) = 
\hat{\Sigma}_{h}^{(1)}\left( p^2 \right) +
\hat{\Sigma}_{h}^{(2')} +
\hat{\Sigma}_{h}^{\rm resum} \; .
\end{equation}
In the partial two-loop contributions $\hat{\Sigma}_{h}^{(2')}$ we take
over the corrections of
\order{\alpha_s\alpha_t,\alpha_s\alpha_b,\alpha_t^2,\alpha_t\alpha_b}, 
assuming that the MSSM-like corrections are also valid in the
\mnSSM. This assumption 
is reasonable since the only difference between the squark sector of the
\mnSSM\ in comparison to the MSSM are the terms proportional to 
$Y^\nu_i v_{iL}$ in the non-diagonal element of the up-type squark mass 
matrices (see \refeq{eq:usquarks12})
and the terms proportional to $v_{iL}v_{iL}$ in the diagonal
elements of the up- and down-type squark mass matrices
(see \refeq{eq:usquarks11}, \refeq{eq:usquarks22},
\refeq{eq:dsquarks11} and \refeq{eq:dsquarks22}),
which numerically will always be negligible in
realistic scenarios since $v_{iL}\ll v_d,v_u,v_R$.
Furthermore,. in \citere{Drechsel:2016ukp} the quality of the MSSM
approximation was tested in the NMSSM, showing that the genuine NMSSM
contributions are in most cases sub-leading. The same is expected for
the contributions stemming from the resummation of large logarithmic
terms given by $\hat{\Sigma}_{h}^{\rm resum}$.

%%%%%%%%%%%%%%%%%%%%%%%%%%%%%%%%%%%%%%%%%%%%%%%%%%%%%%%%%%%%%%%%%%%%%%%%%%%%%%%
%%%%%%%%%%%%%%%%%%%%%%%%%%%%%%%%%%%%%%%%%%%%%%%%%%%%%%%%%%%%%%%%%%%%%%%%%%%%%%%

\section{Numerical analysis}
\label{sec:numanal}

In the following we present for the first time the full one-loop
corrections to the scalar masses in the \mnSSM, 
with one generation of right-handed neutrinos obtained in the
Feynman-diagrammtic approach, taking into account all parameters of the
model and the complete dependence on the external momentum, which
includes a consistent treatment of the imaginary parts of the scalar
self-energies. 
Our results extend the known ones in the literature of
the MSSM and the NMSSM to a model, which has a
rich and unique phenomenology through explicit $R$-parity breaking. The
one-loop results are supplemented by known higher-loop results from the
MSSM (see the previous section) to reproduce the Higgs mass value of 
$\sim 125 \gev$~\cite{Aad:2015zhl}. 
Here the theory uncertainty must be kept in mind. In the
MSSM it is estimated to be at the level of 
$2-3 \gev$~\cite{Degrassi:2002fi,Bahl:2017aev}, and in extended models
it is naturally slightly larger. 

We will present results in several different scenarios, in
all of which one scalar with the correct SM-like Higgs mass
is reproduced.
To get an estimation of the significance of quantum
corrections to the Higgs masses that are unique for the \mnSSM, we
compare the results to the corresponding ones in the NMSSM.
The results in the NMSSM are obtained by a calculation based on
\citere{Drechsel:2016ukp}, but with slightly changed renormalization
conditions to be as close as possible to the calculation in the
\mnSSM. While Ref.~\cite{Drechsel:2016ukp} uses the mass squared of
the charged Higgs mass as input parameter and renormalizes it as OS
parameter we instead use $\DRbar$ conditions for $A^\lambda$.

The benchmark points used in the following were not tested in detail
against experimental bounds including the $R$-parity violating effects of
the \mnSSM. They have been chosen to exemplify the potential magnitude
of unique \mnSSM-like corrections. Nevertheless, the values
we picked for the free parameters should be close to realistic and
experimentally allowed scenarios: the parameters in the scalar sector
are taken over from calculations in the NMSSM~\cite{Drechsel:2016ukp}, and
unique \mnSSM\ parameters are chosen in a range to reproduce
neutrino masses of the correct order of magnitude. That means that the
neutrino Yukawas $Y^\nu_i$ should be of the order $10^{-6}$ to generate
neutrino masses of the order less than $1\,\, \mathrm{eV}$. 
For the left-handed sneutrino
vevs this directly implies $v_{iL}\ll v_d, v_u$ so
that the tadpole coefficients vanish at tree-level
\cite{Escudero:2008jg}. We will leave a more detailed discussion of
numerical results for a future publication, in which we will also include
three generations of right-handed neutrinos.

%%%%%%%%%%%%%%%%%%%%%%%%%%%%%%%%%%%%%%%%%%%%%%%%%%%%%%%%%%%%%%%%%%%%%%%%%%%%%%%

\subsection{NMSSM-like crossing point scenario}
\label{sec:nmssmpoint}

The first scenario we want to analyze is one studied in the NMSSM with a
singlet becoming the LSP in the region of $\lambda > \kappa$
taken from Ref.~\cite{Drechsel:2016ukp}. 
This scenario was tested therein against the experimental limits 
implemented in \texttt{HiggsBounds 4.1.3}~\cite{Bechtle:2008jh,Bechtle:2011sb,Bechtle:2013gu,Bechtle:2013wla,Bechtle:2015pma}.
It has the nice feature that there is a crossing point when
$\lambda\approx\kappa$ in the neutral scalar sector, in which the masses
of the singlet and the SM-like Higgs become degenerate and NMSSM-like
loop corrections become significant \cite{Drechsel:2016jdg}. 

In \refta{tab:crossingpoint} we list the values chosen for the
parameters. The SM-like parameters from the electroweak sector and
the lepton and quark masses are given in appendix \ref{app:smvalues} 
in \refta{tab:smsum}. The
parameters present in the \mnSSM\ and the NMSSM are of course chosen
equally in both models. 
The region $\lambda < 0.026$ is excluded
because the left-handed sneutrinos become tachyonic at tree-level. The
flavor-changing non-diagonal elements in the slepton sector are
zero. The value for $A^\lambda$ is chosen to correspond to a mass of
$m_{H^\pm}= 1000 \gev$ for the charged Higgs mass in the NMSSM with
$m_{H^\pm}$ renormalized OS and $A^\lambda$ not being a free
parameter. $A^\kappa$ should be chosen to be negative in our convention
(when $\kappa$ is positive) to avoid false vacua~\cite{Escudero:2008jg}
or tachyons in the pseudo-scalar sector~\cite{Ghosh:2014ida}. It should
be kept in mind 
that the diagonal soft scalar masses in the neutral sector are extracted
from the values for
$v_{iL}$, $\tb$ and
$\mu$ via the tadpole equations, and their non-diagonal,
flavor-violating elements are always set to zero at tree-level.
This is of crucial importance for the comparison of the
scalar masses
in the \mnSSM\ and the NMSSM, since in the NMSSM the soft slepton masses
$m_{\widetilde{L}}^2$ are independent parameters, while in the \mnSSM\ the
diagonal elements are dependent parameters fixed by the tadpole
\refeqs{eq:tp4}, 
when the vevs are used as input.
%This is more convenient 
The latter strategy is particularly convenient since the order
of magnitude of the vevs is roughly fixed through the electroweak seesaw mechanism
by demanding neutrino masses below the eV scale,
while the soft scalar masses are not directly related to any physical observable.
Consequentially, for each parameter point
calculated in the \mnSSM, the corresponding values that have to be chosen for
$m_{\widetilde{L}}^2$ in the NMSSM have to be adjusted accordingly,
defined as a function of all the free parameters appearing in the
Higgs potential.

%%%%%%%%%%%%%%%%%%%%%%%%%% T A B L E %%%%%%%%%%%%%%%%%%%%%%%%%%%%%%%%%%%%%%%%%%
\begin{table}
\renewcommand{\arraystretch}{1.7}
\centering
\begin{tabular}{c c c c c c c c c c c}
 $v_{iL}/\sqrt{2}$ & $Y^\nu_i$ & $A^\nu_i$ & $\tb$ & $\mu$ & $\lambda$ &
 	$A^\lambda$ & $\kappa$ & $A^\kappa$ & $M_1$ & \\
 \hline
 $10^{-4}$ & $10^{-6}$ & $-1000$ & $8$ & $125$ & $[0.026;0.3]$ &
 	$897.61$ & $0.2$ & $-300$ & $143$ & \\
 \hline
 \hline
 $M_2$ & $M_3$ & $m_{\widetilde{Q}_{iL}}^2$ &
 	$m_{\widetilde{u}_{iR}}^2$ & $m_{\widetilde{d}_{iR}}^2$ &
 	$A^u_3$ & $A^u_{1,2}$ & $A^{d}_{1,2,3}$ & $(m_{\widetilde{e}}^2)_{ii}$ &
 	$A^e_{33}$ & $A^e_{11,22}$ \\
 \hline
 $300$ & $1500$ & $1500^2$& $1500^2$ & $1500^2$ & $-2000$ & $-1500$ &
 	$-1500$ & $200^2$ & $-1500$ & $-100$ 
\end{tabular}
%\vspace{-0.5cm}
\caption{Input parameters for the NMSSM-like crossing point scenario;
  all masses and values for trilinear parameters are in GeV.}
\label{tab:crossingpoint}
\renewcommand{\arraystretch}{1.0}
\end{table}
%%%%%%%%%%%%%%%%%%%%%%%%%% T A B L E %%%%%%%%%%%%%%%%%%%%%%%%%%%%%%%%%%%%%%%%%%

\begin{figure}
    \centering
    \includegraphics[width=0.7\textwidth]{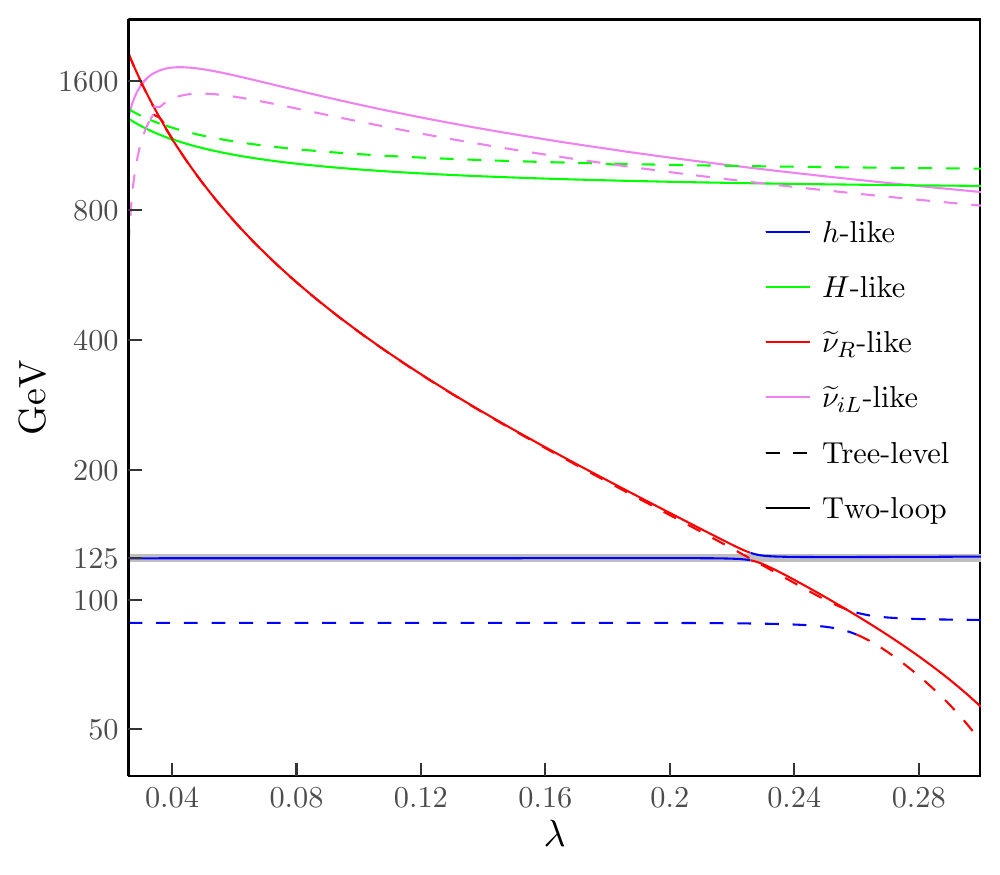}
    \vspace{-0.4cm}
    \caption{Spectrum of $\cp$-even scalar masses in NMSSM-like
      crossing point scenario. The three left-handed sneutrinos
      $\widetilde{\nu}_{iL}$ are degenerate.\protect\footnotemark} 
    \label{fig:crossingspec}
\end{figure}

In \reffi{fig:crossingspec} we show the resulting spectrum of the
$\cp$-even scalars at tree-level and including the full one-loop and
two-loop contributions.%
\footnote{Here and in the following we denote with ``two-loop'' result the
one-loop plus partial two-loop plus resummation corrected masses.}%
~The standard model Higgs mass value
is reproduced accurately when the quantum corrections are included. The
heavy MSSM-like Higgs $H$ and the left-handed sneutrinos are at the
TeV-scale and rather decoupled from the SM-like Higgs boson. The three
left-handed sneutrinos are degenerate because the \mnSSM-like parameters
are set equal for all flavors. 
%%%%%%%%%%%%%%%%%%%%%%%%%%%% F I G U R E %%%%%%%%%%%%%%%%%%%%%%%%%%%%%%%%%%%%%%
\footnotetext{All plots
      have been produced using \texttt{ggplot2}~\cite{ggplot2}
      and \texttt{tikzDevice}~\cite{tikzDevice} in \texttt{R}~\cite{Rlanguage}.}
%%%%%%%%%%%%%%%%%%%%%%%%%%%% F I G U R E %%%%%%%%%%%%%%%%%%%%%%%%%%%%%%%%%%%%%%
The singlet-like scalar mass heavily depends on $\lambda$,
because when $\mu$
is fixed, increasing $\lambda$ leads to
a smaller value for $v_R$ (see \refeq{eq:mudef}).
As was observed in \citere{Drechsel:2016ukp}, the loop-corrected
mass of the singlet becomes smaller than the SM-like Higgs boson mass
at about $\lambda\approx\kappa$.
We observe non-negligible loop-corrections to the singlet in the
region of $\lambda$ where the singlet is the lightest neutral scalar.

\begin{figure}
  \centering
  \includegraphics[width=0.6\textwidth]{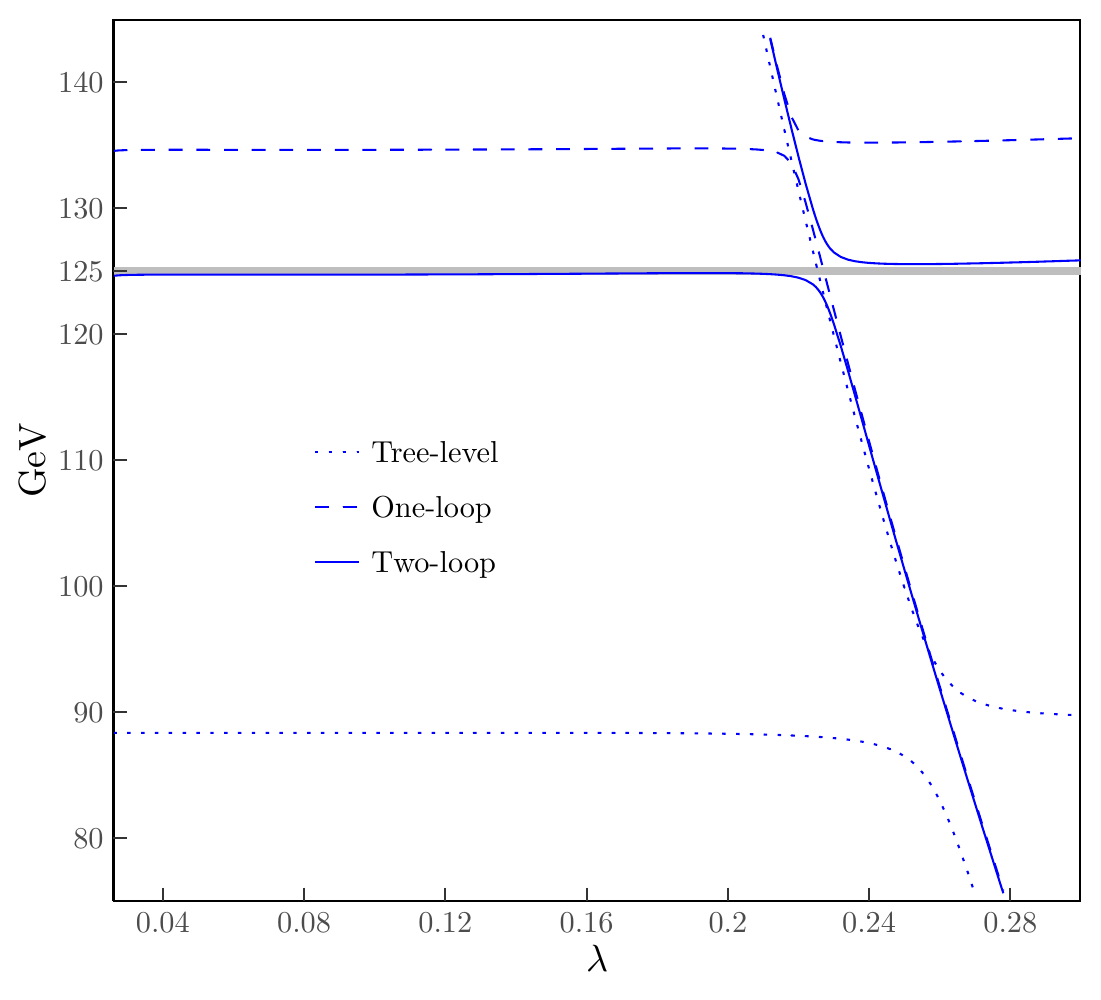}
  \vspace{-0.5cm}
  \caption{Tree-level, one-loop and two-loop corrected
    masses of the SM-like Higgs boson in the \mnSSM\ in the NMSSM-like
    crossing point scenario.} 
  \label{fig:nmssmdif}
\end{figure}

Due to the similarity of the Higgs sectors of the
NMSSM and the \mnSSM, the masses of the doublet-like Higgs bosons and
the right-handed sneutrino will be of comparable size as the masses
predicted for the doublet-like Higgses and the singlet in the
NMSSM.
In \reffi{fig:nmssmdif} we show the tree-level and the one- and
two-loop corrected mass of the SM-like Higgs boson in the
crossing-point scenario. One can see that, as expected, 
the two-loop corrections are crucial to predict a SM-like Higgs mass of $125\gev$.
Indeed, our analysis confirmed that differences in the prediction of the SM-like
Higgs boson mass are negligible compared to the current experimental
uncertainty~\cite{Aad:2015zhl}
and the anticipated
experimental accuracy of the ILC of about
$\lsim 50 \mev$~\cite{Moortgat-Picka:2015yla}, even when there is a
substantial mixing between left-handed sneutrinos and the SM-like
Higgs at tree-level or one-loop.
Apart from that, they
are clearly exceeded by the (future) parametric uncertainties in the Higgs-boson
mass calculations. Consequently, the
Higgs sector alone will not be sufficient to distinguish the
\mnSSM\ from the NMSSM. On the other hand, we can regard
the theoretical uncertainties
in the NMSSM and the \mnSSM\ to be at the same level of accuracy.
%%%%%%%%%%%%%%%%%%%%%%%%%%%% F I G U R E %%%%%%%%%%%%%%%%%%%%%%%%%%%%%%%%%%%%%%
%%%%%%%%%%%%%%%%%%%%%%%%%%%% F I G U R E %%%%%%%%%%%%%%%%%%%%%%%%%%%%%%%%%%%%%%

%%%%%%%%%%%%%%%%%%%%%%%%%%%%%%%%%%%%%%%%%%%%%%%%%%%%%%%%%%%%%%%%%%%%%%%%%%%%%%%

\subsection{\protect\boldmath Light $\tau$-sneutrino scenario}
\label{sec:lighttaulsp}

In the previous scenario we observed that, in
a scenario where the left-handed sneutrinos where practically 
decoupled from the SM-like
Higgs boson, the unique \mnSSM-like
corrections do not account for a substantial deviation of the SM-like Higgs
mass prediction compared to the NMSSM.
In this section we will investigate a scenario in which one of
the left-handed sneutrinos has a small mass close to SM Higgs boson
mass. The phenomenology of such a spectrum was recently studied in
detail,
including a comparison of its predictions with the
LHC searches~\cite{Ghosh:2017yeh,Lara:2018rwv}. It was found that a light
left-handed sneutrino as the LSP can give rise to distinct
signals for the \mnSSM\ (for instance, final states with
diphoton plus missing energy, diphoton plus leptons and multileptons).
%%%%%%%%%%%%%%%%%%%%%%%%%% T A B L E %%%%%%%%%%%%%%%%%%%%%%%%%%%%%%%%%%%%%%%%%%
\begin{table}
\renewcommand{\arraystretch}{1.7}
\centering
\begin{tabular}{c c c c c c c c c c c}
 $v_{1,2L}/\sqrt{2}$ & $v_{3L}/\sqrt{2}$ & $Y^\nu_i$ & $A^\nu_i$ & $\tb$ & $\mu$ & $\lambda$ &
 	$A^\lambda$ & $\kappa$ & $A^\kappa$ \\
 \hline
 $10^{-5}$ & $4\cdot 10^{-4}$ & $5\cdot 10^{-7}$ & $-400$ & $10$ & $270$ & $[0.19;0.3]$ &
 	$1000$ & $0.3$ & $-1000$
\end{tabular}
%\vspace{-0.5cm}
\caption{Input parameters for the light $\tau$-sneutrino scenario;
		 all masses and values for trilinear parameters are in GeV.}
\label{tab:lsptaupoint}
\renewcommand{\arraystretch}{1.0}
\end{table}
%%%%%%%%%%%%%%%%%%%%%%%%%% T A B L E %%%%%%%%%%%%%%%%%%%%%%%%%%%%%%%%%%%%%%%%%%

In \refta{tab:lsptaupoint} we list the relevant parameters that were
chosen to
obtain a light left-handed $\tau$-sneutrino. The parameters not shown
here are chosen to be the same as in the previous case, shown in
\refta{tab:crossingpoint}. One can see that the vev
$v_{3L}$ (corresponding to $\widetilde{\nu}_{3L}$) was increased w.r.t.\
the NMSSM-like scenario. The reason for this becomes clear when one
extracts the leading terms of the diagonal tree-level mass matrix
element of the left-handed sneutrinos,
\begin{equation}\label{eq:approxnumass}
m_{\widetilde{\nu}_{iL}^{\mathcal{R}} \widetilde{\nu}_{iL}^{\mathcal{R}}}^{2}
\approx 
\frac{Y^\nu_i v_R v_u}{2v_{iL}}
\left( - \sqrt{2}A^\nu_i-\kappa v_R 
				+\frac{\sqrt{2}\mu}{\tan\beta} \right) 	
				\; .
\end{equation}
The tree-level masses of the left-handed sneutrinos are roughly
proportional to the inverse of their vev. We also
decreased $A^\nu_3$ in comparison to the previous scenario, keeping it
negative, so that it is of order $\kappa v_R$ and the sum in the brackets of \refeq{eq:approxnumass} becomes small.

%%%%%%%%%%%%%%%%%%%%%%%%%%%% F I G U R E %%%%%%%%%%%%%%%%%%%%%%%%%%%%%%%%%%%%%%
\begin{figure}[!]
    \centering
    \hspace{1cm}\includegraphics[width=0.8\textwidth]{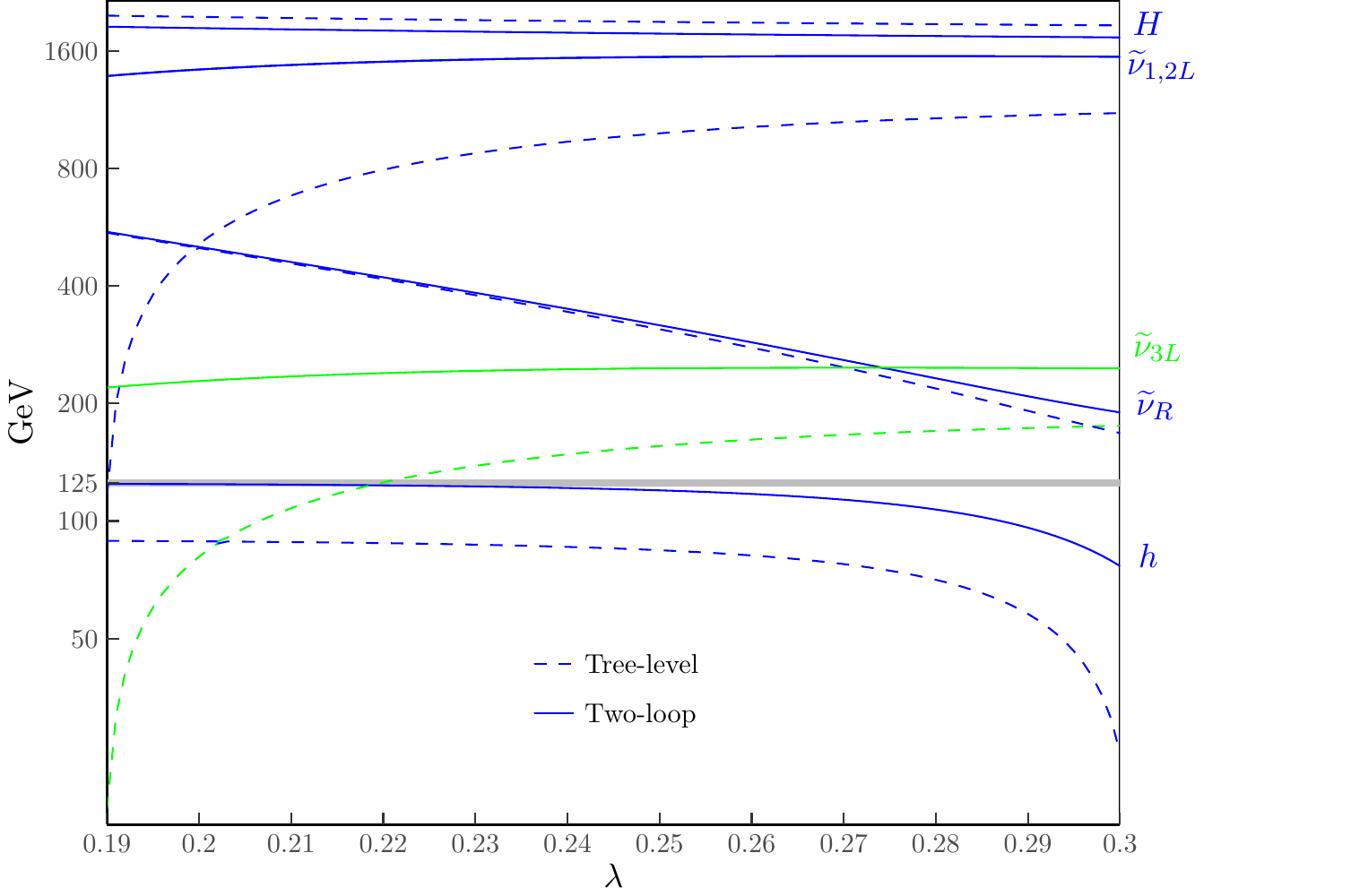}
    \vspace{-0.5cm}
    \caption{$\cp$-even scalar mass spectrum of the \mnSSM\ in the light
      $\tau$-sneutrino scenario, see \refta{tab:lsptaupoint}.
      On the right side we state the dominant composition
      of the mass eigenstates.}
    \label{fig:sntaulspspec}
\end{figure}
%%%%%%%%%%%%%%%%%%%%%%%%%%%% F I G U R E %%%%%%%%%%%%%%%%%%%%%%%%%%%%%%%%%%%%%%

In \reffi{fig:sntaulspspec} we show the tree-level and loop-corrected
spectrum of the scalars in the region of $\lambda$ where there are no
tachyons at tree-level.
For too small $\lambda$ the tree-level mass of
$\widetilde{\nu}_{3L}$ becomes tachyonic,
because when $\mu=(v_R \lambda)/ \sqrt{2}$ is
fixed $v_R$ has to grow and the second term in the bracket
of \refeq{eq:approxnumass} will grow larger than the sum of the first and the
third term.
For too large $\lambda$,
the tree-level mass of the SM-like Higgs boson
becomes tachyonic.
In particular, it starts to mix with the tree-level singlet mass, which
becomes tachyonic because $v_R$ decreases when $\lambda$
increases.
The central value of the
SM Higgs boson mass is reproduced in this scenario
up to values of $\lambda\leq\num{0.22}$. However, 
considering the theoretical uncertainty even higher values of $\lambda$ 
can be viable. For $\lambda=0.236$ the prediction for the SM-like 
Higgs mass decreases below $m_{h_1}\approx 122\gev$. 
As discussed in the introduction we assume a theory uncertainty of
$\sim 3 \gev$ on the mass evaluation, 
so we consider in this scenario the region $\lambda\leq0.236$ to be valid 
regarding the SM Higgs boson mass.
An interesting observation is
that the masses of light left-handed sneutrinos are
mainly induced via quantum corrections, while the tree-level mass
approaches 0 for small values of $\lambda$.
This indicates that a consistent treatment of
quantum corrections to light sneutrino masses is of
crucial importance.

The large upward shift of the left-handed sneutrino masses through
the one-loop 
corrections is due to the fact that in the \mnSSM\ the sneutrino fields
are part of the Higgs potential, each with an associated
tadpole coefficient $T_{\widetilde{\nu}_{iL}}$. To ensure the stability of the
vacuum \wrt quantum corrections, the tadpoles are renormalized OS, absorbing
all finite corrections into the counterterms $\delta T_{\widetilde{\nu}_{iL}}$
(see \refse{sec:condis}). In the mass counterterms for the left-handed
sneutrinos the finite parts $\delta T_{\widetilde{\nu}_{iL}}^{\rm fin}$
introduce the main finite contribution in the form
\begin{equation}
\delta m_{\widetilde{\nu}_{iL}^{\mathcal{R}}
  \widetilde{\nu}_{iL}^{\mathcal{R}}}^{2\;{\rm fin}}
    = -\frac{\delta T_{\widetilde{\nu}_{iL}}^{\rm fin}}{v_{iL}} + \cdots \; ,
\end{equation}
which is enhanced by the inverse of the vev of $\widetilde{\nu}_{iL}$.
It is these terms inside the counterterms of the renormalized self-energies
$\hat{\sum}_{\widetilde{\nu}_{iL}^{\mathcal{R}}
  \widetilde{\nu}_{iL}^{\mathcal{R}}}^{(1)}$ that shift the poles of the
propagator matrix and increase the masses of the left-handed sneutrinos,
especially in cases where the tree-level masses are small.

%%%%%%%%%%%%%%%%%%%%%%%%%%%% F I G U R E %%%%%%%%%%%%%%%%%%%%%%%%%%%%%%%%%%%%%%
% \begin{figure}[!]
%     \centering
%     \hspace{-2cm}\includegraphics[width=0.8\textwidth]{sntauLSPcompare.pdf}
%     \vspace{-0.5cm}
%     \caption{Light $\tau$-sneutrino scenario, see \refta{tab:lsptaupoint}.
%     In the shaded region the prediction for the SM-like Higgs mass is 
%     below $122\gev$.
% \textit{Upper left:} Masses of the SM-like Higgs, the left-handed
% $\tau$-sneutrino and the right-handed sneutrino in the \mnSSM\ at
% tree-level and one-loop. \textit{Upper right:} Masses of the SM-like
% Higgs and the singlet in the NMSSM at tree-level and
% one-loop. \textit{Lower left:} Mass differences
% $\Delta\mhsm=\mhsm^{\mu\nu\text{SSM}}-\mhsm^{\text{NMSSM}}$ of the SM-like
% Higgs at tree-level and one-loop. \textit{Lower right:} Mass differences
% $\Delta m_{\widetilde{\nu}_R}=m_{\widetilde{\nu}_R}^{\mu\nu\text{SSM}}
% -m_{s}^{\text{NMSSM}}$ of the
% singlet-like Higgs at tree-level and one-loop.} 
%     \label{fig:sntaulspcompare}
% \end{figure}
\begin{figure}[!]
     \centering
     \includegraphics[width=0.9\textwidth]{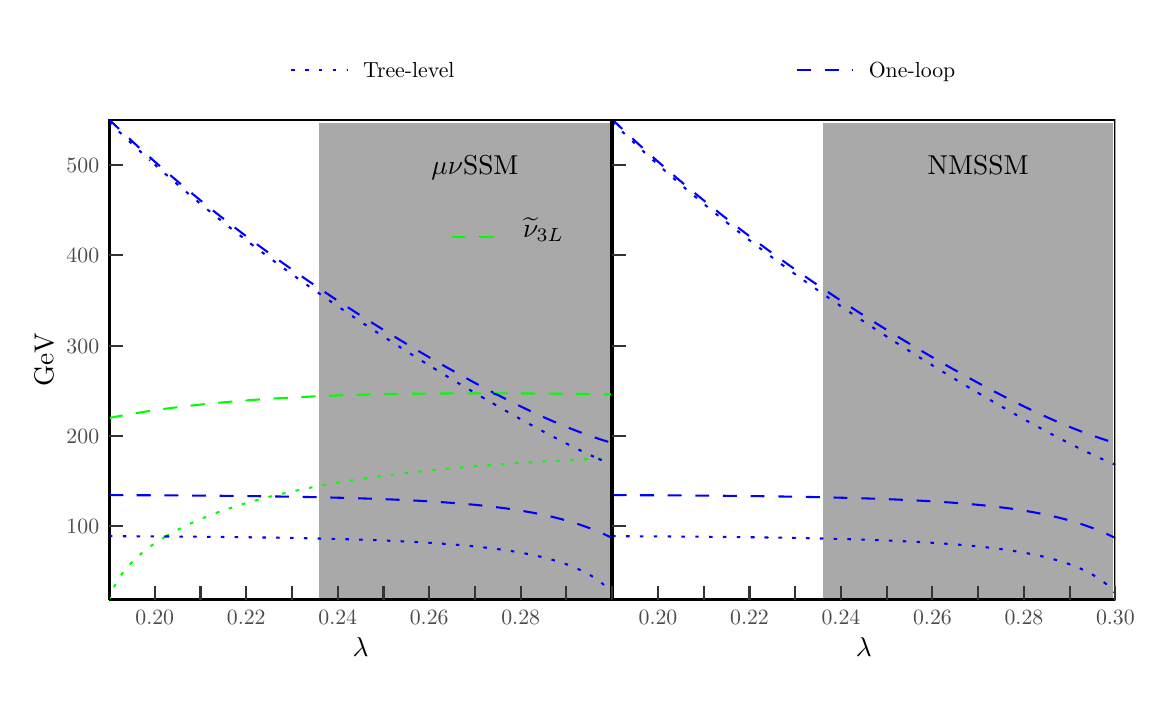}
     \vspace{-1.0cm}
     \caption{Light $\tau$-sneutrino scenario, see \refta{tab:lsptaupoint}.
     In the shaded region the prediction for the SM-like Higgs mass is 
     below $122\gev$. \textit{Left:} Masses of the SM-like Higgs, the left-handed
     $\tau$-sneutrino and the right-handed sneutrino in the \mnSSM\ at
     tree-level and one-loop. \textit{Right:} Masses of the SM-like
     Higgs and the singlet in the NMSSM at tree-level and
     one-loop.}
     \label{fig:sntaulspcompare}
\end{figure}
%%%%%%%%%%%%%%%%%%%%%%%%%%%% F I G U R E %%%%%%%%%%%%%%%%%%%%%%%%%%%%%%%%%%%%%%

This behavior is a peculiarity of the \mnSSM, meaning that the leptonic
sector and the Higgs sector are mixed through the breaking of $R$-parity.
The relations between the vevs $v_{iL}$ and the soft masses
$m_{\widetilde{L}}^2$ via the tadpole equations automatically lead to
dependences between the sneutrino masses and, for instance, the neutrino
or the Higgs sector.
In the NMSSM, on the other hand, the sneutrinos are not part of the
Higgs potential, since the fields are protected by lepton-number conservation.
There, the soft masses $m_{\widetilde{L}}^2$ are, without further assumptions,
free parameters that can be chosen without taking into account
any leptonic observable (such as neutrino masses and mixings).
In principle, the additional dependences of the
\mnSSM\ scalar (neutrino) masses on the neutrino sector could be
used (e.g.\ when all neutrino masses and mixing angles will be
known with sufficient experimental accuracy)
to restrict the possible range of $m_{\widetilde{L}}^2$, and thus the
possible values for the left-handed sneutrino masses. However, with
our current experimental knowledge on the neutrino masses,
the possible values for the vevs $v_{iL}$, and hence the
possible range of left-handed sneutrino masses, are
effectively not yet constrained.

It should be noted as well, that the soft masses
$m_{\widetilde{L}}^2$ also appear 
in the mass matrix of the charged scalars (see \refeq{eq:mlinmch})
and the pseudoscalars (see \refeq{eq:mlinmps}).
In many cases they are the dominant
term in the tree-level masses of the left-handed sleptons and
sneutrinos, so the values of the masses of charged sleptons and
sneutrino of the same family will be close. A precise treatment of
quantum corrections of the size observed in \reffi{fig:sntaulspspec} is
extremely important in those cases, since 
they might easily change the relative sign of their
mass differences. This can result in a complete change of the
phenomenology of the corresponding benchmark point, for instance when
either the neutral (pseudo)scalar or the charged scalar is the
LSP~\cite{Ghosh:2017yeh,Lara:2018rwv}.

We compare the relevant spectrum of the \mnSSM\ to the
corresponding one in the NMSSM in \reffi{fig:sntaulspcompare}.
We show the tree-level
and one-loop corrected masses of the light scalars in the \mnSSM,
and the masses of the SM-like Higgs boson and the singlet in the
NMSSM on the right, with parameters set accordingly.
We shade in grey the region of $\lambda$ where the the prediction for
the SM-like Higgs boson mass is below $122\gev$ if two-loop corrections
are included. As expected, the SM-like Higgs-boson mass and
the mass of 
the singlet turn out to be equal in both models. Even in regions
where there is a substantial mixing
of the SM-like Higgs boson with the left-handed sneutrinos, something that
cannot occur in the NMSSM, the differences in the SM-like Higgs
mass prediction are not larger than a few keV.

\begin{figure}[t]
    \centering
    \includegraphics[width=1.0\textwidth]{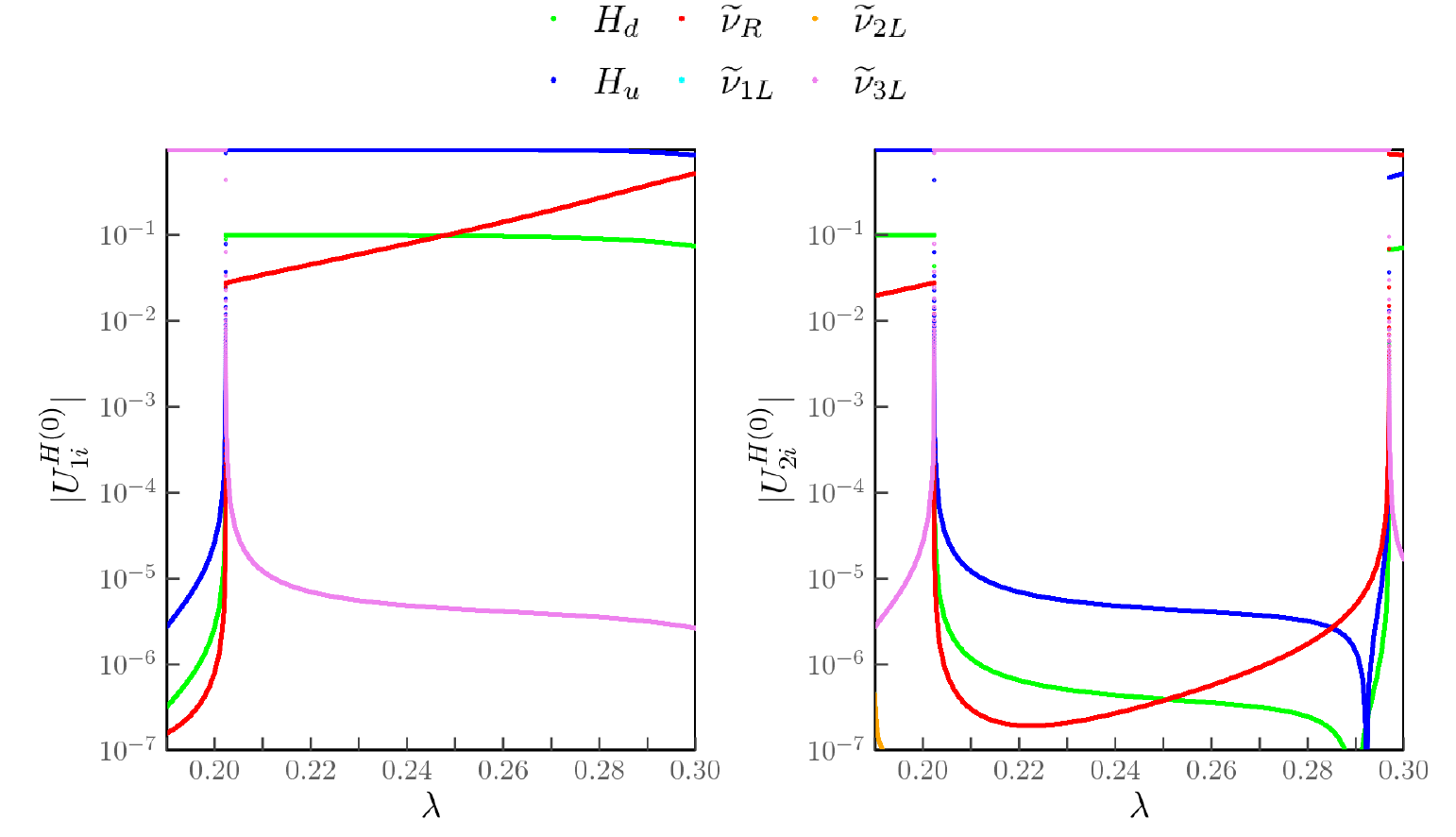}
    \vspace*{-1.0cm}
    \caption{Light $\tau$-sneutrino scenario, see \refta{tab:lsptaupoint}.
             We show the absolute values of the mixing matrix elements at
             tree-level $|U_{1i}^{H(0)}|$ (left) and
             $|U_{2i}^{H(0)}|$ (right), whose squared value
             define the admixture of the two-lightest
             $\cp$-even scalar mass eigenstate $h_{1,2}$ with the fields
             $\varphi_i=(H_d,H_u,\widetilde{\nu}_R,\widetilde{\nu}_{1L},
             \widetilde{\nu}_{2L},\widetilde{\nu}_{3L})$ in the
             interaction basis. A substantial mixing of the $\tau$-sneutrino
             $\widetilde{\nu}_{3L}$ with the SM-like Higgs boson $h^{125}$ and
             with the singlet $\widetilde{\nu}_R$ is present in the narrow region
             where the corresponding tree-level masses are degenerate
             (for example in the \textit{ right} plot at
             $\lambda\sim 0.20237$ and $\lambda\sim0.29692$).}
    \label{fig:sntaulsmixing}
\end{figure}
%%%%%%%%%%%%%%%%% F I G U R E %%%%%%%%%%%%%%%%%%%%%%%%%%%%%%%%%%%%%%%%%%%%%%%%%

It is rather surprising that the SM-like Higgs masses coincide this precisely
in both models, considering the fact
that a substantial mixing with the sneutrino is
possible at tree-level, as we show in \reffi{fig:sntaulsmixing}.
We individually plot the mixing matrix elements of
the two lightest $\cp$-even scalars, whose
squared values define the composition of each mass eigenstate
at tree-level. In the cross-over point of the $\tau$-sneutrino and the
SM-like Higgs boson the lightest scalar results to be a mixture
of $\widetilde{\nu}_\tau$ and the doublet-components $H_u$ and $H_d$, as
one can see in the upper left plot of \reffi{fig:sntaulsmixing}.
For example, if we fine-tune $\lambda = 0.20237$ we find that the lightest Higgs boson
is composed of approximately
\begin{align}
H_d \quad \rightarrow &\quad |U_{11}^{H(0)}|^2\sim 1\%  \; , \\
H_u \quad \rightarrow &\quad |U_{12}^{H(0)}|^2\sim 80\% \; , \\
\widetilde{\nu}_{3L} \quad \rightarrow &\quad |U_{16}^{H(0)}|^2\sim 19\% \; .
\end{align}
Nevertheless, due to the upward shift, as explained before,
the one-loop corrections break the degeneracy and no trace
on the SM-like Higgs mass remains, which would deviate it from the NMSSM
prediction.

\subsection{\protect\boldmath The \mnSSM\ and the CMS $\ga\ga$ excess at
  $96 \gev$}
\label{sec:gaga95}
In this section we will investigate a scenario in which the SM-like Higgs
boson is not the lightest $\cp$-even scalar. This is inspired by
the reported excesses of LEP~\cite{Barate:2003sz}
and CMS~\cite{CMS:2015ocq,CMS:2017yta}
in the mass range around
$\sim 96\gev$, that (as we will show) can be explained
simultaneously by the presence
of a light scalar in this mass window.
While in the NMSSM the light scalar can be interpreted as the $\cp$-even
scalar singlet and can accommodate both excesses at $1\sigma$ level
without violating any known experimental
constraints~\cite{Cao:2016uwt,Domingo:2017},\footnote{Other possible
explanations of the CMS excess were analyzed in
\citere{Cacciapaglia:2016tlr,Mariotti:2017vtv,Crivellin:2017upt}. On the other hand, in the MSSM the CMS excess cannot be
realized~\cite{Bechtle:2016kui}.}
we will interpret the
light scalar as the $\cp$-even right-handed sneutrino of the \mnSSM.
Since the singlet of the NMSSM and the right-handed sneutrino of
the \mnSSM\ are both gauge-singlets, they share very similar properties.
However, the explanation of the excesses in the \mnSSM\ avoids
bounds from direct detection experiments, because $R$-parity is broken
in the \mnSSM\ and the dark matter candidate is not a neutralino as in
the NMSSM but a gravitino with a lifetime longer than the age of
the universe~\cite{Munoz:2016vaa}. This is important because the direct
detection measurements were shown to be very constraining in the NMSSM
while trying to explain the dark matter abundance on top of the
excesses from LEP and CMS~\cite{Cao:2016uwt}.

\begin{table}
\renewcommand{\arraystretch}{1.7}
\centering
\begin{tabular}{c c c c c c c c c c}
 $v_{iL}/\sqrt{2}$ & $Y^\nu_i$ & $A^\nu_i$ & $\tb$ & $\mu$ & $\lambda$ &
 	$A^\lambda$ & $\kappa$ & $A^\kappa$ & $M_1$ \\
 \hline
 $10^{-5}$ & $10^{-7}$ & $-1000$ & $2$ & $[413;418]$ & $0.6$ &
 	$956$ & $0.035$ & $[-300;-318]$ & $100$ \\
 \hline
 \hline
 $M_2$ & $M_3$ & $m_{\widetilde{Q}_{iL}}^2$ &
 	$m_{\widetilde{u}_{iR}}^2$ & $m_{\widetilde{d}_{iR}}^2$ &
 	$A^u_i$ & $A^{d}_i$ & $(m_{\widetilde{e}}^2)_{ii}$ &
 	$A^e_{33}$ & $A^e_{11,22}$ \\
 \hline
 $200$ & $1500$ & $800^2$& $800^2$ & $800^2$ & $0$ &
 	$0$ & $800^2$ & $0$ & $0$ 
\end{tabular}
%\vspace{-0.5cm}
\caption{Input parameters for the scenario featuring the right-handed
		 sneutrino in the mass range of the LEP and CMS excesses and
		 a SM-like Higgs boson as next-to-lightest $\cp$-even scalar;
  		 all masses and values for trilinear parameters are in GeV.}
\label{tab:lepcms}
\renewcommand{\arraystretch}{1.0}
\end{table}%%% T

In \refta{tab:lepcms} we list the values of the parameters we used
to account for the lightest $\cp$-even scalar as the right-handed
sneutrino and the second lightest one the SM-like Higgs boson.
$\lambda$ is chosen to be large to account for a sizable mixing of
the right-handed sneutrino and the doublet Higgses.
In the regime where the SM-like Higgs boson is not the lightest scalar,
one does not need large quantum corrections to the Higgs boson mass,
because the tree-level mass is already well above $100\gev$. This is
why $\tan\beta$ can be low and the soft trilinears $A^{u,d,e}$
are set to zero. The values of $A^{\lambda}$ and $\left|A^\nu\right|$
are chosen to be around $1\tev$ to get masses for the
heavy MSSM-like Higgs and the left-handed sneutrinos of this order, so
they do not play an important role in the following discussion.
On the other hand, $\kappa$ is small to bring the mass
of the right-handed sneutrino below the SM-like Higgs boson mass.
Finally, the two parameters that are varied are $\mu$ and $A^\kappa$.
By increasing $\mu$ the mixing of the right-handed sneutrino with
the SM-like Higgs boson is increased, which is needed to couple
the gauge-singlet to quarks and gauge-bosons. At the same time we
used the value of $A^\kappa$ to keep the mass of the
right-handed sneutrino in the correct range. Accordingly, the
results in this chapter will all be displayed in the scanned
$A^\kappa$-$\mu$ plane.

The process measured at LEP was the production of a Higgs boson
via Higgstrahlung associated with the Higgs decaying to
bottom-quarks:
\begin{equation}
\mu_{\rm LEP}=\frac{\sigma\left( e^+e^- \to Z h_1 \to Zb\bar{b} \right)}
			   {\sigma^{SM}\left( e^+e^- \to Z h 
			   		\to Zb\bar{b} \right)}
			  = 0.117 \pm 0.057 \; ,
\end{equation}
where $\mu_{\rm LEP}$ is called the signal strength, which is the
measured cross section normalized to the standard model expectation,
with the SM Higgs boson mass at $\sim 96\gev$.
The value for $\mu_{\rm LEP}$ was extracted in \citere{Cao:2016uwt}
using methods described in \citere{Azatov:2012bz}.
We can find an
approximate expression for $\mu_{\rm LEP}$ factorizing the production
and the decay of the scalar and expressing it in terms of couplings
to the massive gauge bosons $C_{h_1VV}$ and the up- and down-type
quarks $C_{h_1u\bar{u}}$ and $C_{h_1d\bar{d}}$, respsectively, normalized to
the SM predictions for the corresponding couplings (where with
\mbox{}$^{\mu\nu}$ we denote the \mnSSM\ prediction, and $\Ga$ is the
Higgs-boson decay width):

\begin{align}
\mu_{\rm LEP}^{\mu\nu}&=\frac{\sigma^{\mu\nu}\left(Z^*\to Zh_1 \right)}
          {\sigma^{\rm SM}\left(Z^*\to Zh \right)}\times
      \frac{\br^{\mu\nu}\left( h_1\to b \bar b\right)}
            {\br^{\rm SM}\left( h\to b \bar b\right)} 
            \notag \\
     &\approx\left| C_{h_1VV} \right|^2\times
       \frac{\Gamma^{\mu\nu}_{b\bar{b}}}{\Gamma^{\rm SM}_{b\bar{b}}}
          \times
       \frac{\Gamma^{\rm SM}_{\rm tot}}{\Gamma^{\mu\nu}_{\rm tot}} 
       \notag \\
     &\approx \frac{
     \left| C_{h_1VV} \right|^2\times\left| C_{h_1d\bar{d}} \right|^2
          }{
      \left| C_{h_1d\bar{d}} \right|^2
          (\br^{\rm SM}_{b\bar{b}}+
           \br^{\rm SM}_{\tau\bar{\tau}}) +
           \left| C_{h_1u\bar{u}} \right|^2
            (\br^{\rm SM}_{gg}+
           \br^{\rm SM}_{c\bar{c}})} \; .
\label{eq:mulep}
\end{align}

\begin{figure}
  \centering
  \includegraphics[width=0.6\textwidth]{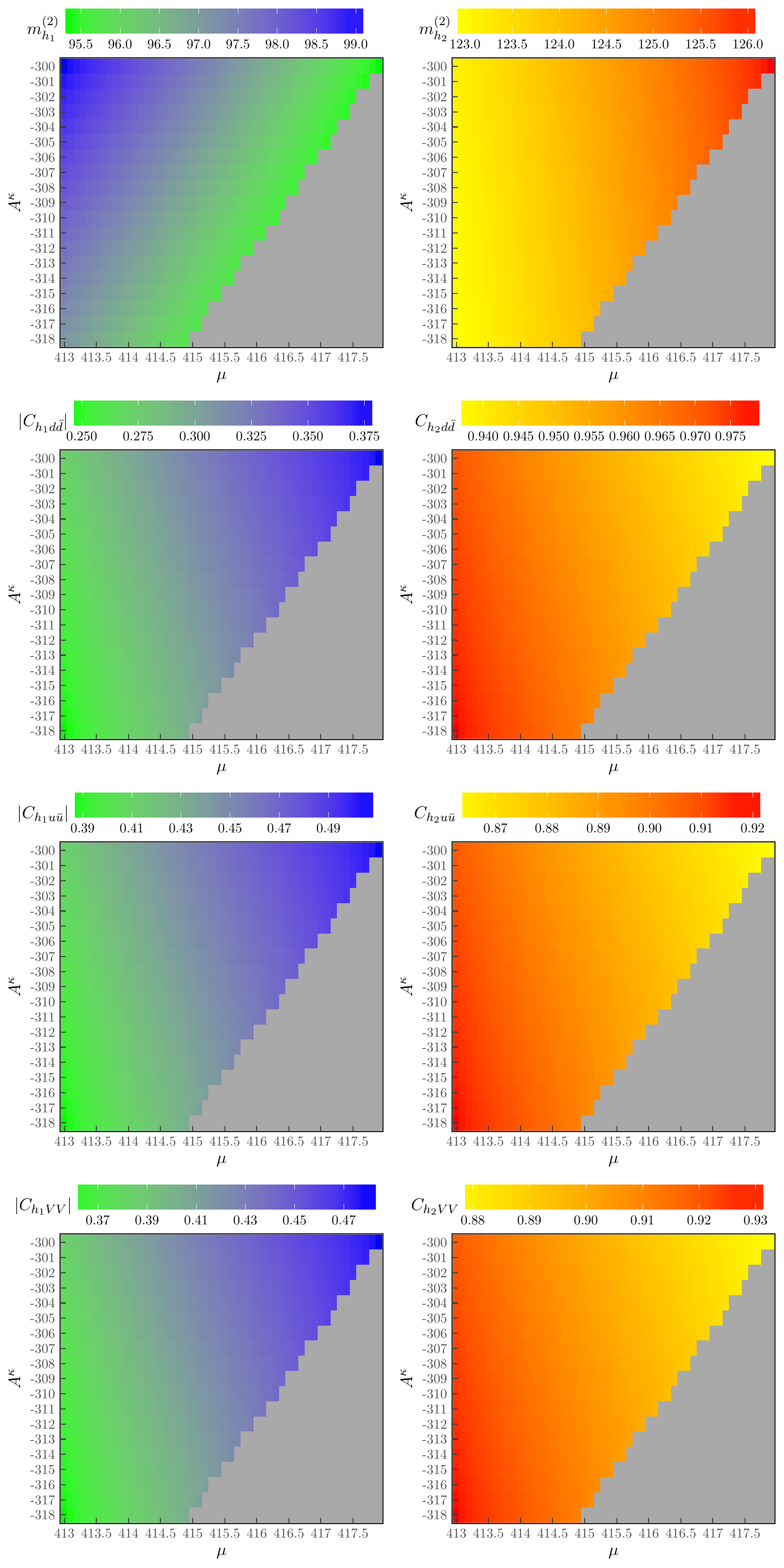}
  \vspace{-0.5cm}
  \caption{Properties of the lightest (left) and next-to-lightest (right)
         $\cp$-even scalar in the $\mu$--$A^\kappa$ plane.
         The couplings are normalized to the SM-prediction
         of a Higgs particle of the same mass. The gray area is
         excluded because the right-handed sneutrino becomes
         tachyonic at tree-level.
         \textit{First row:} two-loop masses,
         \textit{second row:} coupling to down-type quarks,
         \textit{third row:} coupling to up-type quarks,
         \textit{fourth row:} coupling to massive gauge bosons.}
  \label{fig:lep1}
\end{figure}

The SM branching ratios dependent on the Higgs boson mass can be
obtained from \citere{Heinemeyer:2013tqa}.
The denominator is the ratio of the total decay width of $h_1$
in the \mnSSM\ and $h$ in the SM when all SM
branching ratios larger than $1\%$ are considered.
The off-shell decay to $W$ and $Z$ bosons is in
principle also possible, but the BRs are very small for a SM Higgs
boson with a mass around $95\gev$ ($\br^{\rm SM}_{WW}\sim 0.5\%$ 
and $\br^{\rm SM}_{ZZ}\sim 0.06\%$)~\cite{Denner:2011mq,Heinemeyer:2013tqa}.
It is worth noticing that although the right-handed neutrino mass
is small, $m_{\nu_R}\sim62-63\gev$,
in the investigated parameter region, it is nevertheless
larger than half of
the SM-like Higgs boson mass
in all benchmark points, so the decay of the Higgs
to the right-handed neutrino is kinematically forbidden and
cannot spoil the properties of the SM-like Higgs.
Neglecting the vevs $v_{iL}$ the normalized couplings of the
scalars are given at
leading order by the admixture of the mass eigenstate $h_i$ with
the doublet like Higgs $H_d$ and $H_u$ via
\begin{equation}
C_{h_i d \bar{d}}=\frac{U^{H,(2')}_{i1}}{\cos\beta} \; , \quad
C_{h_i u \bar{u}}=\frac{U^{H,(2')}_{i2}}{\sin\beta} \; , \quad
C_{h_i V V}=U^{H,(2')}_{i1}\cos\beta+U^{H,(2')}_{i2}\sin\beta \; ,
\end{equation}
where the partial two-loop plus resummation corrected mixing matrix elements $U^{H,(2')}_{ij}$
were calculated
in the approximation of vanishing momentum,
see the discussion in \refse{sec:higherorders}.
We show in \reffi{fig:lep1} the masses (top row) and the normalized
couplings ($|C_{h_1 d \bar d}|$ second row, $|C_{h_1 u \bar b}$| third
row, $|C_{h_1 VV}|$ lowest row) 
of the lightest and the next-to-lightest $\cp$-even scalar. The lower
right corner (marked in gray) results in the right-handed sneutrino
becoming tachyonic (at tree-level).
The largest mixing of the right-handed sneutrino and the SM-like
Higgs boson is achieved where $\mu$ is largest
and $|A^\kappa|$ is smallest. The mass of $h_2$ is in the allowed
region for a SM-like Higgs boson at $\sim 125\gev$ if we assume a
theory uncertainty of up to $3\gev$ (see the previous subsections).
The LHC measurements of the SM-like Higgs boson
couplings to fermions and massive gauge bosons are still not
very precise~\cite{Khachatryan:2016vau}, with uncertainties
between 10 and $20\%$ at the $1\sigma$ confidence level
(obtained with
the assumption that no beyond-the-SM decays modify the total
width of the SM-like Higgs boson). Therefore, it would be
challenging to exclude parts of the parameter space by considering
the deviations of the normalized couplings of $h_2$.
However, possible future lepton colliders
like the ILC could measure these couplings to a
$\%$-level~\cite{Dawson:2013bba,Moortgat-Picka:2015yla}, which
could exclude (or confirm) most of the parameter space presented here.
Seen from a more optimistic perspective, the precise measurement of
the SM-like Higgs boson couplings at future colliders could
be used to make predictions for the properties of the
lighter right-handed sneutrino in this scenario.
%%%%%%%%%%%%%%%%%%%%%%%%%%%% F I G U R E %%%%%%%%%%%%%%%%%%%%%%%%%%%%%%%%%%%%%%
%%%%%%%%%%%%%%%%%%%%%%%%%%%% F I G U R E %%%%%%%%%%%%%%%%%%%%%%%%%%%%%%%%%%%%%%

The CMS excess was observed in the diphoton channel
with a signal strength of~\cite{Shotkin:2017}
\begin{equation}
\mu_{\rm CMS}=\frac{\sigma\left( gg\to h_1 \to \gamma\gamma \right)}
         {\sigma^{\rm SM}\left( gg\to h \to \gamma\gamma \right)}
     = 0.6 \pm 0.2 \; .
\end{equation}
We calculate the signal strength using the approximation that the
Higgs production via gluonfusion is described at leading order
exclusively by the loop-diagram with a
top quark running in the loop, and that
the diphoton decay is described by the diagrams with $W$~bosons or a
top quark in the loop, which is sufficient in the investigated mass
range of $h_1$. One can then write
\begin{align}
\mu_{\rm CMS}^{\mu\nu}&=\frac{\sigma^{\mu\nu}\left( gg\to h_1 \right)}
         {\sigma^{\rm SM}\left( gg\to h \right)} \times
      \frac{\br^{\mu\nu}_{\gamma\gamma}}
           {\br^{\rm SM}_{\gamma\gamma}} \notag \\
     &\approx \left|C_{h_1 u \bar{u}}\right|^2 \times
        \frac{\Gamma^{\mu\nu}_{\gamma\gamma}}
             {\Gamma^{\rm SM}_{\gamma\gamma}} \times
      \frac{\Gamma^{\rm SM}_{\rm tot}}{\Gamma^{\mu\nu}_{\rm tot}} \notag \\
     &\approx\frac{\left|C_{h_1 u \bar{u}}\right|^2 \times
         \left|C^{\rm eff}_{h_1 \gamma\gamma}\right|^2}
            {\left| C_{h_1d\bar{d}} \right|^2
          (\br^{\rm SM}_{b\bar{b}}+
           \br^{\rm SM}_{\tau\bar{\tau}}) +
           \left| C_{h_1u\bar{u}} \right|^2
            (\br^{\rm SM}_{gg}+
           \br^{\rm SM}_{c\bar{c}})} \; .
\label{eq:mucms}
\end{align}
The effective coupling of the neutral scalars to photons
$C^{\rm eff}_{h_i\gamma\gamma}$ has to be
calculated in terms of the couplings to the $W$ boson and the up-type
quarks. In the SM the dominant contributions to the decay to photons
can be written as~\cite{Djouadi:2005gi}
\begin{equation}
\Gamma^{\rm SM}_{\gamma\gamma} =
\frac{G_\mu\,\alpha^2 \, m_h^3}{128\, \sqrt{2}\, \pi^3}
\left| \frac{4}{3} A_{1/2}\left( \tau_t \right) +
    A_{1}\left( \tau_W \right) \right|^2 \; ,
\end{equation}
where $G_\mu$ is the Fermi-constant and the form factors $A_{1/2}$
and $A_{1}$ are defined as
\begin{align}
A_{1/2}\left(\tau\right)&=2\left( \tau+\left(\tau -1\right)
      \arcsin^2\sqrt{\tau} \right)\tau^{-2} \; , \\
A_{1}\left(\tau\right)&=-\left( 2\tau^2+3\tau
      +3\left( 2\tau-1 \right)
      \arcsin^2\sqrt{\tau} \right)\tau^{-2} \; ,
\end{align}
for $\tau \leq 1$, and the arguments of these functions are
$\tau_t = m_h^2/(4 m_t^2)$ and $\tau_W = m_h^2/(4 M_W^2)$.
In our approximation the only difference between the \mnSSM\ and the
SM will be
that the couplings of $h_i$ to the top quark and the $W$ boson is
modified by the factors $C_{h_i t\bar{t}}$ and $C_{h_i VV}$, so the
effective coupling of the Higgses to photons in the \mnSSM\
normalized to the SM predictions
can be written as
\begin{equation}
\left|C^{\rm eff}_{h_i \gamma\gamma}\right|^2=
\frac{\left| \frac{4}{3} C_{h_i t\bar{t}} A_{1/2}\left( \tau_t \right) +
    C_{h_i VV} A_{1}\left( \tau_W \right) \right|^2}
    {\left| \frac{4}{3} A_{1/2}\left( \tau_t \right) +
    A_{1}\left( \tau_W \right) \right|^2} \; .
\end{equation}

\begin{figure}
  \centering
  \includegraphics[width=0.9\textwidth]{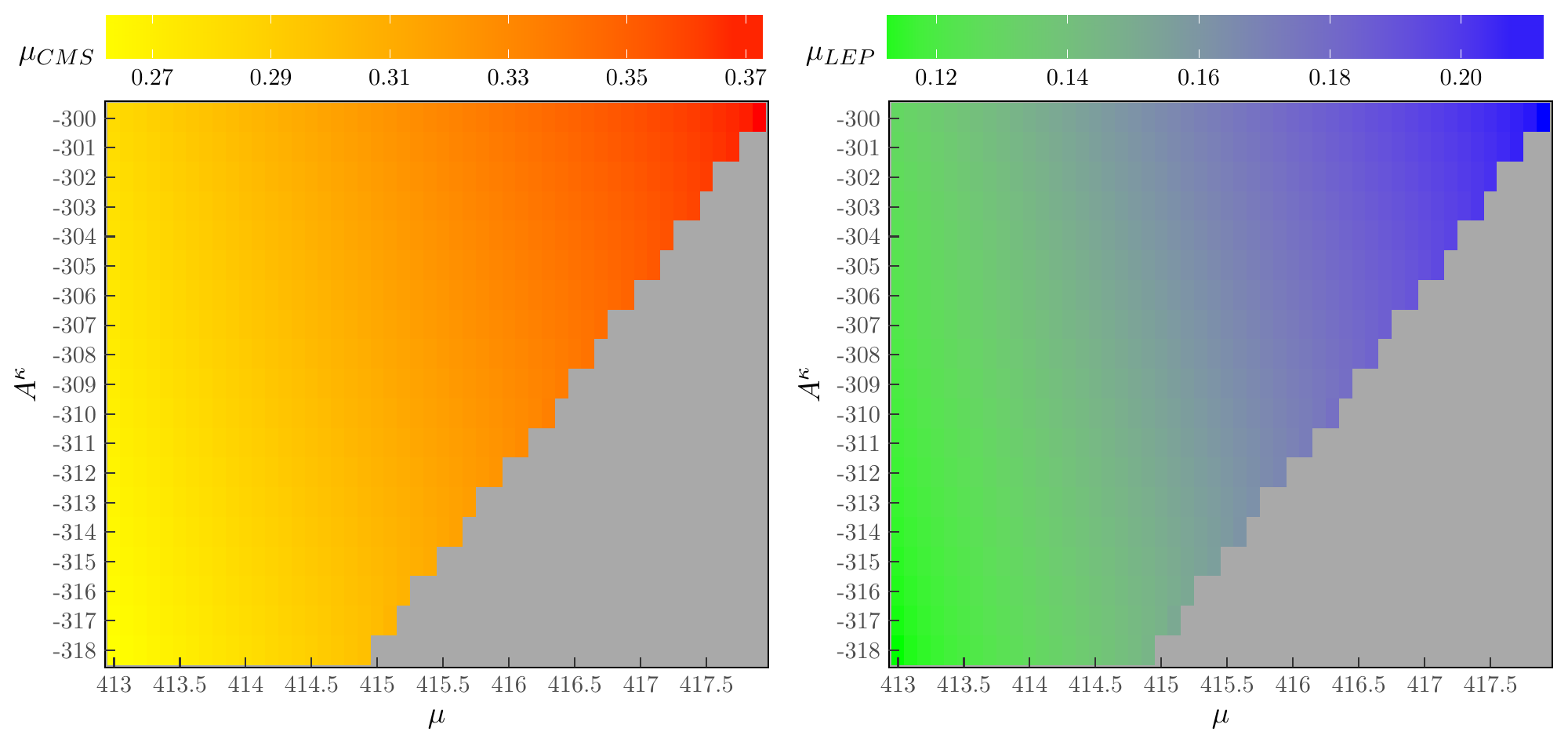}
  \vspace{-0.5cm}
  \caption{Signal strengths for the lightest $\widetilde{\nu}_R$-like
         neutral scalar at CMS ($pp\to h_1 \to \gamma\gamma$)
         (\textit{left}) and
         LEP (${e^+e^-\to h_1 Z \to b\bar{b} Z}$)
         (\textit{right}) in the
         $\mu$-$A^\kappa$ plane. The gray area is
         excluded because the right-handed sneutrino becomes
         tachyonic at tree-level.}
  \label{fig:lep2}
\end{figure}

Using \refeq{eq:mulep} and \refeq{eq:mucms} we can
calculate the two signal strengths. 
The result are shown in \reffi{fig:lep2}, the LEP (left) and the CMS
excesses (right) in the $\mu$--$A^{\kappa}$ plane.
While the LEP excess is easily reproduced
in the observed parameter space, we cannot achieve the central
value for $\mu_{\rm CMS}$, but only slightly smaller values.
As already observed in \citere{Cao:2016uwt},
the reason for this is that for explaining the LEP excess
a sizable coupling to the bottom quark is needed.
On the contrary, the CMS
excess demands a small value for $C_{h_1 d\bar{d}}$ so that the
denominator in \refeq{eq:mucms} becomes small and $\mu_{\rm CMS}$
is enhanced. Nevertheless, considering the large
experimental uncertainties in $\mu_{\rm CMS}$ and $\mu_{\rm LEP}$,
the scenario presented in this section
accommodates both excesses comfortably well
(at approximately $1\sigma$),
and it is a good motivation to keep on
searching for light Higgses in the allowed mass window
below the SM-like Higgs mass. Apart from that, this scenario
illustrates the importance of an accurate calculation of the
loop-corrected scalar masses and mixings, since already small changes
in the parameters can have a big impact on the production and the
decay modes of the $\cp$-even Higgs bosons.

%%%%%%%%%%%%%%%%%%%%%%%%%%%% F I G U R E %%%%%%%%%%%%%%%%%%%%%%%%%%%%%%%%%%%%%%
%%%%%%%%%%%%%%%%%%%%%%%%%%%% F I G U R E %%%%%%%%%%%%%%%%%%%%%%%%%%%%%%%%%%%%%%

%%%%%%%%%%%%%%%%%%%%%%%%%%%%%%%%%%%%%%%%%%%%%%%%%%%%%%%%%%%%%%%%%%%%%%%%%%%%%%%
%%%%%%%%%%%%%%%%%%%%%%%%%%%%%%%%%%%%%%%%%%%%%%%%%%%%%%%%%%%%%%%%%%%%%%%%%%%%%%%

\section{Conclusion and Outlook}
\label{sec:concl}
The \mnSSM\ is a simple SUSY extension of the SM that is capable of predicting neutrino physics in agreement with experimental data. As in other SUSY models, higher-order corrections are crucial to reach a
theoretical uncertainty at the same level of (anticipated) experimental
accuracy. So far, higher-order corrections in the \mnSSM\ had been restricted
to \DRbar\ calculations, which suffer from the disadvantage that they cannot
be directly connected to (possibly future observed) new BSM particles.

In this paper we have performed the complete one-loop renormalization of the
neutral scalar sector of the \mnSSM\ with one generation of right-handed
neutrinos in a mixed on-shell/\DRbar\ scheme. The renormalization procedure
was discussed in detail for each of the free parameters appearing in the
\mnSSM\ Higgs sector. We have emphasized the conceptual differences to the MSSM and the NMSSM regarding the field
renormalization and the treatment of non-flavor-diagonal soft mass parameters, which have their
origin in the breaking of $R$-parity in the \mnSSM. However, we have ensured that the renormalization of the relevant (N)MSSM
parts in the \mnSSM\ are in agreement with previous calculations in those
models. Consequently, numerical differences found can directly be attributed
to the extended structure of the \mnSSM.
The derived renormalization can be applied
to any higher-order correction in the \mnSSM.
The one-loop counterterms derived in this paper are implemented into
the \FA\ model file, so the computation of these corrections
can be done fully automatically.

We have applied the newly derived renormalization to the calculation of the full one-loop corrections to the neutral scalar masses
of the \mnSSM, where we found that all UV-divergences cancel.
In our numerical analysis the newly derived full one-loop contributions are
supplemented by available MSSM higher-order corrections as provided by the
code \fh\ (leading and subleading fixed-order corrections as well as resummed large logarithmic
contributions obtained in an EFT approach.)
We investigated various representative scenarios, in which we obtained
numerical results for a SM-like Higgs boson mass consistent with experimental bounds. We compared our results to predictions of the various
neutral scalars in the NMSSM to
investigate the relevance of genuine \mnSSM-like contributions.
We find negligible
corrections w.r.t.\ the NMSSM,
indicating that the Higgs boson mass calculations in the \mnSSM\ are at 
the same level of
accuracy as in the NMSSM.

Finally we showed that the \mnSSM\ can accommodate a right-handed
($\cp$-even) scalar neutrino 
in a mass regime of $\sim 96 \gev$, where the full Higgs sector is in
agreement with the Higgs-boson measurements obtained at the LHC, as well as
with the Higgs exclusion bounds obtained at LEP, the Tevatron and the LHC.
This includes in particular a SM-like Higgs boson at $\sim 125 \gev$.
We have demonstrated that the light right-handed sneutrino can explain
an excess of $\ga\ga$ events at $\sim 96 \gev$ as reported 
recently by CMS in their Run~I and Run~II date. It can
simultaneously describe the $2\,\sig$ excess of $b \bar b$ events
observed at LEP at a similar mass scale. We are eagerly awaiting the 
corresponding ATLAS Higgs-boson search results.

\subsection*{Acknowledgements}
We thank F.~Domingo for helpful discussions.
This work was supported in part by the Spanish Agencia Estatal de
Investigaci\'on through the grants
FPA2016-78022-P MINECO/FEDER-UE (TB and SH) and
FPA2015-65929-P MINECO/FEDER-UE (CM), and IFT Centro de
Excelencia Severo Ochoa SEV-2016-0597. The work of TB was
funded by Fundaci\'on La Caixa under `La Caixa-Severo Ochoa' international
predoctoral grant. We also acknowledge the support of the MINECO's
Consolider-Ingenio 2010 Programme under grant MultiDark CSD2009-00064.

%%%%%%%%%%%%%%%%%%%%%%%%%%%%%%%%%%%%%%%%%%%%%%%%%%%%%%%%%%%%%%%%%%%%%%%%%%%%%%%
%%%%%%%%%%%%%%%%%%%%%%%%%%%%%%%%%%%%%%%%%%%%%%%%%%%%%%%%%%%%%%%%%%%%%%%%%%%%%%%

\appendix

\section{Mass matrices}

Here we state the entries of the
following scalar mass matrices.

\subsection{\protect\boldmath $\cp$-even scalars}
\label{app:cpeven}

In the interaction basis $\varphi^T=(H_d^{\mathcal{R}},H_u^{\mathcal{R}},
\widetilde{\nu}_{R}^{\mathcal{R}},\widetilde{\nu}_{iL}^{\mathcal{R}})$
the mass matrix for the $\cp$-even scalars $m_{\varphi}^2$ is defined by:
\begin{align}
m_{H_{d}^{\mathcal{R}}H_{d}^{\mathcal{R}}}^{2}=&m_{H_d}^2+
\frac{1}{8}\left(g_1^2+g_2^2\right)\left( 3v_d^2+v_{iL}v_{iL}-v_u^2\right)+
\frac{1}{2}\lambda^2\left( v_R^2+v_u^2\right) \; , \\[4pt]
m_{H_{u}^{\mathcal{R}}H_{u}^{\mathcal{R}}}^{2}=&m_{H_u}^2+
\frac{1}{8}\left(g_1^2+g_2^2\right)\left(3v_u^2-v_d^2-v_{iL}v_{iL}\right)+
\frac{1}{2}\lambda^2\left(v_R^2+v_d^2\right) \notag \\
&+\frac{1}{2}v_R^2Y^\nu_iY^\nu_i-
v_d\lambda v_{iL}Y^\nu_i \; , \\[4pt]
m_{H_{u}^{\mathcal{R}}H_{d}^{\mathcal{R}}}^{2}=&
-\frac{1}{4}\left(g_1^2+g_2^2\right)v_dv_u-
\frac{1}{2}v_R^2\kappa\lambda+
v_dv_u\lambda^2-
v_u\lambda v_{iL}Y^\nu_i-
\frac{1}{\sqrt{2}}T^\lambda \; , \\[4pt]
m_{\widetilde{\nu}^{\mathcal{R}}H_{d}^{\mathcal{R}}}^{2}=&
-v_Rv_u\kappa\lambda+v_dv_R\lambda^2-v_R\lambda v_{iL}Y^\nu_i-
\frac{1}{\sqrt{2}}v_uT^\lambda \; , \\[4pt]
m_{\widetilde{\nu}^{\mathcal{R}}H_{u}^{\mathcal{R}}}^{2}=&
-v_dv_R\kappa\lambda +
v_Rv_u\lambda^2+
v_R\kappa v_{iL}Y^\nu_i+
v_Rv_u Y^\nu_iY^\nu_i+
\frac{1}{\sqrt{2}}v_{iL}T^\nu_i-
\frac{1}{\sqrt{2}}v_dT^\lambda \; , \\[4pt]
m_{\widetilde{\nu}^{\mathcal{R}}\widetilde{\nu}^{\mathcal{R}}}^{2}=&
m_{\widetilde{v}_R}^2+
3v_R^2\kappa^2-
v_dv_u\kappa\lambda + 
\frac{1}{2}\lambda^2\left( v_d^2+v_u^2\right)+
v_u\kappa v_{iL}Y^\nu_i-
v_d\lambda v_{iL}Y^\nu_i \notag \\
&+\frac{1}{2}v_u^2Y^\nu_iY^\nu_i+
\frac{1}{2}\left(v_{iL}Y^\nu_i\right)^2+
\sqrt{2}v_RT^\kappa \; , \\[4pt]
m_{\widetilde{\nu}_{iL}^{\mathcal{R}} H_{d}^{\mathcal{R}}}^{2}=&
\left(m_{H_d\widetilde{L}}^2\right)_i+
\frac{1}{4}\left(g_1^2+g_2^2\right)v_dv_{iL}-
\frac{1}{2}v_R^2\lambda Y^\nu_i-
\frac{1}{2}v_u^2\lambda Y^\nu_i \; , \\[4pt]
m_{\widetilde{\nu}_{iL}^{\mathcal{R}} H_{u}^{\mathcal{R}}}^{2}=&
-\frac{1}{4}\left(g_1^2+g_2^2\right)v_uv_{iL}+
\frac{1}{2}v_R^2\kappa Y^\nu_i-
v_dv_u\lambda Y^\nu_i+
v_uv_{jL}Y^\nu_j Y^\nu_i+
\frac{1}{\sqrt{2}}v_RT^\nu_i \; , \\[4pt]
m_{\widetilde{\nu}_{iL}^{\mathcal{R}} \widetilde{\nu}^{\mathcal{R}}_{R}}^{2}=&
-\lambda v_d v_R Y^\nu_i+\frac{1}{\sqrt{2}}v_u T^\nu_i+
v_R v_u \kappa Y^\nu_i+v_R Y^\nu_i v_{jL}Y^\nu_j \; , \\[4pt]
m_{\widetilde{\nu}_{iL}^{\mathcal{R}} \widetilde{\nu}_{jL}^{\mathcal{R}}}^{2}=&
\left(m_{\widetilde{L}}^2\right)_{ij}+
\frac{1}{8}\delta_{ij}\left(g_1^2+g_2^2\right)\left(v_d^2+v_{kL}v_{kL}\right)
+\frac{1}{4}\left(g_1^2+g_2^2\right)v_{iL}v_{jL}+
\frac{1}{2}\left(v_R^2+v_u^2\right)Y^\nu_iY^\nu_j
\end{align}

%%%%%%%%%%%%%%%%%%%%%%%%%%%%%%%%%%%%%%%%%%%%%%%%%%%%%%%%%%%%%%%%%%%%%%%%%%%%%%%

\subsection{\protect\boldmath $\cp$-odd scalars}
\label{app:cpodd}

In the interaction basis $\sigma^T=(H_d^{\mathcal{I}},H_u^{\mathcal{I}},\widetilde{\nu}_{R}^{\mathcal{I}},\widetilde{\nu}_{iL}^{\mathcal{I}})$ the mass matrix for the $\cp$-odd scalars $m_{\sigma}^2$ is defined by:
\begin{align}
m_{H_{d}^{\mathcal{I}}H_{d}^{\mathcal{I}}}^{2}=&
m_{H_d}^2+
\frac{1}{8}\left(g_1^2+g_2^2\right)\left(v_d^2+v_{iL}v_{iL}-v_u^2\right)+
\frac{1}{2}\lambda^2\left( v_R^2+v_u^2\right) \; , \\[4pt]
m_{H_{d}^{\mathcal{I}}H_{d}^{\mathcal{I}}}^{2}=&
m_{H_u}^2+
\frac{1}{8}\left(g_1^2+g_2^2\right)\left(v_u^2-v_d^2-v_{iL}v_{iL}\right)+
\frac{1}{2}\lambda^2\left( v_R^2+v_d^2\right)+
v_d\lambda v_{iL}Y^\nu_i \notag \\
&+\frac{1}{2}v_R^2Y^\nu_iY^\nu_i+
\frac{1}{2}\left( v_{iL}Y^\nu_i\right)^2 \; , \\[4pt]
m_{H_{u}^{\mathcal{I}}H_{d}^{\mathcal{I}}}^{2}=&
\frac{1}{2}v_R^2\kappa\lambda+
\frac{1}{\sqrt{2}}v_R T^\lambda \; , \\[4pt]
m_{\widetilde{{\nu}}^{\mathcal{I}}_{R} H_{d}^{\mathcal{I}}}^{2}=&
v_Rv_u\kappa\lambda -\frac{1}{\sqrt{2}}v_uT^\lambda \; , \\[4pt]
m_{\widetilde{{\nu}}^{\mathcal{I}}_{R} H_{u}^{\mathcal{I}}}^{2}=&
v_dv_R\kappa\lambda-
v_R\kappa v_{iL}Y^\nu_i+
\frac{1}{\sqrt{2}}v_{iL}T^\nu_i-
\frac{1}{\sqrt{2}}v_d T^\lambda \; , \\[4pt]
m_{\widetilde{{\nu}}^{\mathcal{I}}_{R} \widetilde{{\nu}}^{\mathcal{I}}_{R}}^{2}=&
m_{\widetilde{\nu}}^2+
v_R^2\kappa^2+
v_dv_u\kappa\lambda+
\frac{1}{2}\lambda^2\left(v_d^2+v_u^2\right)-
v_u\kappa v_{iL}Y^\nu_i-
v_d\lambda v_{iL}Y^\nu_i \notag \\
+&\frac{1}{2}v_u^2 Y^\nu_iY^\nu_i+
\frac{1}{2}\left(v_{iL}Y^\nu_i\right)^2-
\sqrt{2}v_R T^\kappa \; , \\[4pt]
m_{\widetilde{\nu}_{iL}^{\mathcal{I}} H_{d}^{\mathcal{I}}}^{2}=&
\left(m_{H_d\widetilde{L}}^2\right)_i-
\frac{1}{2}v_R^2\lambda Y^\nu_i-
\frac{1}{2}v_u^2\lambda Y^\nu_i \; , \\[4pt]
m_{\widetilde{\nu}_{iL}^{\mathcal{I}} H_{u}^{\mathcal{I}}}^{2}=&
-\frac{1}{2}v_R^2\kappa Y^\nu_i-
\frac{1}{\sqrt{2}}v_R T^\nu_i \; , \\[4pt]
m_{\widetilde{\nu}_{iL}^{\mathcal{I}} \widetilde{\nu}^{\mathcal{I}}_{R}}^{2}=&
-v_Rv_u\kappa Y^\nu_i+
\frac{1}{\sqrt{2}}v_u T^\nu_i \; , \\[4pt]
m_{\widetilde{\nu}_{iL}^{\mathcal{I}} \widetilde{\nu}_{jL}^{\mathcal{I}}}^{2}=&
\left(m_{\widetilde{L}}^2\right)_{ij}+
\frac{1}{8}\delta_{ij}\left(g_1^2+g_2^2\right)\left(v_d^2+v_{kL}v_{kL}-v_u^2\right)
+\frac{1}{2}\left(v_R^2+v_u^2\right)Y^\nu_iY^\nu_j \; . \label{eq:mlinmps}
\end{align}

\subsection{Charged scalars}\label{app:charged}
In the eigenstate basis $C^T=
({H^-_d}^*,{H^+_u},\widetilde{e}_{iL}^*,\widetilde{e}_{jR}^*)$ the entries
of $m_{H^+}^2$ are given by:
\begin{align}
m_{H_{d}^-{H^{-}_d}^{*}}^{2}=&m_{H_d}^2+
\frac{1}{8}g_1^2\left( v_d^2+v_{iL}v_{iL}-v_u^2\right)+
\frac{1}{8}g_2^2\left( v_d^2-v_{iL}v_{iL}+v_u^2\right)+
\frac{1}{2}v_R^2\lambda^2 Y^e_{ij}Y^e_{ik}v_{jL}v_{kL} \; , \\[4pt]
m_{{H_{u}^+}^* {H_d^-}^*}^{2}=&m_{H_u}^2+
\frac{1}{8}g_1^2\left( v_u^2-v_d^2-v_{iL}v_{iL}\right)+
\frac{1}{8}g_2^2\left( v_u^2+v_d^2+v_{iL}v_{iL}\right)+
\frac{1}{2}v_R^2\lambda^2+
\frac{1}{2}v_R^2 Y^\nu_iY^\nu_i \; , \\[4pt]
m_{H_{d}^-{H^{+}_u}}^{2}=&\frac{1}{4}g_2^2v_dv_u+
\frac{1}{2} v_R^2\lambda\kappa-\frac{1}{2} v_d v_u\lambda^2+
\frac{1}{2}v_u\lambda v_{iL}Y^\nu_i+\frac{1}{\sqrt{2}}v_R T^\lambda \; , \\[4pt]
m_{\widetilde{e}_{iL} {{H^{-}_d}}^{*}}^2=&
\left(m_{H_d\widetilde{L}}^2\right)_i+
\frac{1}{4}g_2^2v_dv_{iL}-
\frac{1}{2}v_d Y^e_{ji}Y^e_{jk}v_{kL}-
\frac{1}{2}v_R^2\lambda Y^\nu_i \; , \\[4pt]
m_{\widetilde{e}_{iL} H_u^+}^2=&
\frac{1}{4}g_2^2v_u v_{iL}-
\frac{1}{2}v_R^2\kappa Y^\nu_i+
\frac{1}{2}v_d v_u \lambda Y^\nu_i-
\frac{1}{2}v_u v_{jL}Y^\nu_j Y^\nu_i-
\frac{1}{\sqrt{2}}v_R T^\nu_i \; , \\[4pt]
m_{\widetilde{e}_{iR} {{H^{-}_d}}^{*}}^2=&
-\frac{1}{\sqrt{2}}v_{jL}T^e_{ij}-
\frac{1}{2}v_R v_u Y^e_{ij}Y^\nu_j \; , \\[4pt]
m_{\widetilde{e}_{iR} H_u^+}^2=&
-\frac{1}{2}v_R\lambda v_{jL}Y^e_{ij}-
\frac{1}{2}v_d v_R Y^e_{ij}Y^\nu_j \; , \\[4pt]
m_{\widetilde{e}_{iL} \widetilde{e}_{jL}^{*}}^{2}=&
\left(m_{\widetilde{L}}^2\right)_{ij}+
\frac{1}{8}\delta_{ij}\left(g_1^2-g_2^2\right)\left( v_d^2-v_u^2+v_{kL}v_{kL}\right)+
\frac{1}{4}g_2^2v_{iL}v_{jL}+
\frac{1}{2}v_d^2 Y^e_{ki}Y^e_{kj} \notag \\
&+\frac{1}{2}v_R^2 Y^\nu_i Y^\nu_j \; , \label{eq:mlinmch} \\[4pt]
m_{\widetilde{e}_{iR} \widetilde{e}_{jR}^{*}}^{2}=&
\left(m_{\widetilde{e}}^2\right)_{ij}+
\frac{1}{4}\delta_{ij}g_1^2\left( v_u^2-v_d^2-v_{kL}v_{kL}\right)+
\frac{1}{2}v_d^2Y^e_{ik}Y^e_{jk}+
\frac{1}{2}v_{kL}v_{lL}Y^e_{ik}Y^e_{jl} \; , \\[4pt]
m_{\widetilde{e}_{iR} \widetilde{e}_{jL}^{*}}^{2}=&
\frac{1}{\sqrt{2}}v_d T^e_{ij}-
\frac{1}{2}v_R v_u  \lambda Y^e_{ij} \; .
\end{align}

%%%%%%%%%%%%%%%%%%%%%%%%%%%%%%%%%%%%%%%%%%%%%%%%%%%%%%%%%%%%%%%%%%%%%%%%%%%%%%%

%%%%%%%%%%%%%%%%%%%%%%%%%%%%%%%%%%%%%%%%%%%%%%%%%%%%%%%%%%%%%%%%%%%%%%%%%%%%%%%
%%%%%%%%%%%%%%%%%%%%%%%%%%%%%%%%%%%%%%%%%%%%%%%%%%%%%%%%%%%%%%%%%%%%%%%%%%%%%%%

\section{Explicit expressions for counterterms}

In this section we will state the one-loop counterterms that were calculated diagrammatically in the $\DRbar$ scheme and checked against master formulas for the one-loop beta functions and anomalous dimensions of soft SUSY breaking parameters \cite{Martin:1993zk,Yamada:1994id,Luo:2002ti}, superpotential parameters \cite{Machacek:1983fi,Luo:2002ti}, vacuum expectation values \cite{Sperling:2013eva} and wave-functions with kinetic mixing \cite{Machacek:1983tz,Fonseca:2013bua}. The master formulas were evaluated using the mathematica package SARAH \cite{Staub:2010jh}.

%%%%%%%%%%%%%%%%%%%%%%%%%%%%%%%%%%%%%%%%%%%%%%%%%%%%%%%%%%%%%%%%%%%%%%%%%%%%%%%

\subsection{Field renormalization counterterms}
\label{app:fieldcounters}

We list the field renormalization counterterms defined in \refeq{eq:drfieldcounterdef} in the $\DRbar$ scheme in the interaction
basis $(H_d,H_u,\widetilde{\nu}_{R},\widetilde{\nu}_{1L},
\widetilde{\nu}_{2L},\widetilde{\nu}_{3L})$:
\begin{align}
\delta Z_{11}&=-\frac{\Delta}{16 \pi^2}
	\left( \lambda^2 + Y^e_{ij}Y^e_{ij}+3\left(Y^d_iY^d_i\right) \right) \; , \\
\delta Z_{1,3+i}&=\frac{\Delta}{16 \pi^2}
	\lambda Y^\nu_i \; , \\
\delta Z_{22}&=-\frac{\Delta}{16 \pi^2}
	\left( \lambda^2+ Y^\nu_iY^\nu_i+3\left( Y^u_iY^u_i\right) \right) \; , \\
\delta Z_{33}&=-\frac{\Delta}{16 \pi^2}
	\left( \lambda^2+\kappa^2+Y^\nu_iY^\nu_i \right) \; , \\
\delta Z_{3+i,3+j}&=-\frac{\Delta}{16 \pi^2}
	\left( Y^e_{ki}Y^e_{kj} + Y^\nu_i Y^\nu_j \right) \; .
\end{align}
We checked that the coefficients of the divergent part of the field renormalization counterterms are equal to the one-loop anomalous dimensions of the corresponding superfields $\gamma_{ij}^{(1)}$, neglecting the terms proportional to the gauge couplings $g_1$ and $g_2$, and divided by the loop factor $16 \pi^2$, i.e.,
\begin{equation}
\delta Z_{ij}=\left.\frac{\gamma_{ij}^{(1)}\Delta}{16
  \pi^2}\right|^{g1,g2\to 0}
 \; ,
\end{equation} 
which is the same relation that holds in the (N)MSSM.

%%%%%%%%%%%%%%%%%%%%%%%%%%%%%%%%%%%%%%%%%%%%%%%%%%%%%%%%%%%%%%%%%%%%%%%%%%%%%%%

\subsection{Parameter counterterms}
\label{app:paracounters}

We list the explicit form of the counterterms of the free parameters renormalized in the $\DRbar$ scheme:
\begin{align}
\delta\mu=&
\frac{\mu\Delta}{32\pi^2}\left(
-\frac{4\pi\alpha\left( \SW^2+3\CW^2\right)}{\SW^2\CW^2}+
2\lambda^2+3\left( Y^u_iY^u_i+Y^d_iY^d_i\right)+
Y^e_{ij}Y^e_{ij}+2Y^\nu_iY^\nu_i \right) \; ,
\label{eq:dmue}\\
\delta\kappa=&
\frac{3\kappa\Delta}{16\pi^2}\left(
\kappa^2+\lambda^2+Y^\nu_iY^\nu_i \right) \; , \\
\delta\lambda=&
\frac{\lambda\Delta}{32\pi^2}\left(
-\frac{4\pi\alpha\left( \SW^2+3\CW^2\right)}{\SW^2\CW^2}+
4\lambda^2+2\kappa^2+3\left( Y^u_iY^u_i+Y^d_iY^d_i\right)+
Y^e_{ij}Y^e_{ij}+4Y^\nu_iY^\nu_i \right) \; , \\
\delta A^\kappa=&
\frac{3\Delta}{8\pi^2}\left(
A^\kappa \kappa^2 + A^\lambda \lambda^2 +
A^\nu_i {Y^\nu_i}^2 \right) \; , \\
\delta A^\lambda=&
\frac{\Delta}{32\pi^2}\left(
\frac{8\pi\alpha\left( 3\CW^2 M_2 + \SW^2 M_1 \right)}{\CW^2\SW^2}+
4 A^\kappa \kappa^2 + A^\lambda \left( 8\lambda^2 + Y^\nu_iY^\nu_i \right)
\right. \notag \\
&+\left. 6 \left( A^u_i{Y^u_i}^2+A^d_i{Y^d_i}^2 \right)
+ 7 A^\nu_i{Y^\nu_i}^2 + 2 A^e_{ij} {Y^e_{ij}}^2
\vphantom{\frac{8\alpha\pi\left( 3\CW^2 M_2 + \SW^2 M_1 \right)}{\CW^2\SW^2}}\right) \; , \\
\delta v^2 =&
\frac{\Delta}{32\pi^2}\left(
\frac{4\pi\alpha v^2 \left( \SW^2+3\CW^2\right)}{\SW^2\CW^2} - 
2 \left( v_d^2 \left( 3 Y^d_iY^d_i + Y^e_{ij}Y^e_{ij} \right) + 
v_u^2 \left( 3Y^u_iY^u_i + Y^\nu_iY^\nu_i \right) \right. \right.
\notag \\
&+ \left. \left. Y^e_{ij}Y^e_{ik} v_{jL}v_{kL} \right) +
2 v_{iL} Y^\nu_i \left( 2\lambda v_d - v_{jL}Y^\nu_j  \right)
\vphantom{\frac{4\pi\alpha v^2 \left( \SW^2+3\CW^2\right)}{\SW^2\CW^2}}\right) \; , \\
\delta v_{iL}^2 =&
\frac{v_{iL} \Delta}{32\pi^2} \left(
\frac{4\pi\alpha v_{iL}\left( \SW^2+3\CW^2\right)}{\SW^2\CW^2}+
2\left( v_d \lambda Y^\nu_i - v_{kL}Y^e_{ji}Y^e_{jk} -
v_{jL}Y^\nu_iY^\nu_j \right) \right) \; , \\
\delta Y^\nu_i=&
\frac{\Delta}{32\pi^2}\left(
-\frac{4\pi\alpha Y^\nu_i\left( \SW^2+3\CW^2 \right)}{\SW^2\CW^2}+
Y^\nu_i\left( 3Y^u_iY^u_i + 2\kappa^2 + 4\lambda^2 + 4 Y^\nu_jY^\nu_j 
\right)+
Y^e_{ji}Y^e_{jk}Y^\nu_k
\right) \; , \\
\delta\tb=&
\frac{\tb\Delta}{32\pi^2}\left(
3\left( Y^d_iY^d_i-Y^u_iY^u_i \right) + Y^e_{ij}Y^e_{ij} -
Y^\nu_iY^\nu_i - \frac{\lambda}{v_d} v_{iL}Y^\nu_i
\right) \label{eq:dtbdrbar} \; , \\
\delta A^\nu_i=&
\frac{\Delta}{32\pi^2 Y^\nu_i} \left(
\frac{8\pi\alpha Y^\nu_i\left( M_1 \SW^2+3 M_2 \CW^2\right)}{\SW^2\CW^2}+
Y^e_{ji}Y^e_{jk}T^\nu_k + 2T^e_{ji}Y^e_{jk}Y^\nu_k+Y^\nu_i\left(
7Y^\nu_jT^\nu_j
\right. \right. \notag \\
&+\left. \left. 6Y^u_{ij}T^u_{ij}+4\kappa^2 A^\kappa +
7\lambda^2 A^\lambda\right) +A^\nu_i \left(
\left(\lambda^2+Y^\nu_jY^\nu_j\right)Y^\nu_i -
Y^e_{ji}Y^e_{jk}Y^\nu_k
\right)
\vphantom{\frac{8\pi\alpha Y^\nu_i\left( M_1 \SW^2+3 M_2
    \CW^2\right)}{\SW^2\CW^2}}
\right)
\label{eq:danu} \; , \\
\delta\left(m_{H_d\widetilde{L}}^2\right)_i =&
-\frac{\lambda\Delta}{32\pi^2} \left(
Y^\nu_i \left( 2m_{\widetilde{\nu}}^2 + 2A^\lambda A^\nu_i +
m_{H_d}^2 + 2m_{H_u}^2 \right)
+ Y^\nu_j \left(m_{\widetilde{L}}^2\right)_{ji}
\right)
\label{eq:dmhl2} \; , \\
\delta \left(m_{\widetilde{L}}^2\right)_{ij} =&
\frac{\Delta}{32\pi^2} \left(
2 m_{H_d} Y^e_{ki}Y^e_{kj} + 2 T^e_{ki}T^e_{kj} +
Y^e_{li}Y^e_{lk}\left(m_{\widetilde{L}}^2\right)_{jk} +
2 Y^e_{ki}Y^e_{lj}\left(m_{\widetilde{e}}^2\right)_{kl}
\right. \notag \\
&+\left. \left(m_{\widetilde{L}}^2\right)_{ki}Y^e_{lk}Y^e_{lj} -
\lambda \left(m_{H_d\widetilde{L}}^2\right)_j Y^\nu_i +
Y^\nu_i \left(m_{\widetilde{L}}^2\right)_{jk} Y^\nu_k +
\left(m_{\widetilde{L}}^2\right)_{ki} Y^\nu_k Y^\nu_j
\right. \notag \\
&+ \left. Y^\nu_i Y^\nu_j \left(
2 m_{H_u}^2 + 2 m_{\widetilde{\nu}}^2 \right) +
2 T^\nu_i T^\nu_j
\right) \quad \text{for} \quad i\neq j \; .
\label{eq:dml2}
\end{align}
The counterterms in \refeqs{eq:dmue}-(\ref{eq:danu}) were all
calculated diagrammatically in this form and afterwards checked
to fulfill the one-loop relation
\begin{equation}
\delta X = \frac{\beta_X^{(1)}\Delta}{32\pi^2} \; ,
\end{equation}
where $\delta X$ stands for one of the counterterms just mentioned, and $\beta_X^{(1)}$ is the one-loop coefficient of the beta function of the parameter $X$, which could be obtained by the help of the mathematica package SARAH \cite{Staub:2010jh}.

On the contrary, the counterterms of the soft masses stated in 
\refeqs{eq:dmhl2} and (\ref{eq:dml2}) are the ones derived from
the one-loop beta function we obtained with SARAH, which were then
numerically checked to be equal to the counterterms
for $(m_{H_d\widetilde{L}}^2)_i$ and $(m_{\widetilde{L}}^2)_{ij}$ we
calculated diagrammatically in the scalar $\cp$-odd sector.

%%%%%%%%%%%%%%%%%%%%%%%%%%%%%%%%%%%%%%%%%%%%%%%%%%%%%%%%%%%%%%%%%%%%%%%%%%%%%%%
%%%%%%%%%%%%%%%%%%%%%%%%%%%%%%%%%%%%%%%%%%%%%%%%%%%%%%%%%%%%%%%%%%%%%%%%%%%%%%%

\section{Standard model values}\label{app:smvalues}

Table \ref{tab:smsum} summarizes the values for the SM-like parameters we chose in our calculation.
%%%%%%%%%%%%%%%%%%%%%%%%%% T A B L E %%%%%%%%%%%%%%%%%%%%%%%%%%%%%%%%%%%%%%%%%%
\begin{table}[H]
\renewcommand{\arraystretch}{1.7}
\centering
\begin{tabular}{c c c c c c}
 $m_t^{\overline{\text{MS}}}$ & $ m_b$ & $m_c$ & $m_s$ & $m_u$ & $m_d$ \\
 \hline
 $167.48$ & $4.2$ & $1.286$ & $0.095$ & $0.003$ & $0.006$ \\
 \hline
 \hline
 $m_\tau$ & $m_\mu$ & $m_e$ & $\MW$ & $\MZ$ & $v$ \\
 \hline
 $\num{1.7792}$ & $\num{0.105658}$ & $\num{5.10998e-4}$ & $\num{80.385}$ &
 	$\num{91.1875}$ & $\num{246.2196}$ 
\end{tabular}
\caption{Values for parameters of the standard model in GeV.}
\label{tab:smsum}
\renewcommand{\arraystretch}{1.0}
\end{table}
%%%%%%%%%%%%%%%%%%%%%%%%%% T A B L E %%%%%%%%%%%%%%%%%%%%%%%%%%%%%%%%%%%%%%%%%%
\noindent
The value for $v$ corresponds to a value for the Fermi constant of
$G_F=\SI{1.16638e-5}{GeV^{-2}}$. The values for the
gauge boson masses
define the cosine of the weak mixing angle to be
$\CW=\num{0.881535}$. Note that since the SM leptons mix
with the Higgsinos and gauginos in the \mnSSM, the lepton masses are not
the real phyiscal input parameters. However, the mixing is tiny, so there
will always be three mass eigenstates in the charged fermion sector
corresponding to the three standard model leptons, having
approximately the masses $m_e$, $m_\mu$ and $m_\tau$. This is why we use
the values for these masses from \refta{tab:smsum} and then calculate the
real input parameters, which are the Yukawa couplings
\begin{equation}
Y^e_1=\frac{\sqrt{2}m_e}{v_d} \; , \quad Y^e_2=\frac{\sqrt{2}m_\mu}{v_d} \; , 
\quad Y^e_3=\frac{\sqrt{2}m_\tau}{v_d} \; .
\end{equation}

\newpage

\bibliography{munuSSM}

\end{document}